\documentclass[a4paper,11pt]{article}
\pdfoutput=1
\usepackage{jheppub}
\usepackage[T1]{fontenc}
\usepackage{graphicx}
\usepackage{amsmath}
\usepackage{xspace}
\usepackage{subfig}
\usepackage{xcolor}
\usepackage{mathtools}
\usepackage{braket}
\usepackage{multirow}
\usepackage{lipsum}
\usepackage{slashed}

\title{Probing the Nature of Heavy Neutral Leptons in Direct Searches and Neutrinoless Double Beta Decay}

\affiliation[a]{\AddrSISSA}
\affiliation[b]{\AddrLjubljana}
\affiliation[c]{\AddrUCL}
\affiliation[d]{\AddrPitts}
\affiliation[e]{\AddrMitchell}

\author[a,b]{Patrick D. Bolton,} 
\emailAdd{patrick.bolton@ijs.si}
\author[c]{Frank F. Deppisch,} 
\emailAdd{f.deppisch@ucl.ac.uk}
\author[d,e]{Mudit Rai,} 
\emailAdd{muditrai@tamu.edu}
\author[c]{Zhong Zhang} 
\emailAdd{zhong.zhang.19@ucl.ac.uk}

\newcommand{\AddrSISSA}{SISSA, International School for Advanced Studies, \\
	INFN, Sezione di Trieste, Via Bonomea 265, I-34136 Trieste, Italy}

\newcommand{\AddrLjubljana}{Jozef Stefan Institute, Jamova 39, Ljubljana, 1000, Slovenia}

\newcommand{\AddrUCL}{Department of Physics and Astronomy, University College London,\\London WC1E 6BT, United Kingdom}

\newcommand{\AddrPitts}{Pittsburgh Particle Physics, Astrophysics, and Cosmology Center, \\
	Department of Physics and Astronomy,
	University of Pittsburgh, Pittsburgh, USA}

\newcommand{\AddrMitchell}{Mitchell Institute for Fundamental Physics and Astronomy,\\
	Department of Physics and Astronomy, Texas A\&M University, College Station, TX 77843, USA}

\abstract{Heavy Neutral Leptons (HNLs) are a popular extension of the Standard Model to explain the lightness of neutrino masses and the matter-antimatter asymmetry through leptogenesis. Future direct searches, such as fixed target setups like DUNE, and neutrinoless double beta decay are both expected to probe the regime of active-sterile neutrino mixing in a standard Seesaw scenario of neutrino mass generation for HNL masses around $m_N \lesssim 1$~GeV. Motivated by this, we analyse the complementarity between future direct searches and neutrinoless double beta decay to probe the nature of HNLs, i.e., whether they are Majorana or quasi-Dirac states, and CP-violating phases in the sterile neutrino sector. Following an analytic discussion of the complementarity, we implement a generic fixed target experiment modelling DUNE. We perform a statistical study in how a combined search for HNLs in direct searches and neutrinoless double beta decay, using DUNE and LEGEND-1000 as representative examples, can probe the nature of sterile neutrinos.}


\makeatletter
\gdef\@fpheader{\phantom{a}}
\makeatother

\begin{document}
\maketitle
\flushbottom
\newpage
\section{Introduction}
\label{sec:intro}

Heavy Neutral Leptons (HNLs) emerge in popular extensions of the Standard Model (SM). They naturally act as \emph{sterile neutrinos}, unless forbidden to do so by any additional symmetries, as they are uncharged under the SM gauge groups. This also means that they can have a Majorana mass term, giving rise to the \textit{seesaw} mechanism \cite{Minkowski:1977sc, Mohapatra:1979ia, Weinberg:1979sa, Gellmann:1980vs, Yanagida:1979as, Schechter:1980gr} of neutrino mass generation. Ever since the observation of neutrino oscillations \cite{Tanabashi:2018oca}, such a mechanism has been urgently sought as the light neutrino masses demand the existence of physics beyond the SM, either through the presence of right-handed, i.e., sterile, neutrino states or other New Physics breaking lepton number symmetry.

The conventional seesaw mechanism puts sterile neutrinos at a very high mass scale, which connects the light neutrino masses $m_\nu \sim (100~\text{GeV})^2 / \Lambda \lesssim 0.1$~eV \cite{Tanabashi:2018oca, Aghanim:2018eyx} to physics close to the Grand Unified Scale $\Lambda \sim 10^{15}$~GeV. With the realisation of low-scale leptogenesis scenarios as well as recent efforts to search for long-lived exotic particles, there is a strong interest to look for sterile neutrinos at much lower scales. To keep the active neutrinos sufficiently light, the active-sterile neutrino mixing strength $|U_{\alpha N}|$ is expected to be small,
\begin{align}
	|U_{\alpha N}| \sim \sqrt{\frac{m_\nu}{m_N}} 
	\approx 10^{-6} \sqrt{\frac{100~\text{GeV}}{m_N}}\,,
\end{align}
for a single sterile neutrino with mass $m_N$. This describes a \emph{seesaw floor} that needs to be probed experimentally to test low-scale type-I seesaw scenarios. This does not mean that differing active-sterile mixing strengths are excluded. Weaker strengths are trivially allowed by assuming there is another source of light neutrino mass generation. Stronger mixing is possible by employing specific textures \cite{Pilaftsis:1991ug, Buchmuller:1991ce, Gluza:2002vs, Pilaftsis:2004xx, Kersten:2007vk, Xing:2009in, Gavela:2009cd, He:2009ua, Adhikari:2010yt, Ibarra:2010xw, Deppisch:2010fr, Ibarra:2011xn, Mitra:2011qr} or symmetries \cite{Pilaftsis:2004xx, Shaposhnikov:2006nn, Kersten:2007vk, Deppisch:2010fr, Dev:2013oxa, Chattopadhyay:2017zvs} in a three-generation framework. By having additional copies in the sterile neutrino sector, it is likewise possible to decouple the scale of lepton number violation from the sterile neutrino mass scale. This allows large active-sterile mixing strengths for low-scale sterile neutrino masses, as realised in the inverse \cite{mohapatra:1986aw, Nandi:1985uh, mohapatra:1986bd}, linear \cite{wyler:1983dd, akhmedov:1995ip, akhmedov:1995vm, malinsky:2005bi} and generalised inverse seesaw \cite{Ma:2009du, Dev:2012sg, Dev:2012bd, Deppisch:2016jzl}. The main feature of such scenarios is that the Majorana sterile neutrinos form quasi-Dirac pairs with opposite CP phases and a small mutual mass splitting. This partially cancels their joint contribution to the light neutrino masses and other lepton number violating observables such as neutrinoless double beta ($0\nu\beta\beta$) decay.

\begin{figure}[t!]
	\centering
	\includegraphics[width=0.9\textwidth]{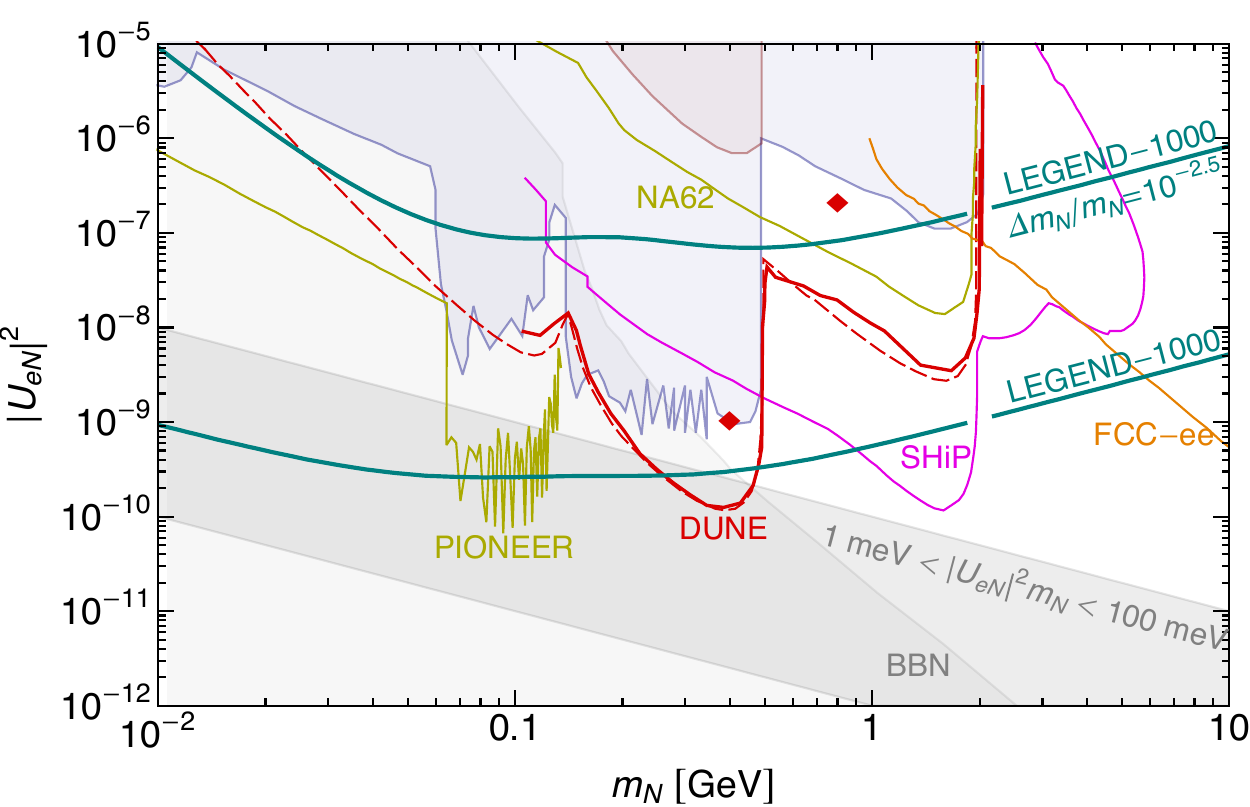}
	\caption{Projected sensitivities of future HNL searches on the electron active-sterile mixing strength $|U_{eN}|^2$ as a function of the HNL mass $m_N$. Shown are projections for the direct searches at PIONEER \cite{PIONEER:2022yag}, NA62 \cite{Drewes:2018gkc}, DUNE (red dashed from \cite{Ballett:2019bgd}, red solid from this work), SHiP \cite{SHiP:2018xqw}, FCC-ee \cite{Blondel:2022qqo} and the $0\nu\beta\beta$ decay experiment LEGEND-1000~\cite{LEGEND:2021bnm}. The latter is shown for the contribution of a single Majorana HNL and a pair of quasi-Dirac HNL with a mass splitting of $\Delta m_N/m_N = 10^{-2.5}$ generating a light neutrino mass of $m_\nu = 10^{-2.5}$~eV. The shaded region is excluded from existing searches \cite{Bolton:2019pcu, Bolton:2022pyf} and the impact of the HNL on Big Bang Nucleosynthesis (BBN) \cite{Ruchayskiy:2012si}. The diagonal band indicates the seesaw floor with the notional neutrino mass $|U_{eN}|^2 m_N$ in the given range. The diamonds indicate benchmark scenarios used in the analysis.}
\label{fig:constraints}
\end{figure}
Following our analysis \cite{Bolton:2019pcu}, we employ a general phenomenological pa\-ra\-metrisation that is agnostic to the nature of the sterile neutrinos and that only assumes that active neutrinos receive their masses at tree level. We extend our parametrisation to include three generations of active neutrinos and two sterile states, all initially described as Weyl fermions. We compare our parametrisation with that of Casas-Ibarra in the appendix and other formulations focusing on quasi-Dirac sterile neutrinos as described in \cite{Das:2012ze, Das:2017nvm}. Motivated by the fact that both future $0\nu\beta\beta$ decay and direct searches are starting to probe the seesaw floor, we analyse the complementarity between the two classes of experiments, with a focus on the mass range $m_N = \mathcal{O}(1)$~GeV. In Fig.~\ref{fig:constraints}, we show the projected sensitivity of select future searches. Considering the constraints from BBN and existing searches, HNL masses must be larger than $m_N \gtrsim 300$~MeV, and DUNE, SHiP and LEGEND-1000 can probe HNLs near the seesaw floor. Neutrinoless double beta decay is a crucial process to determine the presence of Majorana neutrinos and lepton number violation, and it can probe the mediating New Physics and its consequences in various ways \cite{Deppisch:2013jxa, Deppisch:2015yqa, Deppisch:2017ecm, Cepedello:2018zvr, Deppisch:2020mxv, Bolton:2020ncv, Bolton:2021pey}. With respect to our analysis, it is highly sensitive to the nature of HNLs, i.e., whether they are Majorana or quasi-Dirac, and a potential interference with the light neutrino mass contribution. On the other hand, direct searches are largely independent of this, mainly probing the strength of the induced SM-like charged and neutral currents. We thus use the potential of DUNE and LEGEND-1000 to probe the properties of sterile neutrinos, namely the mass splitting of a quasi-Dirac HNL, and its CP phase relative to that of the active neutrinos. 

These properties are crucial in understanding the role of HNLs in generating the matter-antimatter asymmetry of the universe through leptogenesis~\cite{fukugita:1986hr, davidson:2002qv, Canetti:2010aw, Pilaftsis:1997jf, Yamanaka:2022vdt}. In the HNL mass regime of our interest, $m_N = \mathcal{O}(1)$~GeV, cf. Fig.~\ref{fig:constraints}, successful leptogenesis can be achieved in the ARS (Akhmedov, Rubakov, Smirnov) scenario \cite{Akhmedov:1998qx}, where the observed baryon asymmetry can be achieved if the active-sterile mixing strength is in the range $10^{-10} \lesssim \sum_{\ell=e,\mu,\tau} |U_{\ell N}|^2 \lesssim 10^{-5}$ \cite{Drewes:2017zyw}. The mechanism, which relies on $CP$ violating oscillations between two or more HNLs and small active-sterile mixing strengths to keep HNLs sufficiently out-of-equilibrium, can be described consistently \cite{Klaric:2021cpi} in the same framework with that of resonant leptogenesis \cite{Pilaftsis:1997dr}. Different realizations of the scenario are described in \cite{Chun:2017spz, Barrow:2022gsu}, with further references therein. Direct searches and neutrinoless double beta decay have been compared in \cite{Hernandez:2016kel, Hernandez:2022ivz} but in different scenarios and not focussing on the specific parameters extracted from various observational pictures.

The remainder of this paper is organised as follows. In Sec.~\ref{sec:model}, we introduce the phenomenological model used in our analysis. We describe $0\nu\beta\beta$ decay in this model in Sec.~\ref{sec:0vbb} and DUNE as an example of a direct HNL search, in Sec.~\ref{sec:DUNE}. In both cases, we give a general description as well as approximate analytic formulae in the parameter space of interest. Section~\ref{sec:compare} then analyses what we can learn by combining searches in both DUNE and LEGEND-1000. We do so analytically, as well as using a Bayesian analysis incorporating statistical fluctuations. We discuss different scenarios as examples of possible experimental outcomes. In Sec.~\ref{sec:conclusions} we conclude our work with an outlook, and the Appendices~\ref{app:phenom_param} and \ref{app:pheno_vs_CI_param} provide an extension of our parametrisation and a comparison with the Casas-Ibarra parametrisation, respectively.

\section{Phenomenological Model}
\label{sec:model}

In this section we will discuss the addition of two SM-singlet Weyl fermions to the SM. Building on the work of~\cite{Bolton:2019pcu}, we will outline a useful phenomenological parametrisation which can describe the Majorana and quasi-Dirac limits of the HNL pair and their mixing with the light active neutrinos. We will derive the relations between the active-sterile mixing and CP phases that are necessary in order to generate the observed light neutrino oscillation parameters. We keep the number of active generations $\mathcal{N}_A$ general in what follows, for the purpose of comparing $0\nu\beta\beta$ decay and direct searches at DUNE in Sec.~\ref{sec:compare}, and we will consider one or three generations of SM fermions.

\subsection{HNL Lagrangian}

Extending the SM to include $\mathcal{N}_S$ SM-singlet Weyl fermion fields $N_{iR}$, $i = 1,\ldots,\mathcal{N}_S$, the following renormalizable terms can be added to the SM Lagrangian,
\begin{align}
\label{eq:SMplusN}
    \mathcal{L} &= \mathcal{L}_{\text{SM}} + i \bar{N}_{iR}\slashed{\partial}N_{iR} - (Y_\nu)_{\alpha i} \bar{L}_{\alpha} \tilde{H} N_{iR} -\frac{1}{2}(\mathcal{M}_S)_{ij}\bar{N}^c_{iR} N_{jR} + \text{h.c.}\,.
\end{align}
Here, $L_{\alpha} = (\nu_{\alpha L},\ell_{\alpha L})^T$ is the SM lepton doublet of flavour $\alpha  = e,\mu,\tau$, $H = (H^0,H^{-})^T$ is the SM Higgs doublet ($\tilde{H} = i\sigma_2 H$, where $\sigma_2$ is the second Pauli matrix), $Y_\nu$ is the neutrino Yukawa coupling matrix, and $\mathcal{M}_S$ is the Majorana mass matrix between the sterile states. It is convenient to rewrite the Lagrangian in Eq.~\eqref{eq:SMplusN} as
\begin{align}
    \mathcal{L} &= \mathcal{L}_{\text{SM}} + i \bar{N}_{iR}\slashed{\partial}N_{iR} - \frac{1}{2}(\mathcal{M}_{\nu})_{ij}\bar{n}_{iL}n^{c}_{jL} + \text{h.c.}\,,
\end{align}
where $n_{L} \equiv (\nu_{\alpha L}, N_{iR}^c)^{T}$ and the indices $(i,j)$ now run over $\mathcal{N}=\mathcal{N}_A+\mathcal{N}_S$ states. Here, $\mathcal{M}_\nu$ is an $\mathcal{N}\times \mathcal{N}$ complex symmetric matrix,
\begin{align}
\label{eq:Mnu}
    \mathcal{M}_\nu = 
    \begin{pmatrix}
    0  & \mathcal{M}_{D} \\
    \mathcal{M}_{D}^{T} & \mathcal{M}_S
    \end{pmatrix}\,,
\end{align}
where $\mathcal{M}_{D} = \frac{v}{\sqrt{2}}Y_\nu$ and $v = \langle H^0 \rangle \approx 246$~GeV is the Higgs vacuum expectation value. Without loss of generality, it is possible to perform a unitary rotation among the sterile states so that $V^{T}\mathcal{M}_S V$ is diagonal, where $V$ is an $\mathcal{N}_S\times \mathcal{N}_S$ unitary matrix. Alternatively, by exploiting the singular value decomposition of $\mathcal{M}_D$, the rotation among sterile states can be chosen such that $\mathcal{M}_D V = W^{*} \Sigma_D$, where $W$ is an $\mathcal{N}_A\times \mathcal{N}_A$ unitary matrix and $\Sigma_D$ is an $\mathcal{N}_A\times \mathcal{N}_S$ matrix with non-negative real numbers along the diagonal~\cite{Ma:2009du}.

More generally, it is possible to rotate the active and sterile states to the mass basis by performing a unitary rotation on $n_L$, i.e.,
\begin{align}
\label{eq:nL}
    n_L = P_L U (\underbrace{\nu_1, \ldots}_{\mathcal{N}_A}, \underbrace{N_1, \ldots}_{\mathcal{N}_S})^T\,,
\end{align}
where $\nu_i = \nu_{i}^c$ and $N_{\kappa} = N_{\kappa}^c$ are Majorana fields and $U$ is the $\mathcal{N}\times \mathcal{N}$ unitary matrix that diagonalises the mass matrix $\mathcal{M}_\nu$. We are interested in the limit where the $\nu_i$ are \textit{mostly-active} states and $N_{\kappa}$ are \textit{mostly-sterile} states (or HNLs). The former have masses set by the neutrino oscillation data and mix primarily with active neutrino fields $\nu_{\alpha L}$, while the latter have arbitrary masses and mix predominantly with the sterile fields $N^c_{kR}$. 

In the $\mathcal{N}_S = 2$ case, we can express the mass matrix as
\begin{align}
\label{eq:Mnu_2}
    \mathcal{M}_\nu = \begin{pmatrix}
    0  & \mathcal{M}_{D,1} & \mathcal{M}_{D,2} \\
    \mathcal{M}_{D,1}^{T} & \mu_R & m_S \\ 
    \mathcal{M}_{D,2}^{T} & m_S & \mu_S
    \end{pmatrix}
    = U
    \begin{pmatrix}
    m_\nu & 0 & 0 \\ 
    0 & m_N & 0 \\ 
    0 & 0 & m_N(1+r_\Delta)
    \end{pmatrix} U^T\,,
\end{align}
where $(\mathcal{M}_{D,1})_{\alpha} = (\mathcal{M}_{D})_{\alpha 1}$ and $(\mathcal{M}_{D,2})_{\alpha} = (\mathcal{M}_{D})_{\alpha 2}$. Here, $m_\nu$ is an $\mathcal{N}_A\times\mathcal{N}_A$ diagonal matrix containing the observed light neutrino masses, $m_N$ is the mass of the lighter HNL, and $r_\Delta = \Delta m_N/m_N$ (where $\Delta m_N$ is the mass splitting between the HNL pair). We make the usual observation that for $\mathcal{N}_A+\mathcal{N}_S = 3 + 2$, the rank of the matrix $\mathcal{M_\nu}$ is 4, and therefore one of the light neutrinos is massless. This scenario is compatible with the neutrino oscillation data for both normal and inverted neutrino mass ordering.

A non-zero mixing between the active neutrinos and possible sterile states is felt by the SM charged- and neutral-current interactions. For example, taking the charged lepton Yukawa matrix $Y_{\ell}$ to be diagonal, the charged-current can be written as
\begin{align}
\label{eq:charged_current}
    \mathcal{L}_{W^\pm} &= -\frac{g}{\sqrt{2}} \bar{\ell}_{\alpha L}
    \slashed{W}^{\dagger} \nu_{\alpha L} + \text{h.c.} \nonumber \\
    & = -\frac{g}{\sqrt{2}} U_{\alpha i} \bar{\ell}_{\alpha L}
    \slashed{W}^{\dagger} P_L \nu_{i}-\frac{g}{\sqrt{2}} U_{\alpha N_{\kappa}} \bar{\ell}_{\alpha L}
    \slashed{W}^{\dagger} P_L  N_{\kappa} + \text{h.c.}\,,
\end{align}
where we see that the mixing $U_{\alpha N_{\kappa}}$ couples charged leptons to HNLs. This interaction allows HNLs to, for example, mediate $0\nu\beta\beta$ decay and be produced via meson decays.

It is conventional to express an $\mathcal{N}\times \mathcal{N}$ unitary matrix in terms of $\mathcal{N}(\mathcal{N}-1)/2$ mixing angles and $\mathcal{N}(\mathcal{N}+1)/2$ phases. For the mixing matrix $U$, $\mathcal{N}_A$ phases can be eliminated through a re-phasing of the charged lepton fields in Eq.~\eqref{eq:charged_current}. Note that only the upper $\mathcal{N}_A\times\mathcal{N}$ sub-block of $U$ appears in Eq.~\eqref{eq:charged_current}, and so all mixing angles and phases that do not appear in this sub-block are therefore unphysical. This freedom is equivalent to the arbitrary rotation among the sterile states discussed below Eq.~\eqref{eq:Mnu}. The mixing matrix $U$ can be expressed as a combination of real and complex Euler rotations; for example, in the $3 + 2$ scenario, a useful parametrisation of $U$ is~\cite{Schechter:1979bn,Abada:2018qok}
\begin{align}
\label{eq:Uparam}
    U &= W_{N_1 N_2} W_{3 N_2} W_{2 N_2} W_{1 N_2} W_{3 N_1} W_{2 N_1} W_{1 N_1} R_{23} W_{13} R_{12} D\,.
\end{align}
Here, $W_{ij}$ ($R_{ij}$) is a unitary, unimodular matrix describing a complex (real) rotation in the $i$-$j$ plane, which can be written as
\begin{align}
\label{eq:Wij}
    [W_{ij}]_{rs} = \delta_{rs} &+ (\cos\vartheta_{ij}-1)(\delta_{ir}\delta_{is}+\delta_{jr}\delta_{js}) \nonumber\\
    &+ \sin\vartheta_{ij}(e^{-i\eta_{ij}}\delta_{ir}\delta_{js}-e^{i\eta_{ij}}\delta_{ir}\delta_{js})\,,
\end{align}
with $\vartheta_{ij}\in[0,\pi/2]$ and $\eta_{ij}\in[0,2\pi]$, while $R_{ij}$ is given by Eq.~\eqref{eq:Wij} with $\eta_{ij} = 0 $. The angles $\vartheta_{12}$, $\vartheta_{13}$ and $\vartheta_{23}$ and phase $\eta_{13}$ contained in $R_{12}$, $W_{13}$ and $R_{23}$ are taken to be the mixing angles and Dirac CP phase in the PMNS mixing matrix. The matrices $W_{i N_{\kappa}}$ in Eq.~\eqref{eq:Uparam} contain six angles $\vartheta_{i N_{\kappa}}$, which control the active-sterile mixing strengths, and six Dirac CP phases $\eta_{i N_{\kappa}}$; for convenience, we introduce the notation $ \vartheta_{i\kappa}\equiv\vartheta_{iN_\kappa}$, $ c_{i\kappa}\equiv\cos\vartheta_{iN_\kappa}$, $s_{i\kappa}\equiv\sin\vartheta_{iN_\kappa}$ and $\eta_{i\kappa}\equiv \eta_{i N_\kappa}$. Finally, four (Majorana) CP phases which are defined to lie in the range $[0,4\pi]$ are contained in the diagonal matrix
\begin{align}
\label{eq:maj_phases}
    D = \text{diag}(1, e^{i\alpha_{21}/2},e^{i\alpha_{31}/2},e^{i\phi_1/2},e^{i\phi_2/2})\,,
\end{align}
where $\alpha_{21}$ and $\alpha_{31}$ correspond to the usual Majorana phases in the PMNS mixing matrix. In the $3+2$ model where $m_{\text{lightest}} = 0$, only a single Majorana phase appearing in the PMNS mixing matrix is physically relevant; this is the combination $(\alpha_{21}-\alpha_{31})$ in the normal ordering (NO) scenario, where $m_1 = 0$, and $\alpha_{21}$ in the inverted ordering (IO) scenario, where $m_{3} = 0$. Both scenarios are covered by setting $\alpha_{31} = 0$.

In the neutrino mass matrix $\mathcal{M}_{\nu}$ in Eq.~\eqref{eq:Mnu_2}, the upper-left $\mathcal{N}_{A}\times \mathcal{N}_{A}$ sub-block is zero because a mass term of the form $\bar{\nu}_{\alpha L}\nu_{\beta L}^c$ violates $\text{SU}(2)_L$. Equating the two sides of Eq.~\eqref{eq:Mnu_2} gives the following relation between the light neutrino masses $m_i$ and HNL masses $m_{N_{\kappa}}$,
\begin{align}
    \label{eq:Mnu=0}
    (\mathcal{M}_\nu)_{\alpha\beta} = 0 \quad\Rightarrow\quad 0 = \sum_{i = 1}^{\mathcal{N}_{A}}  m_i U_{\alpha i}U_{\beta i} + \sum_{k = 1}^{\mathcal{N}_{S}} m_{N_{\kappa}} U_{\alpha N_{\kappa}}U_{\beta N_{\kappa}}\,.
\end{align}
However, this relation is only valid at tree level. The light neutrino masses also acquire a contribution at one-loop from self-energy diagrams involving Higgs and $Z$ bosons~\cite{Pilaftsis:1991ug,Grimus:2002nk,Fernandez-Martinez:2015hxa}. For HNL masses $m_{N_{\kappa}}\lesssim 1$~GeV and active-sterile mixing $|U_{eN_{\kappa}}|^2 \lesssim 10^{-6}$ probed by $0\nu\beta\beta$ decay and DUNE, such one-loop corrections can be safely neglected~\cite{Bolton:2019pcu}. 

In the limit $s_{i \kappa}\approx\vartheta_{i \kappa}\ll 1$ and $c_{i \kappa}\approx 1$, the upper-left $\mathcal{N}_{A}\times \mathcal{N}_{A}$ sub-block of $U$ is approximately the PMNS mixing matrix, i.e. $U_{\alpha i} \approx (U_{\text{PMNS}})_{\alpha i}$. Deviations from the unitarity of $U_\nu$ are controlled by the size of the angles $\vartheta_{i \kappa}$, but it is safe to ignore this effect in the regime probed by $0\nu\beta\beta$ decay and DUNE. The active-sterile mixing in Eqs.~\eqref{eq:charged_current} and~\eqref{eq:Mnu=0} is now given by
\begin{align}
   U_{\alpha N_{\kappa}} \approx \Theta_{\alpha \kappa} = |\Theta_{\alpha\kappa}|e^{i\phi_{i\kappa}/2} \equiv s_{i\kappa}e^{i\phi_{i\kappa}/2} \,,
\end{align}
where $\phi_{i \kappa}\equiv \phi_{\kappa}-2\eta_{i \kappa}$. In the limit $\vartheta_{i \kappa}\ll 1$, there is a correspondence between the mass index $i$ and the flavour index $\alpha$, i.e., $e,\mu,\tau\Leftrightarrow 1,2,3$, and so in the remainder of this work we use $s_{\alpha k}$ and $\phi_{\alpha k}$. The phase $\phi_{\alpha \kappa}$ controls the CP-violating nature of the charged-current interaction between $\ell_{\alpha}$ and $N_{\kappa}$ in Eq.~\eqref{eq:charged_current}\footnote{Enforcing the diagonalised neutrino mass matrix to be invariant under CP, the HNL mass eigenstate $N_{\kappa}$ transforms as $\text{U}_{\text{CP}}N_{\kappa}\text{U}_{\text{CP}}^{-1} = \pm i \gamma^0 N^c_{\kappa}$. The charged-current between $\ell_{\alpha}$ and $N_{\kappa}$ in Eq.~\eqref{eq:charged_current} then transforms as $\text{U}_\text{CP}\mathcal{L}_{W^{\pm}} \text{U}_\text{CP}^{-1} \supset \mp\frac{g}{\sqrt{2}} i (\xi_{\ell_\alpha}^{\text{CP}})^*\bar{N}_{\kappa} \slashed{W} P_L U_{\alpha N_{\kappa}} \ell_{\alpha L} + \text{h.c.}$, where $(\xi_{\ell_\alpha}^\text{CP})^*$ is the CP parity of $\ell_{\alpha}$. Taking $\xi_{\ell_\alpha}^{\text{CP}} = i$ implies that CP is conserved for $U_{\alpha N_{\kappa}}^* = \pm U_{\alpha N_{\kappa}}$, or equivalently for $e^{i \phi_{\alpha \kappa}} = \pm 1$. The arbitrariness of $\xi_{\ell_\alpha}^{\text{CP}}$ is eliminated when considering relative phase differences, e.g. $\Delta\phi_{\alpha} = \phi_{\alpha 2} - \phi_{\alpha 1}$. CP is conserved for $e^{i\Delta\phi_{\alpha}} = \pm 1$.}. With these simplifications, the relation in Eq.~\eqref{eq:Mnu=0} can be written in the $\mathcal{N}_S = 2$ case as
\begin{align}
\label{eq:constraint}
	m^{\nu}_{\alpha\beta} + m_N \Theta_{\alpha 1} \Theta_{\beta 1} 
	+ m_N (1+r_\Delta) \Theta_{\alpha 2} \Theta_{\beta 2} = 0\,,
\end{align}
where $m_{\alpha\beta}^{\nu} \equiv \sum^3_i m_i U_{\alpha i}U_{\beta i}$. In the next subsection, we will use this result to derive relations between the active-sterile mixing strengths $s_{\alpha k}$ to reproduce the light neutrino masses, mixing angles and phases contained in $m^{\nu}_{\alpha\beta}$.

\subsection{Active-Sterile Mixing Ratio Formulae}
\label{subsec:mixratio}

If a pair of HNLs exists in nature, their masses ($m_N$, $r_\Delta$), active-sterile mixing ($|\Theta_{\alpha 1}|$, $|\Theta_{\alpha 2}|$) and CP phases ($\phi_{\alpha 1}$, $\phi_{\alpha 2}$), for the flavour $\alpha = e,\mu,\tau$ can in principle be measured experimentally. Furthermore, if the HNL pair is the sole (or dominant) contribution to the light neutrino masses and mixing, Eq.~\eqref{eq:constraint} implies that there are correlations among these measurable quantities. If an HNL pair is found that does not follow Eq.~\eqref{eq:constraint}, it cannot be the only contribution to the light neutrino masses. For example, there may exist more than two HNLs; we generalise Eq.~\eqref{eq:constraint} and the results derived below to the $3+\mathcal{N}_S$ model in Appendix~\ref{app:phenom_param}.

We can insert into the neutrino mass matrix $m_{\alpha\beta}^{\nu}$  the light neutrino masses, mixing angles and Dirac CP phase inferred from the neutrino oscillation data; however, the Majorana phases $\alpha_{21}$ and $\alpha_{31}$ remain unconstrained. Taking the lightest neutrino to be massless, as implied in the $3+2$ model, we have
\begin{align}
    m_{\alpha\beta}^{\nu} = m_{i}U_{\alpha i}U_{\beta i} + m_{j}U_{\alpha j}U_{\beta j}\,,
\end{align}
where $(i,j) = (2,3)$ and $(1,2)$ in the NO and IO scenarios, respectively. Now only a single unconstrained Majorana phase is physically relevant, which we can choose to be $\alpha_{21}$.

We first consider the constraint $(\mathcal{M}_{\nu})_{\alpha\beta} = 0$ for $\alpha = \beta$, i.e., the constraints along the diagonal of Eq.~\eqref{eq:Mnu_2}. The expression in Eq.~\eqref{eq:constraint} can be rearranged to give
\begin{align}
\label{eq:diag}
	\frac{\Theta_{\alpha 2}^2}{\Theta_{\alpha 1}^2} = - \frac{1+x_{\alpha\alpha}^{\alpha}}{1+r_\Delta}\,; \quad x_{\alpha\beta}^{\rho} \equiv \frac{m_{\alpha\beta}^{\nu}}{m_N \Theta_{\rho 1}^2}\,.
\end{align}
Using $\Theta_{\alpha \kappa} = |\Theta_{\alpha \kappa}|e^{i\phi_{\alpha\kappa}/2} \equiv s_{\alpha \kappa}e^{i\phi_{\alpha\kappa}/2}$, it is now straightforward to find
\begin{align}
\label{eq:sratio}
	\frac{|\Theta_{\alpha 2}|^2}{|\Theta_{\alpha 1}|^2} = \frac{|1 +x_{\alpha\alpha}^{\alpha}|}{1+r_\Delta}\,, 
	\quad
	\cos\Delta\phi_{\alpha} = -\frac{\text{Re}[1+x_{\alpha\alpha}^{\alpha}]}{|1+x_{\alpha\alpha}^{\alpha}|}
	= \frac{|x_{\alpha\alpha}^{\alpha}|^2 - 1 - |1+x_{\alpha\alpha}^{\alpha}|^2}{2|1+x_{\alpha\alpha}^{\alpha}|}\,,
\end{align}
where $\Delta\phi_{\alpha} = \phi_{\alpha 2} - \phi_{\alpha 1}$ is the CP phase difference. Written in this way, we see that in order to reproduce the light neutrino data in $m_{\alpha\alpha}^{\nu}$ (up to a value of the Majorana phase $\alpha_{21}$), the values of $|\Theta_{\alpha 2}|$ and $\cos\Delta\phi_{\alpha}$ are fixed by $m_N$, $r_\Delta$, $|\Theta_{\alpha 1}|$ and $\phi_{\alpha 1}$. 

If we instead consider $(\mathcal{M}_{\nu})_{\alpha\beta} = 0$ for $\alpha \neq \beta$, i.e., the off-diagonal constraints of Eq.~\eqref{eq:Mnu_2}, we obtain
\begin{align}
\label{eq:off-diag}
    \frac{\Theta_{\beta 1}}{\Theta_{\alpha 1}} =  \frac{x_{\alpha\beta}^{\alpha} \pm \sqrt{(x_{\alpha\beta}^{\alpha})^2-x_{\alpha\alpha}^{\alpha}x_{\beta\beta}^{\alpha}}\sqrt{1+x_{\alpha\alpha}^{\alpha\phantom{!}}}}{x_{\alpha\alpha}^{\alpha}}\equiv y_{\alpha\beta}^{\alpha}\,,
\end{align}
which can be used to find
\begin{align}
	\frac{|\Theta_{\beta 1}|^2}{|\Theta_{\alpha 1}|^2} 
	= |y_{\alpha\beta}^{\alpha}|^2\,,\quad \cos(\phi_{\beta 1}-\phi_{\alpha 1}) 
	= \frac{\text{Re}\big[(y_{\alpha\beta}^{\alpha})^2\big]}{|y_{\alpha\beta}^{\alpha}|^2}\,.
\end{align}
Combining Eqs.~\eqref{eq:diag} and \eqref{eq:off-diag}, it is also possible to write,
\begin{align}
\label{eq:combined_ratio}
    \frac{\Theta_{\beta 2}}{\Theta_{\alpha 1}} = \pm i\sqrt{\frac{(y_{\alpha\beta}^{\alpha})^2+x_{\beta\beta}^{\alpha}}{1+r_\Delta}}\,,
\end{align}
and therefore all active-sterile mixing strengths $|\Theta_{\beta i}|$ and CP phases $\phi_{\beta i}$ (for $\alpha\neq\beta$) can be written in terms of $m_{\alpha\beta}^{\nu}$ (containing the light neutrino masses, mixing angles and Dirac CP phase from the neutrino oscillation data and a single Majorana phase $\alpha_{21}$ for $m_{\text{light}} = 0$), the HNL masses $m_{N}$ and $m_{N}(1+r_\Delta)$, a single active-sterile mixing $|\Theta_{\alpha 1}|$ and CP phase $\phi_{\alpha 1}$ for a particular flavour $\alpha$. We note that this approach is compatible with the Casas-Ibarra parametrisation~\cite{Casas:2001sr}, which splits the mixing matrix $U$ into sub-blocks and uses the relation $(\mathcal{M}_{\nu})_{\alpha\beta} = 0$ to express the active-sterile mixing strengths in terms of the light neutrino masses, mixing angles and phases, HNL masses and arbitrary angles in a complex orthogonal matrix $R$. Changing the sign of the square root in Eq.~\eqref{eq:combined_ratio} corresponds to an unphysical shift in the CP phases. In Appendix~\ref{app:pheno_vs_CI_param}, we compare the phenomenological approach summarised in this section to the minimal $3+2$ parametrisation of~\cite{Donini:2012tt}.

An interesting limit to consider for Eqs.~\eqref{eq:diag}~and~\eqref{eq:off-diag} is $|x^{\alpha}_{\alpha\beta}|\ll 1$, or equivalently $|\Theta_{\alpha 1}|^2 \gg |m_{\alpha\beta}^{\nu}|/m_N$, which is exactly the \textit{inverse seesaw} limit. With the lightest neutrino being massless in the $3+2$ model, we obtain
\begin{align}
\label{eq:sratio_ISS}
	|\Theta_{\alpha 2}|^2 \approx \frac{|\Theta_{\alpha 1}|^2}{1+r_\Delta}\,, \quad
	\cos\Delta\phi_{\alpha} 
	= -1\,,\quad \frac{\Theta_{\beta 1}}{\Theta_{\alpha 1}} 
	\approx \frac{\sqrt{m_i}\,U_{\beta i}\mp i\sqrt{m_j}\,U_{\beta j}}{\sqrt{m_i}\,U_{\alpha i}\mp i\sqrt{m_j}\,U_{\alpha j}}\,,
\end{align}
where $(i,j) = (2,3)$ and $(1,2)$ in the NO and IO scenarios, respectively. As discussed in Appendix~\ref{app:phenom_param}, the choice of sign in Eqs.~\eqref{eq:off-diag} and \eqref{eq:combined_ratio}, and therefore $y_{\alpha\beta}^{\nu}$ in Eq.~\eqref{eq:sratio_ISS}, is arbitrary. In Eqs.~\eqref{eq:combined_ratio} and \eqref{eq:sratio_ISS}, we choose the positive sign and allow the physically relevant Majorana phase $\alpha_{21}$ in the $U_{\alpha i}$ to lie in the range $[0,4\pi]$, which takes into account both values of of the sign.

\begin{figure}[t!]
	\centering
	\includegraphics[width=0.50\textwidth]{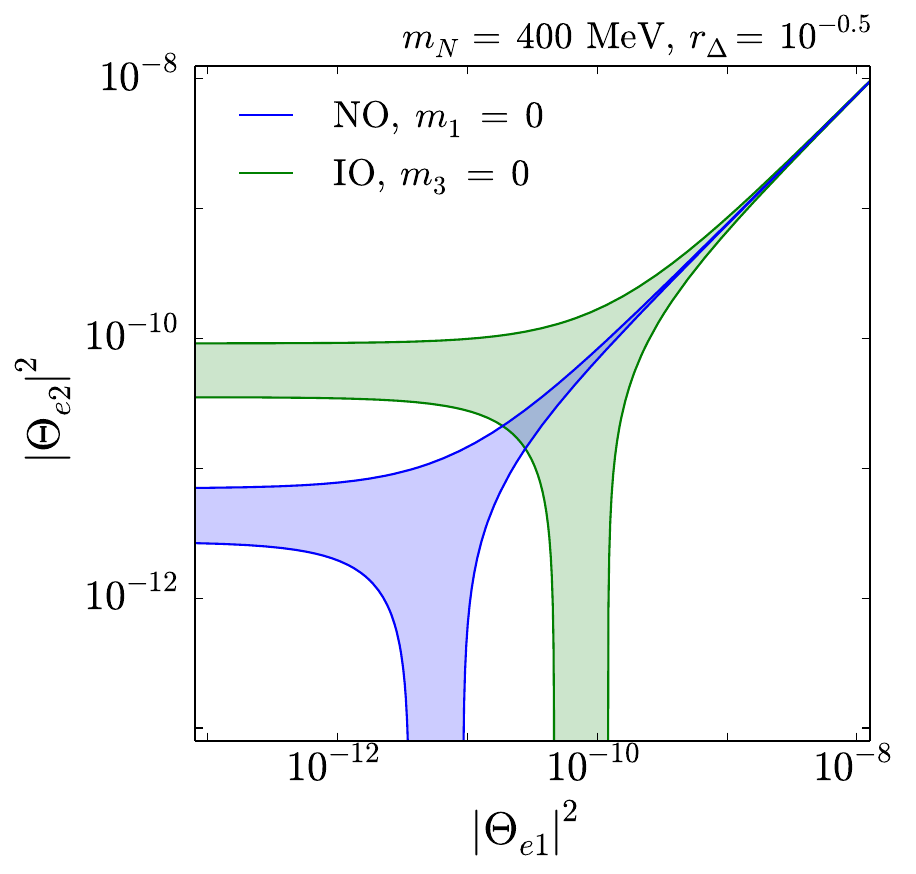}
	\includegraphics[width=0.49\textwidth]{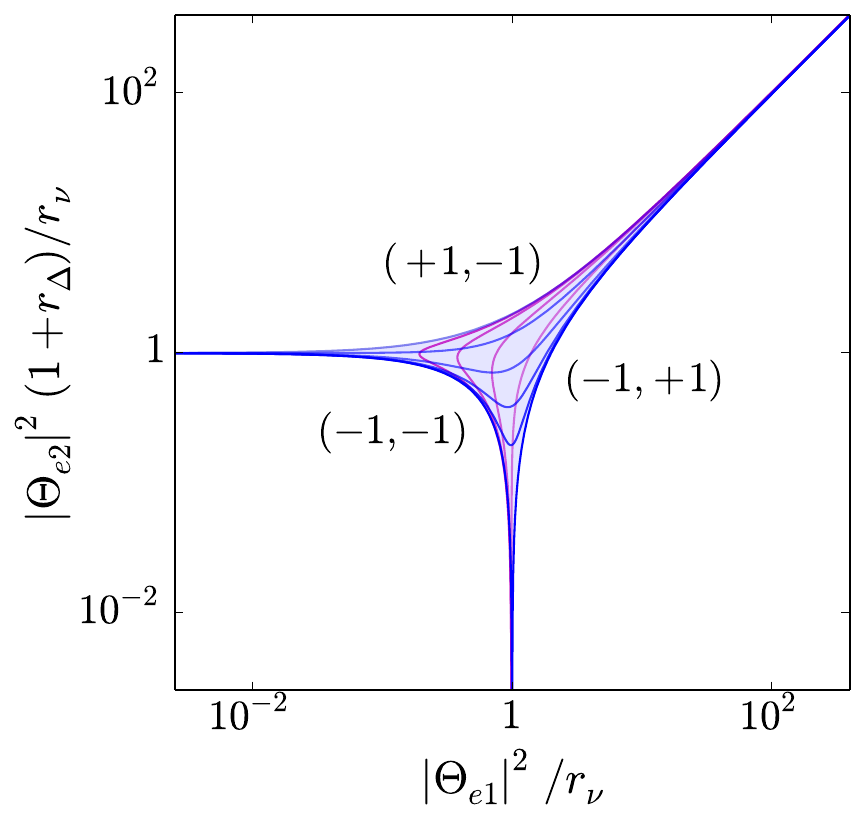}
	\caption{(Left) Regions in the $(|\Theta_{e1}|^2, |\Theta_{e2}|^2)$ plane that are compatible with Eq.~\eqref{eq:diag} in the $3+2$ model, showing the NO (blue) and IO (green) cases for $m_N = 400$~MeV, $r_\Delta = 10^{-0.5}$ and $m_{\text{lightest}} = 0$. (Right) Region in the $(|\Theta_{e1}|^2/r_{\nu}, |\Theta_{e2}|^2(1+r_\Delta)/r_{\nu})$ plane that is compatible with Eq.~\eqref{eq:diag} in the simple $1+2$ model. The blue and magenta lines are contours of constant $\cos\phi_{e1}$ and $\cos\phi_{e2}$, respectively; their values on the edges of the allowed region are shown in parentheses.}
\label{fig:se1_se2_plot}
\end{figure}

In Fig.~\ref{fig:se1_se2_plot} (left), we depict two regions in the $(|\Theta_{e1}|^2,|\Theta_{e2}|^2)$ plane that are compatible with the $(\mathcal{M}_\nu)_{ee} = 0$ constraint for fixed $m_{N} = 400$~MeV and $r_{\Delta} = 10^{-0.5}$, but allowing $\alpha_{21}$ and $\phi_{e1}$ to vary. The blue and green shaded regions are for the NO and IO scenarios, respectively. We see that the regions can tend to small $|\Theta_{e1}|^2$ or $|\Theta_{e2}|^2$ values while the other mixing strength remains constant. To the upper right, the region can also tend to large values of both mixing strengths; here, the ratio $|\Theta_{e2}|^2/|\Theta_{e1}|^2$ follows the relation in Eq.~\eqref{eq:sratio_ISS}. These limits will be discussed in more detail in the next subsection.

\subsection{Approximate Ratio Formulae}
\label{subsec:approxmixratio}

We will now examine Eq.~\eqref{eq:diag} in the simple $1+2$ model, with the active neutrino flavour $\alpha = e$. These results will allow a broad strokes comparison of $0\nu\beta\beta$ decay and DUNE in the following sections. The light neutrino mass matrix element $m_{\alpha\beta}^{\nu}$ is simply replaced by $m_{\nu}$, which gives the expression
\begin{align}
\label{eq:se2}
	|\Theta_{e 2}|^2 
	= \frac{|r_{\nu} + |\Theta_{e1}|^2 e^{i\phi_{e1}}|}{1+r_\Delta} 
	= \frac{\sqrt{r_{\nu}^2+|\Theta_{e1}|^4+2r_{\nu}|\Theta_{e1}|^2\cos\phi_{e1}}}{1+r_\Delta}\,,
\end{align}
where we have defined $r_\nu\equiv m_\nu/m_N$. In  Fig.~\ref{fig:se1_se2_plot} (right), we use Eq.~\eqref{eq:se2} to plot the combination $|\Theta_{e2}|^2(1+r_\Delta)/r_\nu$ as a function of $|\Theta_{e1}|^2/r_\nu$ for different values of $\cos\phi_{e1}$. The blue region formed for $\cos\phi_{e1}\in [-1,1]$ indicates the parameter space that is compatible with $(\mathcal{M}_{\nu})_{ee} = 0$. Every point in the parameter space has an associated value of $\cos\phi_{e1}$ and $\cos\phi_{e2}$; in Fig.~\ref{fig:se1_se2_plot} (right), the values of these phases on the extremities of the allowed region are shown. Likewise, the value of $\cos\Delta\phi_e$ is given by
\begin{align}
\label{eq:cosphie}
	\cos\Delta\phi_{e} 
	&= \frac{r_\nu^2 - |\Theta_{e1}|^4 - (1+r_\Delta)^2|\Theta_{e2}|^4}{2(1+r_\Delta)|\Theta_{e1}|^2|\Theta_{e2}|^2} \nonumber\\
	&= \frac{r_\nu^2 - |\Theta_{e1}|^4 - ||\Theta_{e1}|^2 
		+ r_{\nu}e^{-i\phi_{e1}}|^2}{2|\Theta_{e1}|^2||\Theta_{e1}|^2 +r_{\nu}e^{-i\phi_{e1}}|} \nonumber\\
	&= -\frac{|\Theta_{e1}|^2 
		+ r_\nu \cos\phi_{e1}}{\sqrt{r_{\nu}^2+|\Theta_{e1}|^4+2r_{\nu}|\Theta_{e1}|^2\cos\phi_{e1}}}\,.
\end{align}
The relevant limits of Eq.~\eqref{eq:se2} are made clear in the figure; firstly, the thin extensions to the left and bottom of the allowed region indicate that the active-sterile mixing strengths tend to constant values when the other mixing is much smaller than $r_{\nu}$, i.e.,
\begin{align}
	|\Theta_{e1}|^2 &= r_\nu\,,~\cos\phi_{e1} = -1 \hspace{3.1em} (|\Theta_{e2}|^2 \ll r_\nu/(1+r_\Delta))\,, \\
	|\Theta_{e2}|^2 &= \dfrac{r_\nu}{1 + r_\Delta}\,,~\cos\phi_{e2} = -1 \quad (|\Theta_{e1}|^2 \ll r_\nu)\,.
\end{align}
This is exactly the \textit{standard seesaw} scenario, where one of the HNL states generates the light neutrino mass at tree level, while the other decouples. In each case, the value of one phase is constrained, while the phase difference can take any value. Secondly, the thin extension to the upper right of the allowed region suggests that the active-sterile mixing can also tend to values much larger than $r_\nu$. In this limit,
\begin{align}
	|\Theta_{e2}|^2 
	= \dfrac{|\Theta_{e1}|^2}{1 + r_\Delta}\,,~
	\cos\Delta\phi_{e}=-1 \quad (|\Theta_{e1}|^2 \gg r_\nu) \,.
\end{align}
This is the \textit{inverse seesaw} scenario, where both HNL states have a large active-sterile mixing, but a cancellation ensures that their combined contribution to the tree level neutrino mass is small.

\section{Neutrinoless Double Beta Decay}
\label{sec:0vbb}

In this section we will examine the $0\nu\beta\beta$ decay in the phenomenological parametrisation set out in Sec.~\ref{sec:model}. Firstly, we will provide a general formula for the $0\nu\beta\beta$ decay half-life in the $3+2$ model, examining how an experimental measurement of $0\nu\beta\beta$ decay can be used to constrain the active-sterile mixing and phases for a given HNL mass $m_N$ and mass splitting ratio $r_\Delta$. We will then simplify to the $1+2$ scenario, considering only the first generation. Various limits of the resulting formulae will be investigated.

\subsection{In the Phenomenological Model}

The $0\nu\beta\beta$ decay rate $\Gamma_{0\nu}$ and half-life $T^{0\nu}_{1/2}$, taking into the account the exchange of $\mathcal{N}_{A}$ active neutrinos and $\mathcal{N}_S$ sterile neutrinos, can be written as
\begin{align}
\label{eq:0vbb_rate}
    \frac{\Gamma_{0\nu}}{\ln 2} 
    = \frac{1}{T^{0\nu}_{1/2}} 
    = \frac{G_{0\nu}g_A^4}{m_e^2}\bigg|\sum^{\mathcal{N}_A}_i U_{ei}^2m_i \mathcal{M}^{0\nu}(m_i) 
    + \sum^{\mathcal{N}_S}_{\kappa}U_{e N_{\kappa}}^2 m_{N_{\kappa}}\mathcal{M}^{0\nu}(m_{N_{\kappa}})\bigg|^2\,,
\end{align}
where $G_{0\nu}$ is the kinematic phase space factor for the $0\nu\beta\beta$ decay isotope, $g_A$ is the axial coupling strength, $m_e$ is the electron mass, and $\mathcal{M}^{0\nu}(m_i)$ is the nuclear matrix element (NME) of the process, which depends on the mass $m_i$ of the exchanged neutrino and encodes the non-trivial transition between initial and final-state nuclei. 

The typical energy scale of $0\nu\beta\beta$ decay, $k_F \sim 100$~MeV (the Fermi momentum of a nucleus), is much smaller than the electroweak scale. The first step in deriving Eq.~\eqref{eq:0vbb_rate} is therefore to write the interactions between the quarks, outgoing electrons and exchanged neutrino as the effective Fermi interaction, i.e., the low-energy limit of the charged-current interaction in Eq.~\eqref{eq:charged_current}. Due to the non-perturbative nature of QCD, this description of $0\nu\beta\beta$ decay in terms of quarks and gluons breaks down below the~GeV scale. It becomes necessary to use chiral perturbation theory to characterise pions and nucleons as degrees of freedom below the chiral symmetry breaking scale $\Lambda_{\chi} \sim 1$~GeV. In turn, chiral effective field theory (EFT) and the non-relativistic limit must be used to describe the many-nucleon initial and final states. These steps go into calculating the NMEs in Eq.~\eqref{eq:0vbb_rate}; this has been performed numerically using many-body methods such as the quasi-particle random phase approximation (QRPA), shell model and interacting boson model (IBM-2). More recently, \textit{ab initio} methods have been used to calculate the NMEs directly from the chiral EFT~\cite{Cirigliano:2022rmf}.

\begin{table}[t!]
\centering
\renewcommand{\arraystretch}{1.25}
\setlength\tabcolsep{2.5pt}
\begin{tabular}{c|cccccccc}\hline
	NME & $\mathcal{M}_{F}$ & $\mathcal{M}_{GT}^{AA}$ & $\mathcal{M}_{GT}^{\prime AP}$ & $\mathcal{M}_{GT}^{\prime\prime PP}$ & $\mathcal{M}_{GT}^{\prime WW}$ & $\mathcal{M}_{T}^{\prime AP}$ & $\mathcal{M}_{T}^{\prime\prime PP}$ & $\mathcal{M}_{T}^{\prime WW}$ \\ \hline
	$\mathcal{M}_X$ & $-0.780$ & $6.062$ & $0.036$ & $0.00034$ & $0.089$ & $-0.010$ & $-0.00014$ & $-0.035$ \\
	$\mathcal{M}_{X,\text{sd}}$ & $-48.89$ & $170.0$ & $2.110$ & $0.028\phantom{00}$ & $2.945$ & $-1.310$ & $-0.022\phantom{00}$ & $-6.541$ \\\hline
\end{tabular}
\caption{Standard and short-distance Fermi, Gamow-Teller and tensor NMEs of $^{76}$Ge. The resulting \textit{light} and \textit{heavy} NMEs used in the simple interpolation formula of Eq.~\eqref{eq:naive_interp} are $\mathcal{M}^{0\nu}_{\nu} = -5.28$ and $\mathcal{M}^{0\nu}_{N} = -194$, respectively. The values are taken from~\cite{Deppisch:2020ztt}.}
\label{tab:NMEs}
\end{table}

The mass of the exchanged neutrino has a large impact on the above discussion. For $m_i \lesssim Q_{\beta\beta}$, where $Q_{\beta\beta} = E_{I}-E_{F}-2m_e \sim \mathcal{O}(\text{MeV})$ is the $Q$-value of the isotope ($E_{I}$ and $E_{F}$ are the energies of the initial and final nuclei, respectively), the $0\nu\beta\beta$ decay rate is dominated by the exchange of \textit{potential} neutrinos, with $(p^0,\mathbf{p})\sim(0,k_F)$, and \textit{hard} neutrinos, with $p^0\sim\mathbf{p}\sim \Lambda_{\chi}$~\cite{Cirigliano:2018hja, Dekens:2020ttz}. In this limit, the NMEs can be written as
\begin{align}
\label{eq:light_NME}
    \mathcal{M}^{0\nu}(m_{i}\lesssim Q_{\beta\beta}) 
    = \frac{g_V^2}{g_{A}^2}\mathcal{M}_{F} - \frac{2m_e m_p g_{\nu}^{NN}}{g_A^2}\mathcal{M}_{F,\text{sd}} 
    - \mathcal{M}_{GT} + \mathcal{M}_{T}\,,
\end{align}
where $\mathcal{M}_{F}$ is the Fermi NME, while
\begin{align}
\label{eq:MGT}
    \mathcal{M}_{GT}& = \mathcal{M}_{GT}^{AA} - \frac{g_P}{6g_A}\mathcal{M}_{GT}^{\prime AP} 
    + \frac{(g_V+g_W)^2}{6g_A^2}\mathcal{M}_{GT}^{\prime WW} 
    + \frac{g_P^2}{48g_A^2}\mathcal{M}_{GT}^{\prime\prime PP}\,,\\
    \mathcal{M}_{T}& = \frac{g_P}{6g_A}\mathcal{M}_{T}^{\prime AP}
    + \frac{(g_V+g_W)^2}{12g_A^2}\mathcal{M}_{T}^{\prime WW} 
    - \frac{g_P^2}{48g_A^2}\mathcal{M}_{T}^{\prime\prime PP}\,,
\label{eq:MT}
\end{align}
are the Gamow-Teller and tensor NMEs. In Eq.~\eqref{eq:light_NME}, $m_p$ is the proton mass and $g_V = 1$, $g_W = 3.7$ and $g_P = 231$ are the vector, magnetic and pseudoscalar charges, respectively~\cite{Simkovic:1999re}. The potential neutrinos contribute to $\mathcal{M}_F$, $\mathcal{M}_{GT}$ and $\mathcal{M}_T$, while the hard neutrinos induce the term proportional to $g_{\nu}^{NN}\sim 1/(2F_\pi)^2$, $F_\pi = 92.2$~MeV, which is a low-energy constant (LEC) describing the short-range nucleon-nucleon coupling. This term contains the \textit{short-range} Fermi NME $\mathcal{M}_{F,\text{sd}}$, which is found by replacing the long-range neutrino potential in the calculation of the Fermi NME with a short-range potential (we use the normalisation factor of $m_e m_p$ similar to~\cite{Simkovic:1999re,Barea:2013bz} instead of $m_\pi^2$ used by~\cite{Dekens:2020ttz}). All NMEs in Eqs.~\eqref{eq:light_NME}-\eqref{eq:MT} are given in Table~\ref{tab:NMEs} for $^{76}$Ge.

If the exchanged neutrinos are instead much heavier than the chiral symmetry breaking scale, $m_{N_{\kappa}}\gg \Lambda_{\chi}$, they must be integrated out before matching onto the chiral perturbation theory; the exchange of such states is then described by a dimension-nine operator~\cite{Menendez:2017fdf,Dekens:2020ttz,Graf:2022lhj}. Neutrinos in the intermediate mass region $Q_{\beta\beta} \lesssim m_{N_{\kappa}} \lesssim \Lambda_{\chi}$ are more difficult to describe, as loop corrections scaling as $\sim m_{N_{\kappa}}/\Lambda_{\chi}$ become large. Nevertheless, it is possible to approximate this intermediate region by applying an interpolating formula between the well-understood low- and high-mass regions. In the treatment of~\cite{Dekens:2020ttz}, this involves constructing an interpolating formula for each NME $\mathcal{M}_{X}$ appearing in Eqs.~\eqref{eq:light_NME}-\eqref{eq:MT}, for example,
\begin{align}
\label{eq:NME_interp}
    \mathcal{M}^{\text{int}}_{X(,\text{sd})}(m_{N_{\kappa}}) 
    = \mathcal{M}_{X(,\text{sd})}\,\frac{\langle\mathbf{p}_X^2\rangle }{\langle\mathbf{p}_X^2\rangle 
    	+ m_{N_{\kappa}}^2}\,;\quad \langle\mathbf{p}_X^2\rangle 
    \equiv m_e m_p \left|\frac{\mathcal{M}_{X,\text{sd}}}{\mathcal{M}_X}\right|\,,
\end{align}
where $\langle\mathbf{p}_X^2\rangle \sim k_F^2$. An interpolating formula can also be formulated for $g_{\nu}^{NN}$, which connects the LEC for hard neutrino exchange to other LECs from dimension-nine operators. Rather than using an interpolation formula for each NME in Eqs.~\eqref{eq:light_NME}-\eqref{eq:MT}, a more naive approach is to interpolate between the low- and high-mass limits of the whole NME $\mathcal{M}^{0\nu}$. For example,
\begin{align}
\label{eq:naive_interp}
    \widetilde{\mathcal{M}}_{0\nu}^{\text{int}}(m_{N_{\kappa}}) 
    = \mathcal{M}_{\nu}^{0\nu}\,\frac{\langle\mathbf{p}^2\rangle }{\langle\mathbf{p}^2\rangle+m_{N_{\kappa}}^2}\,;\quad 
    \langle\mathbf{p}^2\rangle \equiv 
    m_e m_p \left|\frac{\mathcal{M}^{0\nu}_{N}}{\mathcal{M}^{0\nu}_{\nu}}\right|\,,
\end{align}
where the \textit{light} NME $\mathcal{M}^{0\nu}_{\nu}$ is the expression in Eq.~\eqref{eq:light_NME} and the \textit{heavy} NME $\mathcal{M}^{0\nu}_{N}$ is given by Eq.~\eqref{eq:light_NME} with $g_{\nu}^{NN}\to 0$ and $\mathcal{M}_{X}\to \mathcal{M}_{X,\text{sd}}$. Inserting the interpolation formula in Eq.~\eqref{eq:naive_interp} into Eq.~\eqref{eq:0vbb_rate}, it is now possible to write the inverse $0\nu\beta\beta$ decay half-life as
\begin{align}
    \label{eq:inv_0vbb}
    \frac{1}{T^{0\nu}_{1/2}}=\frac{G_{0\nu}g_A^4 |\mathcal{M}^{0\nu}_{\nu}|^2}{m_e^2}\big|m_{\beta\beta}^{\text{eff}}\big|^2\,,
\end{align}
where the effective Majorana neutrino mass from $0\nu\beta\beta$ decay is
\begin{align}
\label{eq:mbb_eff_full}
	\big|m_{\beta\beta}^{\text{eff}}\big| 
	= \left|\sum^{\mathcal{N}_A}_i U_{ei}^2m_i 
	+ \sum^{\mathcal{N}_S}_{\kappa}U_{e N_{\kappa}}^2 m_{N_{\kappa}}
	\frac{\langle\mathbf{p}^2\rangle 
		\mathcal{F}(m_{N_{\kappa}})}{\langle\mathbf{p}^2\rangle+m_{N_{\kappa}}^2}\right|\,.
\end{align}
Here, $\mathcal{F}(m_{N_{\kappa}}) = \mathcal{M}_{0\nu}^{\text{int}}(m_{N_{\kappa}})/\widetilde{\mathcal{M}}_{0\nu}^{\text{int}}(m_{N_{\kappa}})$ is a correction function to the naive interpolation formula in Eq.~\eqref{eq:naive_interp}. As mentioned above, a more accurate approach would be to include in $\mathcal{M}_{0\nu}^{\text{int}}(m_{N_{\kappa}})$ an interpolating formula for $g_{\nu}^{NN}$ and each NME in Eqs.~\eqref{eq:light_NME}-\eqref{eq:MT}.

\begin{figure}[t!]
	\centering
	\includegraphics[width=0.48\textwidth]{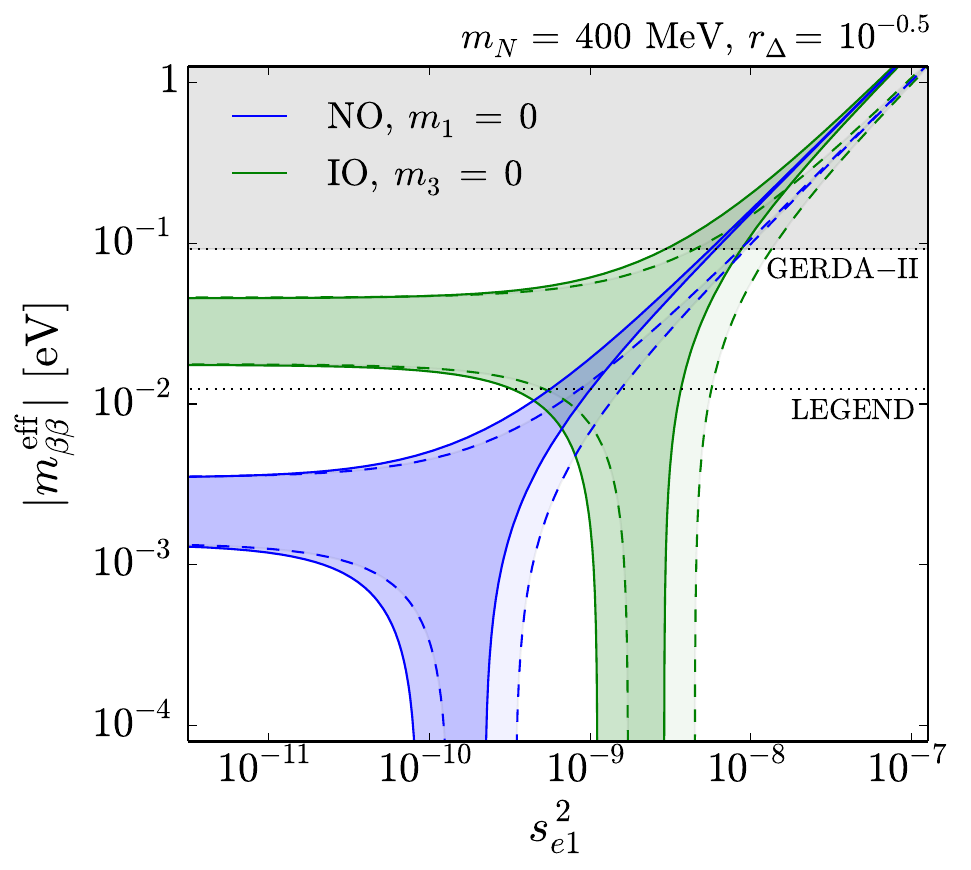}
	\includegraphics[width=0.496\textwidth]{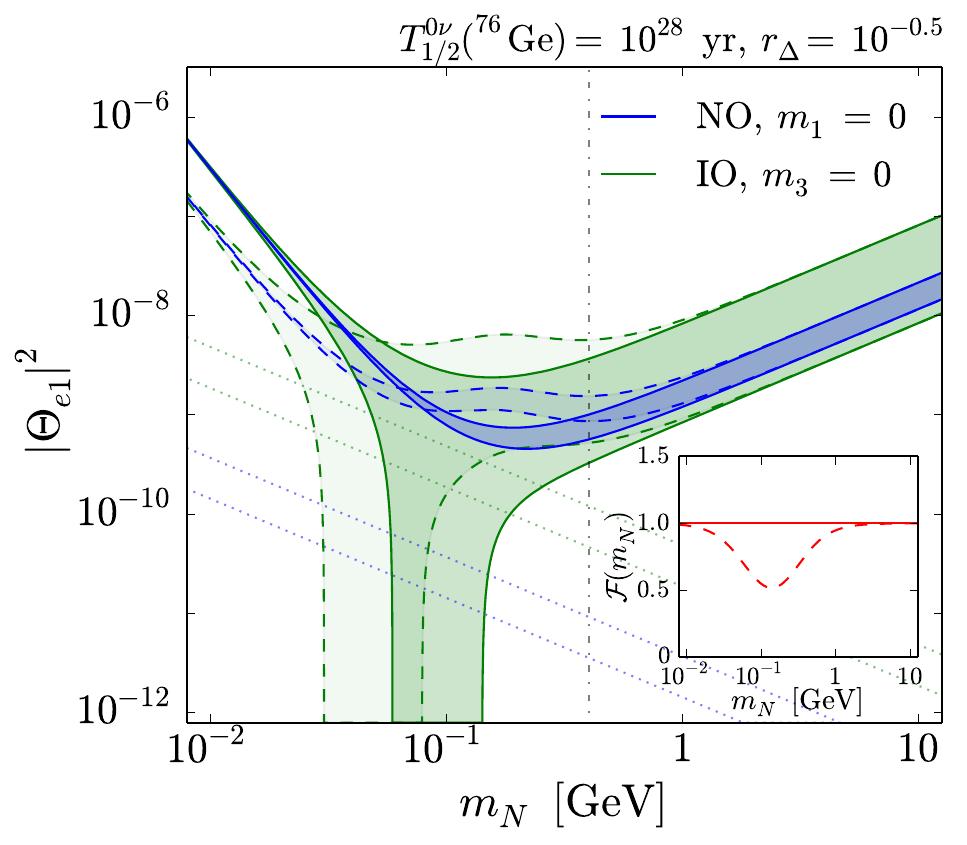}
	\caption{(Left) In the $3+2$ model, the effective Majorana mass $\big|m_{\beta\beta}^{\text{eff}}\big|$ from the $0\nu\beta\beta$ decay of $^{76}$Ge as a function of the active-sterile mixing $|\Theta_{e1}|^2$ for $m_{N} = 400$~MeV, $r_{\Delta} = 10^{-0.5}$, marginalising over $\alpha_{21}$ and $\phi_{e1}$ in the NO (blue shaded) and IO (green shaded) scenarios with $m_{\text{light}} = 0$. (Right) Regions in the $(m_N,|\Theta_{e1}|^2)$ parameter space compatible with a measured $0\nu\beta\beta$ decay half-life of $T_{1/2}^{0\nu}(^{76}\text{Ge})=10^{28}$~yr, for a mass splitting ratio of $r_{\Delta} = 10^{-0.5}$ in the NO (blue) and IO (green) scenarios, marginalising over $\alpha_{21}$ and $\phi_{e1}$. In both plots, solid lines use $\mathcal{F}(m_N) = 1$, dashed lines the more accurate NME interpolating formula (shown in the inset). The corresponding seesaw lines, $|\Theta_{e1}|^2 = |m_{ee}^{\nu}|/m_N$, are shown (dotted).}
\label{fig:mbb-plot}
\end{figure}

In the $3+2$ model we have $U_{eN_{1}} \approx |\Theta_{e1}|e^{i\phi_{e1}/2}$ and $U_{eN_{2}} \approx |\Theta_{e2}|e^{i\phi_{e2}/2}$. Inserting these into Eq.~\eqref{eq:mbb_eff}, it is then possible to use Eq.~\eqref{eq:diag} to eliminate $|\Theta_{e2}|^2$ and $\phi_{e2}$, giving
\begin{align}
\label{eq:mbb_eff}
	\left|m_{\beta\beta}^{\text{eff}}\right| 
	= \left|\alpha m_{ee}^\nu + \beta m_N |\Theta_{e1}|^2 e^{i\phi_{e1}}\right| \,,
\end{align}
where we have introduced
\begin{align}
\label{eq:alphabeta}
	\alpha \equiv 
	1 - \frac{\langle\mathbf{p}^2\rangle  \mathcal{F}\big(m_N(1+r_\Delta)\big)}{\langle\mathbf{p}^2\rangle+m_N^2(1+r_\Delta)^2}\,\quad\beta 
	\equiv \frac{\langle\mathbf{p}^2\rangle\mathcal{F}(m_N)}{\langle\mathbf{p}^2\rangle+m_N^2}-\frac{\langle\mathbf{p}^2\rangle \mathcal{F}\big(m_N(1+r_\Delta)\big)}{\langle\mathbf{p}^2\rangle+m_N^2(1+r_\Delta)^2}\,.
\end{align}
In Fig.~\ref{fig:mbb-plot} (left), we plot the effective Majorana mass from $0\nu\beta\beta$ decay in Eq.~\eqref{eq:mbb_eff} as a function of the active-sterile mixing $|\Theta_{e1}|^2$ for an HNL mass $m_{N} = 400$~MeV and mass splitting ratio $r_\Delta = 10^{-0.5}$. We assume that $\phi_{e1}$ and the Majorana phase $\alpha_{21}$ are not constrained by other means; we therefore marginalise over them, forming a band of possible $\big|m_{\beta\beta}^{\text{eff}}\big|$ values. The blue and green shaded regions depict the NO and IO scenarios, respectively, with a massless lightest neutrino. The solid lines make use of the naive interpolating formula in Eq.~\eqref{eq:naive_interp}, i.e., setting $\mathcal{F}(m_N) = 1$ in Eq.~\eqref{eq:mbb_eff}, while the dashed lines use an interpolating formula for $g_{\nu}^{NN}$ and each NME in Eqs.~\eqref{eq:light_NME}-\eqref{eq:MT}. The regions are compared to the current upper limit on $\big|m_{\beta\beta}^{\text{eff}}\big|$ from the GERDA-II experiment, $T_{1/2}^{0\nu}(^{76}\mathrm{Ge}) > 1.8\times 10^{26}$~yr \cite{GERDA:2020xhi}, and the future sensitivity of the LEGEND-1000 experiment, $T_{1/2}^{0\nu}(^{76}\mathrm{Ge}) > 10^{28}$~yr \cite{Brugnera:2023zgw}.

In Fig.~\ref{fig:mbb-plot} (right), we plot the active-sterile mixing $|\Theta_{e1}|^2$ as a function of $m_N$ found from Eq.~\eqref{eq:mbb_eff} for $r_{\Delta} = 10^{-0.5}$ and a $0\nu\beta\beta$ decay half-life of $T_{1/2}^{0\nu}(^{76}\mathrm{Ge}) = 10^{28}$~yr. Again, we depict the NO (blue shaded) and IO (green shaded) scenarios for a massless lightest neutrino and marginalise over $\phi_{e1}$ and the Majorana phase $\alpha_{21}$. The solid and dashed lines, respectively, use the naive and more accurate interpolating formulae; in the inset, we plot the values of $\mathcal{F}(m_N)$ over the same mass range. The blue and green dotted lines show the ranges of possible seesaw relations, $|\Theta_{e1}|^2 = |m_{ee}^{\nu}|/m_N$, in the NO and IO cases, respectively.

\subsection{Approximate $0\nu\beta\beta$ Decay Rate}
\label{sec:0vbb_approx}

In the $3+2$ model, $0\nu\beta\beta$ decay is induced by the exchange of the light neutrinos and a pair of HNLs. Depending on the parameters of the model, one of these contributions may dominate over the others; alternatively, if the contributions are of similar size, constructive or destructive interference may occur in $\big|m_{\beta\beta}^{\text{eff}}\big|$. In order to investigate this interplay, it is useful to use the simple $1+2$ model as in Sec.~\ref{subsec:approxmixratio}. Replacing $m_{ee}^{\nu}$ with $m_\nu$ in Eq.~\eqref{eq:mbb_eff}, we obtain the simplified effective Majorana mass
\begin{align}
\label{eq:mbb_eff_simp}
    \big|m_{\beta\beta}^{\text{eff}}\big| 
    &= \big|\alpha m_{\nu} + \beta m_N |\Theta_{e1}|^2 e^{i\phi_{e1}}\big| \nonumber\\
    &= m_N\sqrt{\alpha^2 r_\nu^2 + \beta^2 |\Theta_{e1}|^4 + 2 \alpha\beta r_\nu |\Theta_{e1}|^2 \cos\phi_{e1}}\,.
\end{align}
We see that the interplay between the light neutrino and HNL pair is controlled by the active-sterile mixing $|\Theta_{e1}|^2$, light neutrino mass $m_{\nu}$, HNL mass $m_{N}$, mass splitting ratio $r_{\Delta}$ (contained in $\alpha$ and $\beta$) and CP phase $\phi_{e1}$.

We will now examine how $\big|m_{\beta\beta}^{\text{eff}}\big|$ depends on the active-sterile mixing $|\Theta_{e1}|^2$. Firstly, in the \textit{standard seesaw} limit, i.e., when the active-sterile mixing of the HNL pair lies in the thin extensions to the left or bottom of the allowed region in Fig.~\ref{fig:se1_se2_plot} (right), we have
\begin{align}
\big|m_{\beta\beta}^{\text{eff}}\big| \approx \begin{cases}
	(\alpha-\beta)m_{\nu} & (|\Theta_{e1}|^2 = r_\nu)\\
    \alpha m_\nu & (|\Theta_{e1}|^2 \ll r_\nu)
\end{cases}\,.
\end{align}
Here, one HNL decouples while the other has an active-sterile mixing following the seesaw relation and a CP phase $\cos\phi_{e i} = -1$, therefore adding destructively with the light neutrino contribution. Instead, in the \textit{inverse seesaw} limit $|\Theta_{e1}|^{2}\gg r_{\nu}$, where the active-sterile mixing strengths follow $|\Theta_{e2}|^2 = |\Theta_{e1}|^2/(1+r_\Delta)$ and the relative CP phase is $\cos\Delta\phi_{e} = -1$ (and therefore the HNLs add destructively with each other in $\big|m_{\beta\beta}^{\text{eff}}\big|$), there are three interesting cases. If the active-sterile mixing $|\Theta_{e1}|^2$ lies in the range $r_{\nu} \ll |\Theta_{e1}|^2 \ll \alpha r_\nu /\beta$ (note that $\beta < \alpha$ and therefore $\alpha/\beta > 1$ for all positive values of $m_N$ and $r_{\Delta}$), it is not large enough for the HNL pair to dominate over the light neutrino contribution, and again $\big|m_{\beta\beta}^{\text{eff}}\big| \approx \alpha m_\nu$. Instead, for $|\Theta_{e1}|^2 \approx \alpha r_\nu /\beta$, cancellations can occur between the light neutrino and HNL pair. The phase $\phi_{e1}$ is unconstrained in the $|\Theta_{e1}|^{2}\gg r_{\nu}$ limit, so for $|\Theta_{e1}|^2 = \alpha r_\nu /\beta$ we may have any value between $\big|m_{\beta\beta}^{\text{eff}}\big|\approx 0$ for $\cos\phi_{e1}=-1$ and $\big|m_{\beta\beta}^{\text{eff}}\big| \approx 2 \alpha m_{\nu}$ for $\cos\phi_{e1}=1$. Finally, for $|\Theta_{e1}|^2 \gg \alpha r_\nu /\beta$, the HNL pair dominates over the light neutrino contribution, and
\begin{align}
	\big|m_{\beta\beta}^{\text{eff}}\big| \approx \beta m_N |\Theta_{e1}|^2 \quad (|\Theta_{e1}|^2 \gg \alpha r_{\nu}/\beta)\,.
\end{align}
This is nothing but the destructive interference between the HNL contributions. In this limit, $\big|m_{\beta\beta}^{\text{eff}}\big|$ is insensitive to the phase $\phi_{e1}$.

We can now ask the question: what would a measurement of $0\nu\beta\beta$ decay tell us about the available parameter space of the HNL pair? We can equate $\big|m_{\beta\beta}^{\text{eff}}\big|$ in Eq.~\eqref{eq:mbb_eff_simp} to the experimental value of the effective Majorana mass,
\begin{align}
   \big|m_{\beta\beta}^{\text{exp}}\big| &\equiv \frac{m_e}{g_A^2|\mathcal{M}^{0\nu}_{\nu}|\sqrt{G_{0\nu}T_{1/2}^{0\nu}}} \nonumber\\
   &= 1.24\times 10^{-2}~\text{eV}\bigg(\frac{2.36\times 10^{-15}~\text{yr}^{-1}}{G_{0\nu}}\bigg)^{1/2}\bigg(\frac{5.28}{|\mathcal{M}^{0\nu}_{\nu}|}\bigg)\bigg(\frac{10^{28}~\text{yr}}{T_{1/2}^{0\nu}}\bigg)^{1/2}\,,
\end{align}
where $T_{1/2}^{0\nu}$ is the measured $0\nu\beta\beta$ decay half-life. The constraint $\big|m_{\beta\beta}^{\text{eff}}\big| = \big|m_{\beta\beta}^{\text{exp}}\big|$ defines a hypersurface in the $(m_\nu,m_N,r_\Delta,|\Theta_{e1}|^2,\cos\phi_{e1})$ parameter space, described by
\begin{align}
\label{eq:cosphie1_0vbb}
	\cos\phi_{e1} = 
	\frac{\big|m^{\text{exp}}_{\beta\beta}\big|^2 
		- \alpha^2 m_\nu^2-\beta^2 m_N^2 |\Theta_{e1}|^4}{2 \alpha \beta m_\nu m_N |\Theta_{e1}|^2}\,.
\end{align}
We can also use the constraint $\big|m_{\beta\beta}^{\text{eff}}\big| = \big|m_{\beta\beta}^{\text{exp}}\big|$ to define a hypersurface in the parameter space with $\phi_{e1}$ replaced by the relative CP phase $\Delta\phi_{e}$, described by
\begin{align}
\label{eq:cosDeltaphie_0vbb}
	\cos\Delta \phi_{e} = \frac{\alpha^2 m_{\nu}^2 - (2\alpha - \beta)\beta m_N^2 |\Theta_{e1}|^4-|m_{\beta\beta}^{\text{exp}}|^2}{2\sqrt{\alpha\beta^{\phantom{!}}}m_N |\Theta_{e1}|^2\sqrt{|m_{\beta\beta}^{\text{exp}}|^2-(\alpha-\beta)(\alpha m_{\nu}^2-\beta m_N^2 |\Theta_{e1}|^4)}}\,,
\end{align}
which is found by inserting Eq.~\eqref{eq:cosphie1_0vbb} into Eq.~\eqref{eq:cosphie}.

\begin{figure}[t!]
	\centering
	\includegraphics[width=0.99\textwidth]{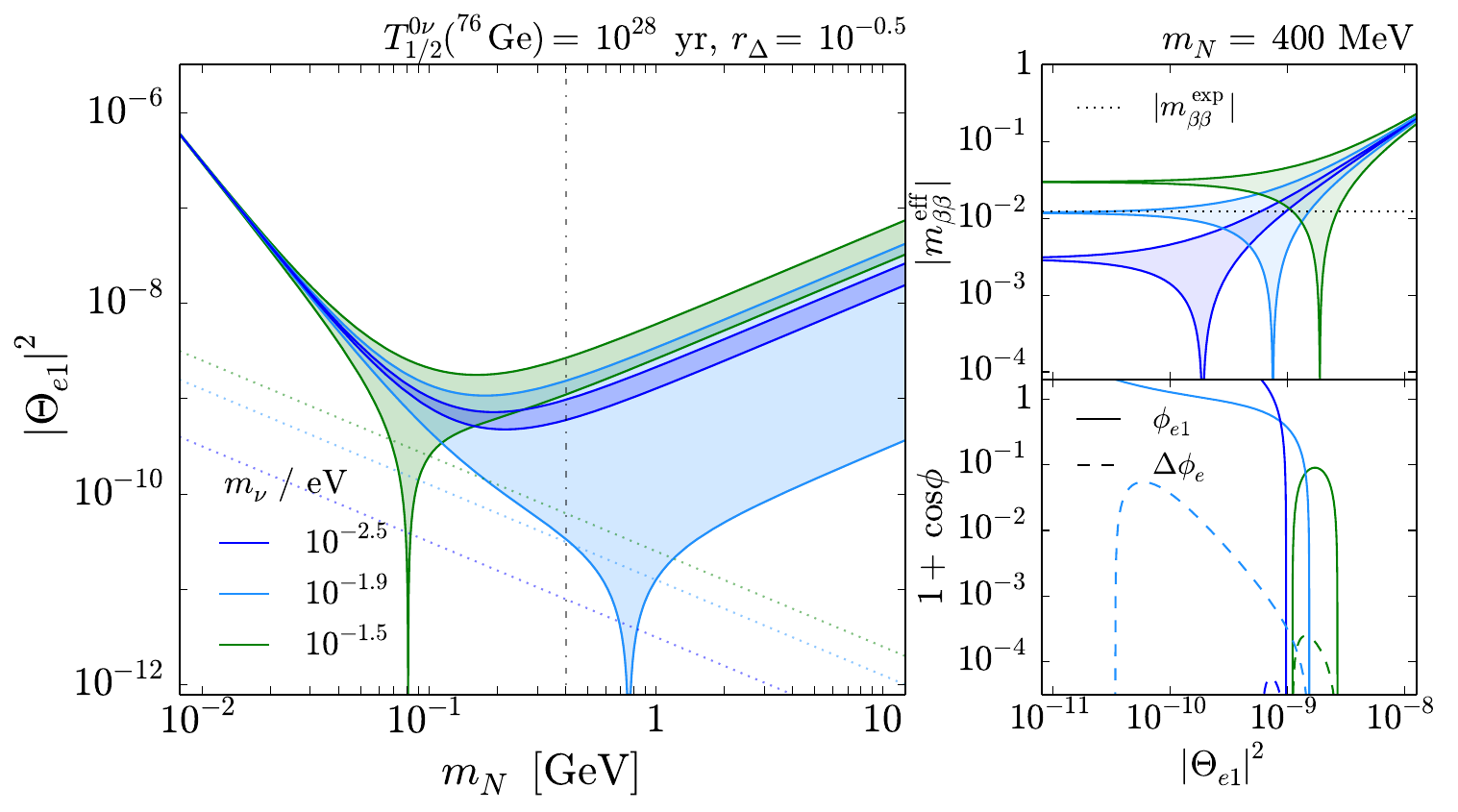}
	\caption{(Left) In the $1+2$ model, regions in the $(m_N,|\Theta_{e1}|^2)$ parameter space compatible with a measured $0\nu\beta\beta$ decay half-life of $T_{1/2}^{0\nu}(^{76}\text{Ge})=10^{28}$~yr, for a mass splitting ratio of $r_{\Delta} = 10^{-0.5}$, an arbitrary CP phase $\phi_{e1}$ and a light neutrino mass $m_{\nu} = 10^{-2.5}$~eV (blue), $10^{-1.9}$~eV (light blue) or $10^{-1.5}$~eV (green). The corresponding seesaw lines, $|\Theta_{e1}|^2 = m_\nu/m_N$, are shown (dotted). (Upper right) For $m_{N} = 400$~MeV, allowed values of $|m^{\text{eff}}_{\beta\beta}|$ as a function of $|\Theta_{e1}|^2$ for the same three $m_\nu$ values, compared to $|m^{\text{exp}}_{\beta\beta}|$ from a $0\nu\beta\beta$ decay half-life of $10^{28}$~yr (black dotted). (Lower right) For $m_{N} = 400$~MeV, $r_{\Delta} = 10^{-0.5}$ and three $m_\nu$ values, $1+\cos\phi_{e1}$ (solid) and $1+\cos\Delta\phi_{e}$ (dashed) as a function of $|\Theta_{e1}|^2$, implied by a measured $0\nu\beta\beta$ decay half-life of $10^{28}$~yr.}
\label{fig:mbb-plot-1gen}
\end{figure}

In Fig.~\ref{fig:mbb-plot-1gen} (left), we plot regions in the $(m_N,|\Theta_{e1}|^2)$ plane that are compatible with a measured $0\nu\beta\beta$ decay half-life of $T_{1/2}^{0\nu}(^{76}\text{Ge})=10^{28}$~yr, a mass splitting ratio of $r_{\Delta} = 10^{-0.5}$ and three different values of the light neutrino mass: $m_{\nu} = 10^{-2.5}$~eV (blue), $10^{-1.9}$~eV (light blue) and $10^{-1.5}$~eV (green). We allow the phase $\cos\phi_{e1}$ to lie anywhere in the range $\cos\phi_{e1} \in [-1,1]$, forming bands of allowed $|\Theta_{e1}|^2$ values. In Fig.~\ref{fig:mbb-plot-1gen} (upper right), we show the effective Majorana mass $\big|m_{\beta\beta}^{\text{eff}}\big|$ as a function of $|\Theta_{e1}|^2$ for $m_{N} = 400$~MeV, $r_{\Delta} = 10^{-0.5}$ and the same three values of $m_\nu$. The observed Majorana mass $\big|m_{\beta\beta}^{\text{exp}}\big|$ for a $0\nu\beta\beta$ decay half-life of $10^{28}$~yr is shown as a horizontal (dotted) line; where this line intersects the $\big|m_{\beta\beta}^{\text{eff}}\big|$ region indicates the allowed range of $|\Theta_{e1}|^2$ along the $m_{N} = 400$~MeV line (dot dashed) in the left plot of Fig.~\ref{fig:mbb-plot-1gen}.

In Fig.~\ref{fig:mbb-plot-1gen} (lower right), we plot the values of $1+\cos\phi_{e1}$ (solid) and $1+\cos\Delta \phi_{e}$ (dashed) as a function of $|\Theta_{e1}|^2$ using Eqs.~\eqref{eq:cosphie1_0vbb} and Eqs.~\eqref{eq:cosDeltaphie_0vbb}, respectively, given a $0\nu\beta\beta$ decay half-life  of $T_{1/2}^{0\nu}(^{76}\text{Ge})=10^{28}$~yr, for $m_{N}=400$~MeV, $r_\Delta = 10^{-0.5}$ and the same three $m_{\nu}$ values. For $m_{\nu} = 10^{-2.5}$~eV, $\cos\phi_{e1}$ can lie anywhere between $-1$ and $1$. This is made clear from Fig.~\ref{fig:mbb-plot-1gen} (upper right); the light neutrino mass does \textit{not} saturate the $0\nu\beta\beta$ decay rate and so $\big|m_{\beta\beta}^{\text{exp}}\big|$ intersects the $\big|m_{\beta\beta}^{\text{eff}}\big|$ region where $\big|m_{\beta\beta}^{\text{eff}}\big|\approx \beta m_N |\Theta_{e1}|^2$ (i.e., the HNL pair dominates), which is insensitive to $\phi_{e1}$. For $m_{\nu} = 10^{-1.5}$~eV, we instead have $\cos\phi_{e1} \approx -1$; now, the light neutrino mass \textit{does} saturate the $0\nu\beta\beta$ decay rate and $\big|m_{\beta\beta}^{\text{exp}}\big|$ intersects the $\big|m_{\beta\beta}^{\text{eff}}\big|$ region where there must be a cancellation between the light neutrino and HNL pair. For the intermediate case $m_{\nu} = 10^{-1.9}$~eV, the light neutrino contribution \textit{only just} saturates the $0\nu\beta\beta$ decay rate and therefore a wider range of $|\Theta_{e1}|^2$ values are allowed. 

For the relative CP phase $\Delta\phi_e$, we see that $\cos\Delta\phi_{e}\approx -1$ for all three $m_{\nu}$ values; this is because the intersection of $\big|m_{\beta\beta}^{\text{exp}}\big|$ and the $\big|m_{\beta\beta}^{\text{eff}}\big|$ region occurs well within the inverse seesaw regime for $m_{N} = 400$~MeV and $r_\Delta = 10^{-0.5}$. This is confirmed in Fig.~\ref{fig:mbb-plot-1gen} (left), where we see that the allowed bands of $|\Theta_{e1}|^2$ are well above the seesaw lines. Nevertheless, large values of $\cos\Delta\phi_{e}$ are still possible in the narrow light blue and green regions where the allowed $|\Theta_{e1}|^2$ values fall below the seesaw lines. For a light neutrino mass $m_\nu > |m^{\text{exp}}_{\beta\beta}|$, this occurs when the HNL mass $m_N$ is such that $\big|m^{\text{exp}}_{\beta\beta}\big| = \alpha m_\nu$, i.e., 
\begin{align}
    m_N = \frac{1}{1+r_\Delta}\sqrt{\frac{|m^{\text{exp}}_{\beta\beta}|\langle \mathbf{p}^2\rangle}{m_\nu - |m^{\text{exp}}_{\beta\beta}|}}\,,
\end{align}
and so $\big|m^{\text{exp}}_{\beta\beta}\big|$ intersects the $\big|m_{\beta\beta}^{\text{eff}}\big|$ region along the thin extension to small $|\Theta_{e1}|^2$ values.

To conclude this section, we consider how the factors $\alpha$ and $\beta$ depend on $m_N$ and $r_{\Delta}$. To do this, it is convenient to write the factors $\alpha$ and $\beta$ (for $\mathcal{F}(m_N) = 1$) as
\begin{align}
\alpha & = 1-\frac{1}{1+r_{p}(1+r_\Delta)^2}\,,\quad \beta = \frac{1}{1+r_{p}}-\frac{1}{1+r_{p}(1+r_\Delta)^2}\,,
\end{align}
where we have defined $r_{p} \equiv m_{N}^2/\langle\mathbf{p}^2\rangle$. The relevant limits of $\alpha$ and $\beta$ are listed in Table~\ref{tab:ablimits}. First, we can consider the scenario where the mass splitting ratio is small ($r_\Delta \ll 1$) and the HNL mass $m_N$ is below the scale $k_F \sim 100$~MeV ($r_{p}\ll 1$). In this limit, $\alpha \approx r_p \ll 1$ and $\beta \approx 2r_pr_\Delta \ll 1$ and therefore the effective Majorana mass is suppressed. This can be understood from Eq.~\eqref{eq:mbb_eff}; the effective Majorana mass is approximately proportional to $(\mathcal{M}_{\nu})_{ee}$ in Eq.~\eqref{eq:Mnu=0} which is zero at tree level, and the light neutrino and HNL pair contributions exactly cancel each other. Another possible situation is that $r_{p}\ll 1$, i.e. the lighter HNL is less massive than $\sim 100$~MeV, but $r_{\Delta}\gg 1/r_p^{1/2}$ (or $\Delta m_N \gg \langle\mathbf{p}^2\rangle^{1/2}$), so that the heavier HNL is more massive than $\sim 100$~MeV. This gives $\alpha \approx \beta \approx 1$ and so the contributions from the light neutrino and lighter HNL dominate the $0\nu\beta\beta$ decay rate, while the contribution from the heavier HNL is suppressed.

\begin{table}[t!]
\centering
\renewcommand{\arraystretch}{1.25}
\setlength\tabcolsep{6pt}
\begin{tabular}{c|c|c}
\hline
	$\alpha$, $\beta$ & $r_p \ll 1$            & $r_p \gg 1$           \\\hline
	$r_\Delta \ll 1$  & $r_p$, $2r_p r_\Delta$ & $1$, $2r_\Delta/r_p$  \\
	$r_\Delta \gg 1$  & $1$, $1$               & $1$, $1/r_p$          \\\hline
\end{tabular}
\caption{Approximate values of the factors $\alpha$ and $\beta$ in the different limits of the ratios $r_p = m_N^2/\langle \mathbf{p}^2\rangle$ and $r_\Delta = \Delta m_N/m_N$.}
\label{tab:ablimits}
\end{table}

We next examine the limit in which both HNL states are heavier than $\sim 100$~MeV ($r_{p}\gg 1$). This gives $\alpha \approx 1$ and $\beta \approx 2r_{\Delta}/r_{p}\ll 1$ if the mass splitting ratio is small ($r_\Delta \ll 1$) and $\alpha \approx 1$ and $\beta \approx 1/r_{p}\ll 1$ if the splitting is large ($r_\Delta \gg 1$). As DUNE probes HNLs in the mass range $100 ~\text{MeV}\lesssim m_N \lesssim 2 ~\text{GeV}$, this is the interesting regime for the comparison of $0\nu\beta\beta$ decay and DUNE. If we are in the small $m_{\nu}$ limit (so the HNL pair dominates $0\nu\beta\beta$ decay),
\begin{align}
\label{eq:mbb_approx}
	\big|m_{\beta\beta}^{\text{eff}}\big| 
	&\approx \beta m_N |\Theta_{e1}|^2 
	\approx \frac{m_N}{r_{p}}\bigg(1-\frac{1}{(1+r_\Delta)^2}\bigg)|\Theta_{e1}|^2 \nonumber\\
	&\approx \begin{cases}
		2m_N r_{\Delta}|\Theta_{e1}|^2/r_{p} & (r_\Delta \ll 1)\\
    	m_N |\Theta_{e1}|^2/r_{p} & (r_\Delta \gg 1)
	\end{cases}\,,
\end{align}
where we first approximate the factor $\beta$ for $r_{p}\gg 1$ and then for the two different limits of $r_{\Delta}$. For $r_{\Delta}\ll 1$, $\big|m_{\beta\beta}^{\text{eff}}\big|$ is given by the difference of the HNL contributions, which is proportional to $r_{\Delta}$. The inverse $0\nu\beta\beta$ decay half-life in Eq.~\eqref{eq:inv_0vbb} therefore scales as $(T_{1/2}^{0\nu})^{-1}\propto (\Delta m_N)^2|\Theta_{e1}|^4/m_N^4$. For $r_{\Delta}\gg 1$, $\big|m_{\beta\beta}^{\text{eff}}\big|$ is just the contribution of the lighter HNL, because the contribution of the heavier HNL is suppressed. The inverse $0\nu\beta\beta$ decay half-life then scales as $(T_{1/2}^{0\nu})^{-1}\propto |\Theta_{e1}|^4/m_N^2$. Equating Eq.~\eqref{eq:mbb_approx} to $\big|m_{\beta\beta}^{\text{exp}}\big|$ and solving for $|\Theta_{e1}|^2$ (keeping an arbitrary value of $r_{\Delta}$) gives 
\begin{align}
\label{eq:se1_0vbb}
    |\Theta_{e1}|^2 \approx \frac{m_N|m_{\beta\beta}^{\text{exp}}|}{\langle \mathbf{p}^2\rangle}\frac{(1+r_\Delta)^2}{r_\Delta(2+r_{\Delta})}\,.
\end{align}
This result will be used to compare $0\nu\beta\beta$ decay and direct searches at DUNE analytically in Sec.~\ref{sec:compare}.

\section{The Deep Underground Neutrino Experiment}
\label{sec:DUNE}

Fixed target experiments are sensitive probes of HNL scenarios, which we intend to use in conjunction with $0\nu\beta\beta$ decay to explore the nature of an HNL pair. As seen in Fig.~\ref{fig:mbb-plot-1gen}, $0\nu\beta\beta$ decay searches have the highest sensitivity to the active-sterile mixing in the 100~MeV to GeV mass range. To explore the complementarity with direct searches, we focus on the upcoming experiment DUNE, which will be able to probe the direct production of HNLs in a similar mass range.

In this section we will first describe the production and subsequent decay of HNLs in fixed target experiments. We will then examine in more detail the sensitivity of the DUNE near detector to HNLs produced from the decays of pions, kaons and $D$ mesons. Using \textsc{Pythia} to simulate the momentum profiles of HNLs produced from these production channels, we will apply simplified geometric cuts on the decays of HNLs in the DUNE near detector to estimate the sensitivity of the experiment. DUNE will initially receive a 1.2 MW proton beam from the main injector accelerator at the Long Baseline Neutrino Facility (LBNF) at FNAL. The $120$~GeV proton beam will impinge on a graphite target, which corresponds to $pp$ collisions with the target at rest. DUNE will use two detectors; a far detector (FD) situated at a distance of $1300$~km from the target and a smaller near detector (ND) at a distance of $574$~m from the target~\cite{Ballett:2019bgd}. 

For our analysis, we will model the ND with a simplified geometry; since our goal is to obtain the number of events from DUNE for a comparison with $0\nu\beta\beta$ decay, we do not include the exact geometry of the fiducial volume, which might result in $\mathcal{O}(1)$ corrections that can be obtained from a detailed analysis~\cite{Ballett:2019bgd}. The basic schematic of our analysis is as follows: the proton beam hits the target, leading to the production of mesons from $pp$ collisions, which travel and decay to SM particles and HNLs. We employ geometric cuts, where we demand that the produced HNL decays to charged tracks inside the fiducial volume, taken to have a length of $5$~m along the beam axis and a conical cross section. We demand that 2.44 events be detected to reject a null event rate with $90 \% $ C.L., assuming no background events and a Poisson-like distribution. We first limit our analysis to the case where the HNL mixes uniquely to the electron flavour ($\alpha = e$), while the more general case of mixing with all the three flavours will be explored later.

\subsection{Meson Production at DUNE}

Meson decays are the dominant source of HNLs in fixed target experiments. A variety of mesons, such as pions, kaons and $D$ mesons are produced at DUNE via $pp$ collisions, which we will now briefly describe in terms of their production fractions and momentum profiles. 

\begin{figure}[t!]
	\centering
	\includegraphics[width=0.34\textwidth]{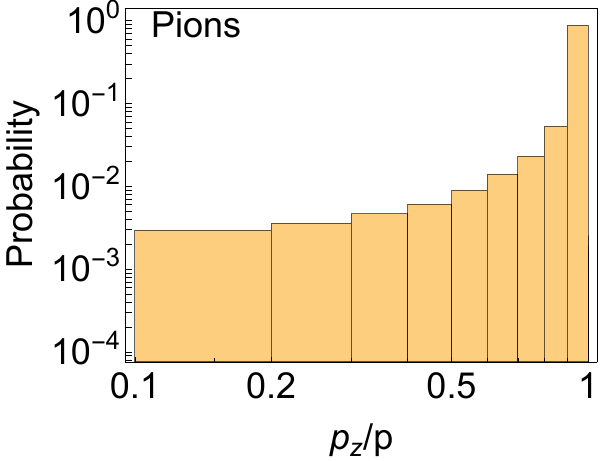}
	\includegraphics[width=0.32\textwidth]{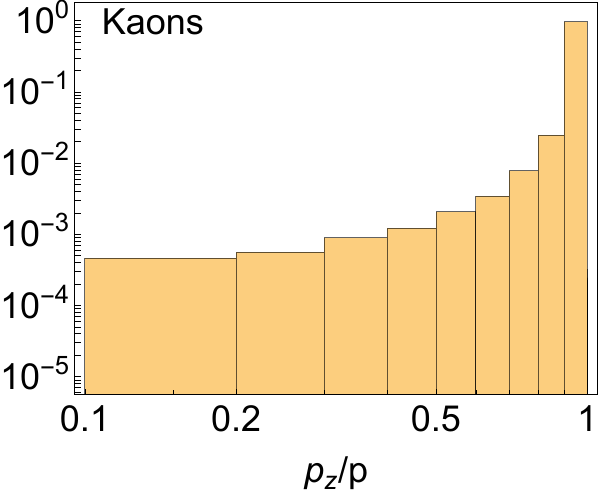}
	\includegraphics[width=0.32\textwidth]{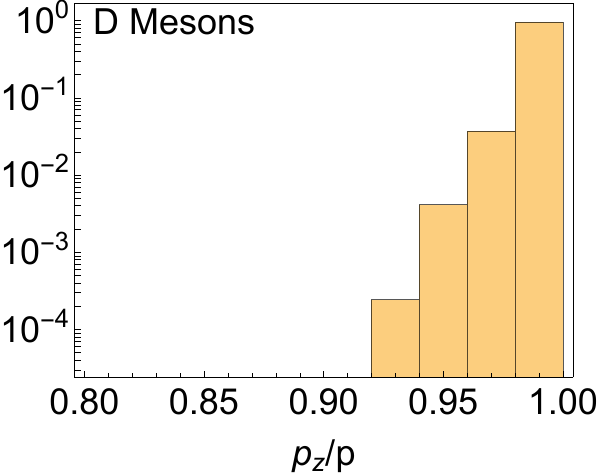}
	\caption{Distributions of momentum fractions along the beam axis for pions (left), kaons (centre) and $D$~mesons (right) from \textsc{Pythia}-generated events.}
\label{fig:mom_distributions}
\end{figure}

\begin{table}[t!]
	\centering
	\renewcommand{\arraystretch}{1.25}
	\setlength\tabcolsep{6pt}
	\begin{tabular}{c|cccccccc|}\hline
		Meson $P$  & $\pi^+$ & $K^+$  & $K_{L,S}^0$ & $D^0$             & $D^+$                & $D_s^+$ \\ \hline
		Mesons/POT & $2.8$   & $0.24$ & $0.18$      & $6\times 10^{-5}$ & $1.2 \times 10^{-5}$ & $ 3.3\times 10^{-6}$ \\ \hline
	\end{tabular}
	\caption{Number of positively-charged and neutral pseudoscalar mesons produced per proton on target (POT) in DUNE, for a $120$~GeV proton beam.}
	\label{tab:MesonDUNE}
\end{table}
Following the approach of Refs.~\cite{Berryman:2019dme,Coloma:2020lgy,Krasnov:2019kdc}, we show the production fractions of mesons at DUNE in Table~\ref{tab:MesonDUNE}, using the values given in the literature when in agreement with the values we extract from the simulation of meson production in \textsc{Pythia} (v.~8.307)~\cite{Sjostrand:2014zea}.\footnote{In the current literature, there exists a slight variation in the production fraction of pions from $pp$ collisions at DUNE, varying roughly by an order of magnitude. A more systematic study would be needed to improve the accuracy, but it would not lead to a drastic change in the overall DUNE sensitivity (roughly at the $10\%$ level), and would not affect our final results.} In \textsc{Pythia}, the momentum distributions for mesons produced in DUNE are also extracted. In Fig.~\ref{fig:mom_distributions}, we show the momentum distribution histograms for pions (left), kaons (centre) and $D$ mesons (right), where it can be seen that the produced mesons are highly boosted along the beam direction.

\subsection{HNL Production from Meson Decays}

After simulating the production of mesons at DUNE, we consider all possible decay modes of pions, kaons and $D$ mesons leading to the production of HNLs. Meson decays to HNLs are calculated in the rest frame of the meson and then boosted to the lab frame using the momentum of the corresponding meson, extracted from \textsc{Pythia}.

Meson decays can be grouped into two categories, the first being purely leptonic two-body decays such as $P^{+} \to e^{+} N$, which have the generic branching fraction
\begin{align}
\label{eq:prod_BR}
	\text{Br}(P^{+} \to e^{+} N) \simeq  
	\tau_P \frac{G_F^2 m_P^3}{8\pi}f_P^2|U_{eN}|^2|V_{q\bar{q}}|^2
	\frac{m_N^2}{m_P^2}\bigg(1-\frac{m_N^2}{m_P^2}\bigg)^2\,, 
\end{align}
where $\tau_P$, $f_P$, $m_P$ and $V_{q\bar{q}}$ correspond to the meson lifetime, decay constant, mass, and CKM mixing matrix element (depending on valence quark content of the meson), respectively~\cite{Batell:2020vqn}. 

\begin{figure}[t!]
	\centering
	\includegraphics[trim={1cm -1cm 0 0},clip,width=0.47\textwidth]{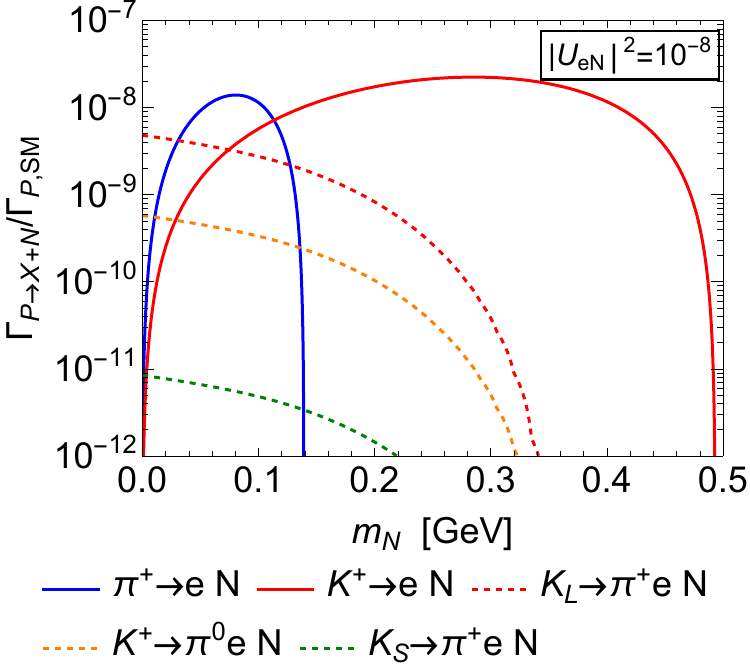}
	\includegraphics[trim={1cm 0 0 0},clip,width=0.47\textwidth]{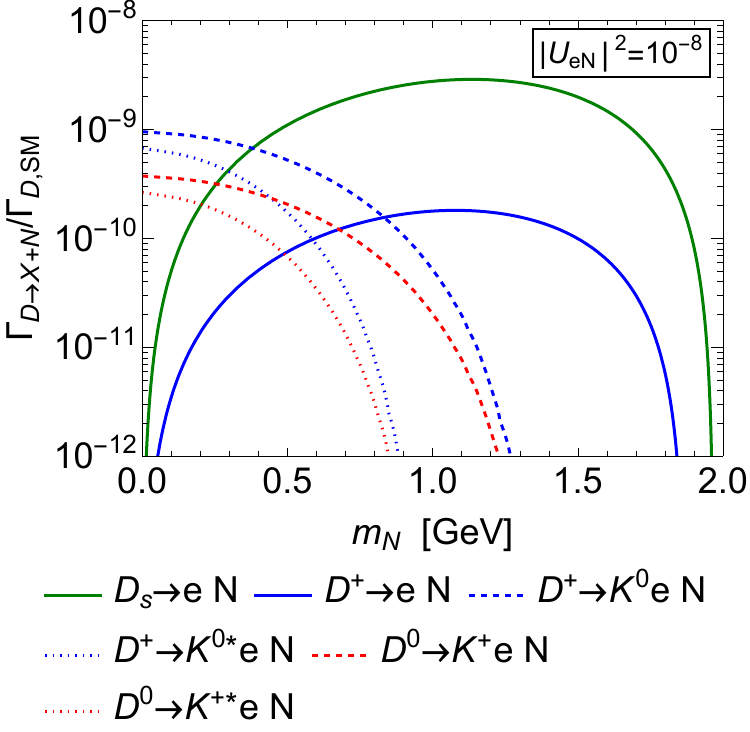}
	\caption{Branching fractions of decays of pions and kaons (left) and $D$ mesons (right) to HNLs, with a value of $|U_{eN}|^2=10^{-8}$ for the active-sterile mixing with the electron flavour.}
	\label{fig:production_channels}
\end{figure}

The second category contains three-body semi-leptonic decays, which can be competitive to the two-body channels, as seen in Fig.~\ref{fig:production_channels}. This is mainly attributed to the absence of CKM or chirality-flip suppressions, which compensates for the suppressed phase space~\cite{Batell:2020vqn, Bondarenko:2018ptm}. The formulae for three-body decays require the use of form factors and depend on whether the daughter meson is a pseudoscalar or vector meson. For the decay into a pseudoscalar meson, the branching fraction is given by
\begin{align}
	\text{Br}(P^+\to P^{\prime 0} e^+ N) \propto 
	\tau_{P}\,\frac{G_F^2 m_P^5}{64 \pi^3} |V_{qq'}|^2 |U_{eN}|^2 f(\mathcal{I}_{P})\,.
\end{align}
For the decay into a vector meson, the branching fraction becomes,
\begin{align}
	\text{Br}(P^+\to V^{0}e^+N) \propto 
	\tau_{P}\,\frac{G_F^2 m_P^7}{64 \pi^3 m_{V}^2}|V_{qq'}|^2 |U_{eN}|^2 f(\mathcal{I}_V)\,,
\end{align}
where $f(\mathcal{I}_{P})$ and $f(\mathcal{I}_V)$ are functions that depend on the meson coupling constants and form factors, respectively~\cite{Bondarenko:2018ptm}.

As shown in Fig.~\ref{fig:production_channels}, for HNL masses below the pion threshold, $\pi^+ \to e^+ N$ (solid blue) is the dominant production channel, which slowly gives way to  $K^+ \to e^+ N$ (solid red) up to the kaon threshold. For heavier HNLs above the kaon but below the charm threshold, production via $D$ meson decays is dominant, with $D_s^+ \to e^+ N$ (solid green) being the most important channel.

\subsection{HNL Decays}

HNLs are unstable, and will therefore decay to SM particles via charged- or neutral-current processes, suppressed via the active-sterile mixing $|U_{eN}|^2$. These decays proceed via off-shell $W^{\pm}/Z$ bosons, which means that they can be long-lived, with decay lengths in the tens of meters or larger, and thus are very well suited to being explored at fixed target experiments. Schematically, the charged-current interactions result in processes such as $N\to \nu e^+e^-$ and $N\to e^- u \bar{d}$, depending on the leptons/quarks. The neutral-current mediated decays lead to processes of the form of $N \to \nu e^{+} e^{-}$ and $N \to \nu q \bar{q}$. Both charged-current and neutral-current mediated processes interfere for processes involving the same generations, viz. the decay modes $N \to 3\nu $ and $N \to \nu e^+ e^-$.  

Below the pion threshold, the invisible decay $N \to 3\nu$ is the dominant decay mode. Above the pion threshold and below $1.5$~GeV, the quarks generally confine into hadronic states, and the resultant single meson hadronic channels, corresponding to $N \to e^- P^+$ (charged-current), $N \to \nu P^0$ and $N \to \nu V^0$ (neutral-current) are important. The decay mode to a single pion makes up a large proportion of the hadronic decay modes.

Here, we provide schematic expressions for the two-body and three-body decay modes, with a more detailed analysis to be found in previous works~\cite{Bondarenko:2018ptm}. For example, the two-body decay of an HNL to an electron and a pseudoscalar meson has the branching fraction
\begin{align}
    \label{eq:det_BR}
    \text{Br}(N\to e^{\mp}P^{\pm}) &\approx \tau_N\frac{G_F^2 m_{N}^3}{8\pi} f^2_{P}|U_{eN}|^2|V_{qq'}|^2  \bigg(1-\frac{m_P^2}{m_{N}^2}\bigg)^2\,,
\end{align}
while the branching fraction for the three-body decay of an HNL into an electron and quark pair is given by
\begin{align}
  \text{Br} (N \to e^{\mp} u d) = \tau_N \frac{G_F^2 m_N^5}{192 \pi^3}
  |U_{eN}|^2 \mathcal{J}(x_u,x_d,x_e)\,,
  \label{eq:Ndec3bdy}
\end{align}
where $x_{u,d,e} = m_{u,d,e}/m_N$,
\begin{align}
  \label{eq:38}
  \mathcal{J}(x_u,x_d,x_e) \equiv 12\!\!\!\!\!\!\!\int\limits_{(x_d + x_e)^2}^{(1-x_u)^2}
  \!\!\!\!\frac{dx}{x} 
  (x - x^2_{e} - x^2_d)(1 + x_u^2 - x) 
  \sqrt{\lambda(x, x_e^2, x_d^2) \lambda(1, x, x_u^2)}\,,
\end{align}
and $\lambda(x, y, z)=x^2 + y^2 + z^2 - 2xy - 2yz - 2zx$ is the  K\"{a}ll\'{e}n function. 

\begin{figure}[t!]
	\centering
	\includegraphics[clip,width=0.50\textwidth]{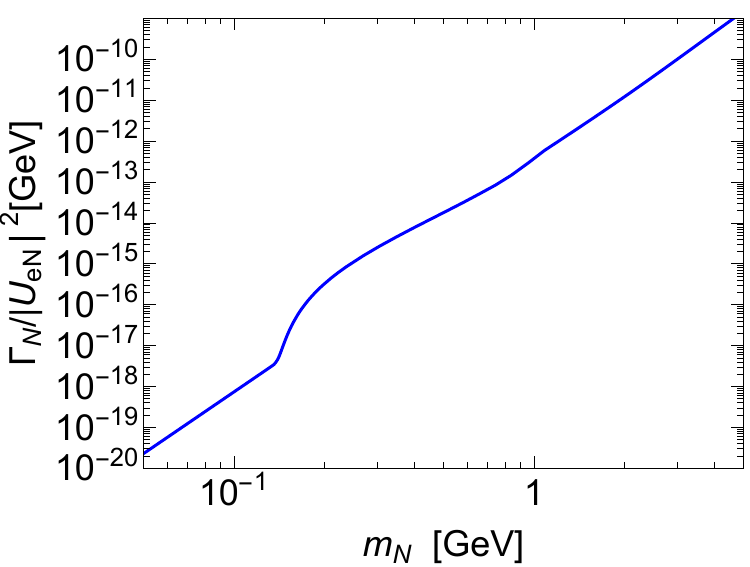}
	\includegraphics[clip,width=0.49\textwidth]{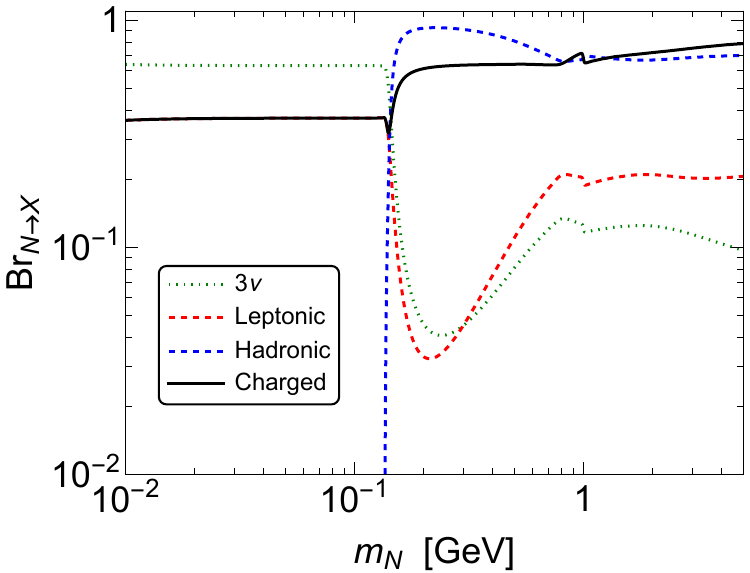}
	\caption{(Left) Total HNL decay width, normalised by the active-sterile mixing $|U_{eN}|^2$. (Right) branching fractions of an HNL to $3\nu$ (dashed green), other combinations of leptons (dashed red), and hadronic states (dashed blue). The black curve corresponds to the total branching fraction of decays to charged tracks (leptonic or semi-leptonic).}
	\label{fig:HNL_decay_plot}
\end{figure}
In the left hand side of Fig.~\ref{fig:HNL_decay_plot}, we depict the total HNL decay width $\Gamma_N$, normalised to the active-sterile mixing $|U_{eN}|^2$. In Fig.~\ref{fig:HNL_decay_plot} (right), we show the relevant branching fractions, including the branching fraction of invisible decays to three neutrinos (solid green) and charged final states (solid black), the latter being a requirement to observe tracks in the DUNE ND. Both the total HNL decay width $\Gamma_N$ and the branching fraction to charged tracks are needed to estimate the DUNE sensitivity.

\subsection{Majorana vs. Quasi-Dirac HNLs at DUNE}

In the literature, the distinction between Majorana and quasi-Dirac HNLs in beam dump and collider experiments has been studied in detail~\cite{Antusch:2017ebe,Tastet:2019nqj,Tastet:2021vwp}. Naively, for Majorana and quasi-Dirac HNLs with the same active-sterile mixing $|U_{eN}|^2$, one would expect the rate for Majorana HNL decays to charged final-states to be twice that for a quasi-Dirac HNL in DUNE. This is because a Majorana HNL produced from the decay $K^{+}\to e^{+}N$ can decay via both the LNC and LNV channels $N\to e^{-}\pi^{+}$ and $N\to e^{+}\pi^{-}$, respectively, while a quasi-Dirac HNL with a vanishing mass splitting can only decay via $N\to e^{-}\pi^{+}$. 

However, when the mass splitting is non-zero, oscillations with a frequency controlled by $\Delta m_N = m_N r_\Delta$ can occur between the Majorana pair forming the quasi-Dirac HNL. The appearance of these oscillations depends on whether the HNL decay is prompt or long-lived. If the HNL decays promptly, i.e., $\Gamma_N\tau\gg 1$, an experiment can only observe the time-integrated rates for the LNC and LNV decay modes. The ratio of LNV to LNC events in the detector is given by
\begin{align}
   R_{ll} = \frac{m_N^2 r_{\Delta}^2}{2\Gamma_N^2 + m_N^2 r_{\Delta}^2} \,,
\end{align}
which can range from $0$ and $1$ depending on the relative sizes of the HNL mass splitting and decay width~\cite{Bray:2007ru,Dev:2013wba}.

To decay inside the DUNE ND, however, the quasi-Dirac HNL must be long-lived. For the baseline $L = 574$~m, we require an HNL lifetime,
\begin{align}
   \tau = \frac{L}{\beta\gamma}\sim \frac{2\times 10^{-6}~\text{s}}{\beta\gamma}\,,
\end{align}
where $\beta\gamma = p_{N_z}/m_N$ is the HNL boost factor. The signal in the detector now depends on the relative sizes of $m_N r_\Delta$ and $\tau$. If the mass splitting is such that $m_N r_{\Delta} \tau \gg 2\pi$, or
\begin{align}
   r_\Delta \gg \frac{2\pi}{m_N}\frac{1}{\tau}\sim 10^{-17}\bigg(\frac{400~\text{MeV}}{m_N}\bigg)\bigg(\frac{574~\text{m}}{L}\bigg)\beta\gamma\,,
\end{align}
the oscillations must be averaged out, yielding an equal rate for LNC and LNV decays. Only when the mass splitting is extremely small will the LNV decay modes be suppressed for quasi-Dirac HNLs. For the mass splitting $r_\Delta$ required to produce an observable $0\nu\beta\beta$ decay rate, as explored in Sec.~\ref{sec:compare}, we must include both LNC and LNV decay modes to estimate the DUNE sensitivity. In the following, we consider the sensitivity of DUNE to a single Majorana HNL. For a quasi-Dirac HNL, we would need to add the contributions of two Majorana states.

\subsection{DUNE Acceptance}

We will now describe our approach to calculating the expected number of signal events in the DUNE ND. The detector is located at a distance of $L = 574$~m from the target, with a transverse cross-section of $A_{\text{det}} = 12~\text{m}^2$ and a depth of $\Delta\ell_{\text{det}} = 5$~m along the beam axis~\cite{Abdullahi:2022jlv}. Instead of using the cuboidal geometry of the detector, we use a simplified yet sufficient conical geometry to model the experimental setup. Keeping the beam axis the same and defining an angular aperture $\theta_{\text{det}}$ for the transverse cross-section,
\begin{align}
   \theta_{\text{det}} = \tan^{-1}\left(\frac{\langle x_{\text{det}} \rangle}{\langle L \rangle}\right) \sim \frac{3.8~\text{m}}{576.5~\text{m}} \sim 7\times 10^{-3}\,,
\end{align}
where $\langle x_{\text{det}} \rangle$ corresponds to the average transverse width of the fiducial volume and $\langle L \rangle$ is the mean distance along the beam axis from the target to the fiducial volume.

As described above, we first use \textsc{Pythia} (v.~8.307)~\cite{Sjostrand:2014zea} to simulate the production of mesons from a $pp$ collision at $\sqrt{s}=15$~GeV, extracting the four-momenta profile to define the lab frame for each event. The HNLs are produced from the decays of these simulated mesons at rest, and then boosted corresponding to the extracted meson momentum. The HNLs with angular distribution smaller than $\theta_{\text{det}}$ will end up in the the fiducial volume, and hence can contribute to the event rate if they decay to a charged final state. The HNLs with larger transverse angles will not enter the fiducial volume and therefore not contribute to the event rate, regardless of the decay mode. Putting this together, the total number of signal events is
\begin{align}
\label{eq:DUNE_sensitivity}
N_{\text{sig}} = N_{P} \cdot \text{Br}(P\to  N) \cdot \text{Br}(N\to \text{charged}) \cdot \epsilon_{\text{geo}}\,,
\end{align}
where $N_P$ is the relevant production fraction in Table~\ref{tab:MesonDUNE} multiplied by the total number of protons on target $N_{\text{POT}}$. The geometrical efficiency is given by
\begin{align}
\label{eq:geo_cut}
\epsilon_{\text{geo}}=\frac{1}{N_{\text{tot}}}\sum\limits_{\text{cut}}e^{-\frac{m_{N}\Gamma_N}{p_{N_z}} \langle L\rangle}\bigg(1-e^{-\frac{m_{N}\Gamma_N}{p_{N_z}}\Delta\ell_{\text{det}}}\bigg)\,,
\end{align}
where $N_{\text{tot}}$ is the total number of simulated events, $p_{N_z}$ is the lab-frame momentum of the HNL along the beam axis, and $\Gamma_N$ is the total decay width of the HNL. We use $N_{\text{POT}}=6.6\times 10^{21}$ protons on target, which corresponds to a run-time of 6 years~\cite{Ballett:2019bgd}. In Eq.~\eqref{eq:geo_cut}, `cut' refers to the HNLs which pass through the fiducial volume, i.e., all HNLs that have an angular momentum profile smaller than the angle $\theta_{\text{det}}$,
\begin{align}
	\frac{p_{N_T}}{p_{N_z}} < \theta_{\text{det}} \sim 7\times 10^{-3}\,,
\end{align}
where $p_{N_T}$ is the lab-frame momentum of the HNL transverse to the beam axis. The geometric efficiency $\epsilon_{\text{geo}}$ therefore corresponds to the requirement that the HNL decays inside the fiducial volume.

The analysis above only considers HNL production and decay via the electron neutrino mixing, but this can be easily generalised to the mixing to all three active neutrinos. As an example, we consider arbitrary mixing strengths to electron, muon and tau neutrinos. The basic schematics of the calculation are similar, with HNL production and decay proceeding via all $|U_{\ell N}|^2$ ($\ell = e, \mu, \tau$). In our simulation, we fix $m_N=800$~MeV as a benchmark scenario and calculate the sensitivity for arbitrary $|U_{\ell N}|^2$. At this HNL mass, the dominant production channels are $D_s^{+}\to \ell^+ N$ ($\ell = e,\mu$) and HNL decays to pions have the largest branching ratios, $N\to \ell^\mp\pi^\pm$ ($\ell = e,\mu$). As expected, the inclusion of the muon channels leads to an enhanced signal rate; the behaviour of the electron and muon channels are also similar, since both charged leptons can be considered massless compared to an HNL of mass 800~MeV. As we will explore in Sec.~\ref{sec:DUNE_3+2}, HNL decays such as $N\to \ell^\mp\pi^\pm$ that only proceed via one of the mixing strengths $|U_{\ell N}|^2$ can be used as observables of the phenomenological model in Sec.~\ref{sec:model}. While we will include the mixing to tau neutrinos when calculating the possible decays in a consistent scenario including the observed three-neutrino oscillations, a charged tau cannot be produced in the dominant meson decays considered. While tau neutrinos will be produced, it will not be possible to identify the specific flavour mixing to the HNL. 

\subsection{Extension to Other Setups}
\begin{table}[t!]
	\centering
	\renewcommand{\arraystretch}{1.25}
	\setlength\tabcolsep{6pt}
	\begin{tabular}{c|cccccccc|}
		\hline
		Experiment & Beam type &  $\Delta A_{\text{det}}~[\text{m}^2]$ & $\Delta\ell_{\text{det}}$ [m] & $\ell_{\text{det}}$ [m] & $N_\text{POT}$  \\ \hline
		DUNE & $120$ GeV, $p$ & $ 12$ & $5$ & $574$ & $6.6\times 10^{21}$  \\
		SHiP & $400$ GeV, $p$ & $50$ & $ 45$ & $50$ & $2.0\times10^{20}$   \\ \hline
	\end{tabular}
	\caption{Parameters of the upcoming fixed target experimental facilities, DUNE and SHiP, considered in our analysis. The beam type corresponds to a proton beam of given energy impinging on the target nuclei \cite{Abdullahi:2022jlv}.}
	\label{tab:fixedtarget}
\end{table}
We emphasise that the methods of this analysis are not restricted to the DUNE setup, and can be extended to other fixed target experimental facilities. The generic scheme would require the relevant proton beam energy and the associated meson production fractions, as well as a new geometrical cut for the specific experimental setup. Putting these together, one can then estimate the sensitivity using Eq.~\eqref{eq:DUNE_sensitivity}. To this end, we have modified our analysis for the well-studied SHiP proposal by changing the incoming proton beam energy to $400$~GeV as shown in Table~\ref{tab:fixedtarget}, the number of protons on target to $2\times 10^{20}$ and adjusting the detector geometry to a distance of $50$~m from the beam dump and a fiducial volume of length $45$~m and transverse cross section of $5\times10~\text{m}^2$~\cite{Abdullahi:2022jlv}. Our estimated upper bound on $|U_{eN}|^2$ is in good agreement with the results of~\cite{SHiP:2018xqw}, and can be improved further by introducing an additional attenuation factor to account for the exact geometrical cut (as the fiducial volume has a much longer length compared to DUNE). Note that we do not use any ad-hoc factors for DUNE since the detector length is small and so our assumed conical geometry works well.

\section{Probing HNLs with DUNE and LEGEND-1000}
\label{sec:compare}

We will begin this section with an analytical comparison of $0\nu\beta\beta$ decay and the direct production of HNLs in DUNE. Using the approximate results for the $0\nu\beta\beta$ decay half-life in Sec.~\ref{sec:0vbb_approx} and the DUNE ND event rate, we will estimate the HNL mass splitting $r_\Delta = \Delta m_N/m_N$ implied by the observation of $0\nu\beta\beta$ decay and an HNL-like DUNE signal. This will be done in the optimistic scenario where the light neutrino mass does not saturate the observed $0\nu\beta\beta$ decay half-life and the DUNE event rate is large. 

Following on from these estimates, we will perform a more detailed Bayesian analysis, using Markov chain Monte Carlo (MCMC) methods to sample the statistical likelihoods, given the HNL pair hypothesis, of positive signals in a $0\nu\beta\beta$ decay experiment and the DUNE ND. This will allow us to identify the 68\% and 95\% credible regions in the $1+2$ model parameter space for four benchmark scenarios. As mentioned, the $1+2$ scenario has been chosen to exemplify the potential of combining results from the two experiments in probing the nature and properties of the HNLs, especially their quasi-Dirac character. We will also consider the sensitivities in the scenario where $0\nu\beta\beta$ decay is observed but no HNL-like events are seen at DUNE, and vice versa. Finally, we will derive excluded regions in the $1+2$ model parameter space if neither $0\nu\beta\beta$ decay nor HNL-like DUNE events are seen. To extend, we will perform a Bayesian analysis implementing the current active neutrino data in the complete $3+2$ model with a massless lightest neutrino.

In the following, we will consider the upcoming search for $0\nu\beta\beta$ decay by the LEGEND experiment~\cite{LEGEND:2017cdu}, which will use a $^{76}\mathrm{Ge}$ detector to provide an excellent energy resolution and low intrinsic background. The next phase of the experiment will be LEGEND-1000~\cite{LEGEND:2021bnm}, which increases the mass of  $^{76}\mathrm{Ge}$ detector to 1 ton and an exposure leading to a sensitivity of $T^{0\nu}_{1/2} = 10^{28}$~y. We assume that DUNE collects data for 6 years in the neutrino beam mode, collecting $N_{\text{POT}} = 6.6\times 10^{21}$.

\subsection{Analytical Comparison}
\label{sec:analytal_comp}

\begin{figure}[t!]
	\centering
	\includegraphics[width=\textwidth]{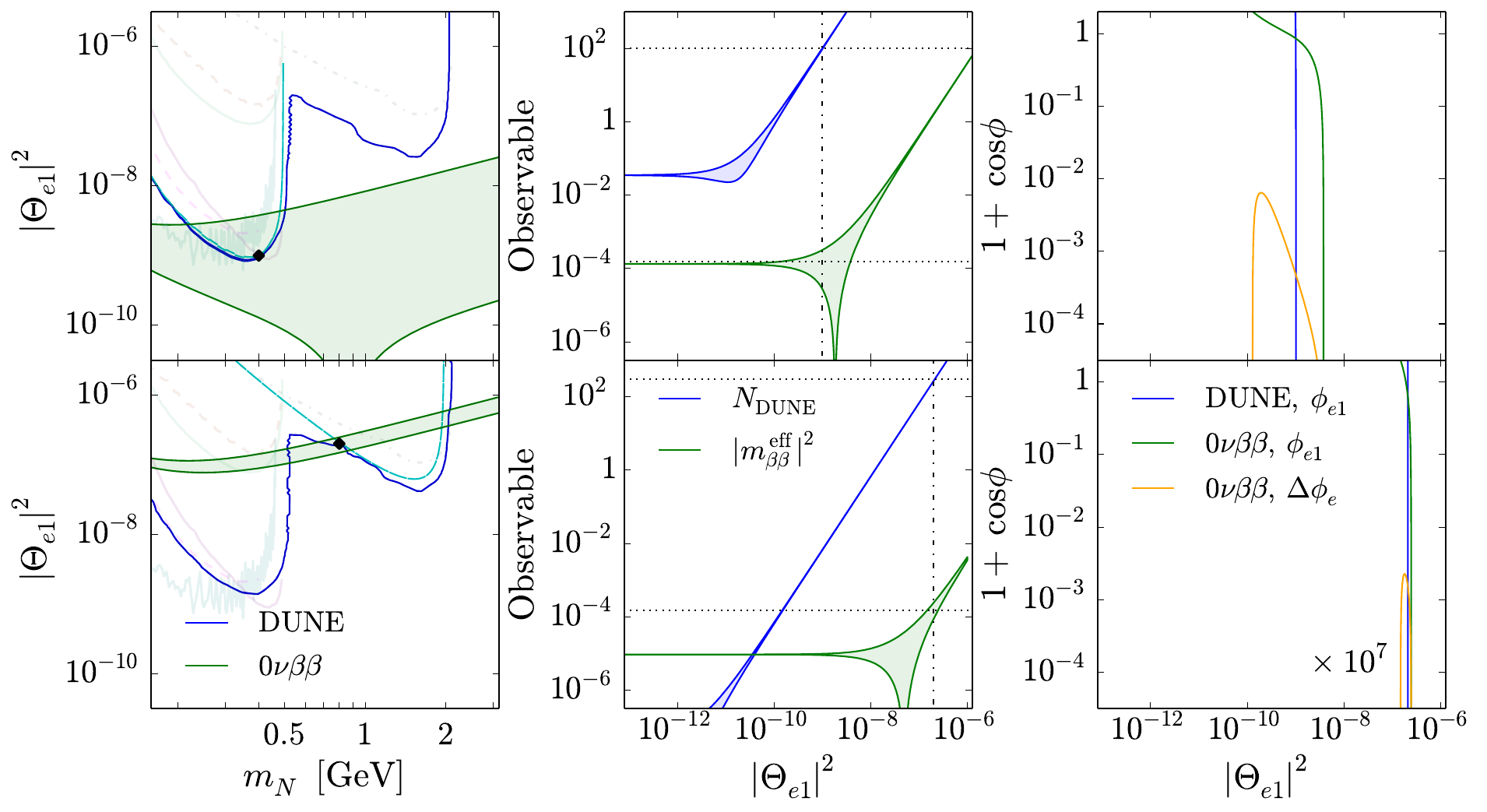}
	\caption{Constraints in the $1+2$ model parameter space if both $0\nu\beta\beta$ decay and HNL-like events at DUNE are observed, for $T_{1/2}^{0\nu} = 10^{28}$~yr, $N_\text{DUNE} = 100$, $r_\Delta = 0.1$ (top) and $T_{1/2}^{0\nu} = 10^{28}$~yr, $N_\text{DUNE} = 300$, $r_\Delta = 1.5\times 10^{-3}$ (bottom). Left: Regions in the $(m_N, |\Theta_{e1}|^2)$ plane from signals at DUNE (blue) and $0\nu\beta\beta$ decay (green), marginalising over $\cos\phi_{e1}$. The regions roughly overlap at the benchmark HNL masses (black diamond) $m_N = 400$~MeV (top) and $m_N = 800$~MeV (bottom). Middle: $N_\text{DUNE}$ (blue) and $|m_{\beta\beta}^\text{eff}|^2$ (green) as a function of $|\Theta_{e1}|^2$. Right: Implied values of $\cos\phi_{e1}$ from DUNE (blue) and $0\nu\beta\beta$ decay (green) signals as a function of $|\Theta_{e1}|^2$. Also shown is the implied value of $\cos\Delta\phi_{e1}$ from $0\nu\beta\beta$ decay (yellow) as a function of $|\Theta_{e1}|^2$.}
	\label{fig:0vbb_DUNE_compare}
\end{figure}
In Fig.~\ref{fig:0vbb_DUNE_compare}, we illustrate two signal scenarios in $0\nu\beta\beta$ decay and direct HNL searches. The plots on the upper (lower) row assume a $0\nu\beta\beta$ decay half-life of $10^{28}$ years from the LEGEND experiment and $100$ ($300$) HNL-like events at DUNE. Considering the constraints on the simple $1+2$ model, on the upper row we take the light neutrino mass to be $m_\nu  = 10^{-1.9}$~eV, which marginally saturates the observed $0\nu\beta\beta$ decay half-life. Below, we instead take $m_\nu  = 10^{-2.5}$~eV, which does not saturate the half-life and therefore implies a dominant HNL contribution, as discussed in Sec.~\ref{sec:0vbb}.

The left-hand plots in Fig.~\ref{fig:0vbb_DUNE_compare} show the regions in the $(m_N,|\Theta_{e1}|^2)$ plane implied by the positive $0\nu\beta\beta$ decay (green region) and DUNE (blue region) signals, with the spread from varying $\cos\phi_{e1}$ (only visible for $0\nu\beta\beta$ decay). In the $1+2$ model, the remaining relevant parameter is the mass splitting $r_\Delta$; in the upper (lower) row, we take $r_{\Delta} = 0.1$ ($r_{\Delta} = 1.5\times 10^{-3}$) which results in the $0\nu\beta\beta$ decay and DUNE regions overlapping at the benchmark point of $m_N = 400$~MeV ($800$~MeV) and $|\Theta_{e1}|^2 = 10^{-9}$ ($2\times 10^{-7}$),  just below the current upper bound from T2K (CHARM). 

We also show the DUNE region using an approximate formula for the number of signal events induced by the HNL pair in the DUNE ND (dashed cyan). For $m_N = 400$~MeV ($800$~MeV), the dominant HNL production channel and decay mode to charged final states are $K^{+}\to e^{+}N$ ($D_s^{+}\to e^{+}N$) and $N\to e^{\mp}\pi^{\pm}$, respectively. We can then estimate the number of events as
\begin{align}
\label{eq:Ndet_HNLpair}
    N_{\text{DUNE}} \approx C_{\text{det}}^P\bigg[\mathcal{A}_{PP'}(m_N)|\Theta_{e1}|^4 + \mathcal{A}_{PP'}\big(m_N(1+r_\Delta)\big)\frac{|r_\nu + |\Theta_{e1}|^2e^{i\phi_{e1}}|^2}{(1+r_\Delta)^2}\bigg]\,,
\end{align}
where we have used the expression for $|\Theta_{e2}|^2$ in the $1+2$ model in Eq.~\eqref{eq:se2} and
\begin{align}
    \mathcal{A}_{PP'}(m_N) \equiv \frac{m_N \Gamma_N}{|U_{eN}|^4}\,\text{Br}(P^+\to e^{+}N)\text{Br}(N\to e^{\pm}P^{\prime\mp})\,.
\end{align}
Here, $\text{Br}(P^+\to e^{+}N)$ and $\text{Br}(N\to e^{\pm}P^{\prime\mp})$ are given by Eqs.~\eqref{eq:prod_BR} and \eqref{eq:det_BR}, respectively. For convenience, we have introduced the factor
\begin{align}
    C_{\text{det}}^P \equiv N_{P}\epsilon_{\text{geo}}\frac{\Delta\ell_{\text{det}} \epsilon_{\text{det}}}{\langle p_{N_z}\rangle} = N_{P}\frac{\langle p_{N_z}\rangle}{\langle p_{N_T}\rangle^2}\frac{V_{\text{det}} \epsilon_{\text{det}}}{L^2}\,,
\end{align}
where $V_{\text{det}}$ is the fiducial volume and we use the estimate $\epsilon_{\text{geo}} = \langle{p_{N_z}}\rangle^2/\langle{p_{N_T}}\rangle^2\times A_{\text{det}}/L^2$ for the geometric efficiency, with $\langle{p_{N_z}}\rangle$ and $\langle{p_{N_T}}\rangle$ the average longitudinal and transvese HNL momenta. For $N_{\text{DUNE}}^{\text{exp}} = 100$ $(300)$, we use Eq.~\eqref{eq:Ndet_HNLpair} to plot the dashed cyan lines in Fig.~\ref{fig:0vbb_DUNE_compare}. The expressions above are simplified if we consider the limit $|\Theta_{e1}|^2 \gg r_{\nu}$, i.e., the \textit{inverse seesaw} regime. This allows to pull out a factor of $|\Theta_{e1}|^4$ in Eq.~\eqref{eq:Ndet_HNLpair} and write
\begin{align}
\label{eq:Ndet_ISS}
    |\Theta_{e1}|^2 &= \sqrt{\frac{N_{\text{DUNE}}}{C_{\text{det}}^P \mathcal{B}_{PP'}}}\quad (|\Theta_{e1}|^2 \gg r_\nu)\,, \nonumber\\
    \mathcal{B}_{PP'} &\equiv \mathcal{A}_{PP'}(m_N) + \frac{\mathcal{A}_{PP'}\big(m_N(1+r_\Delta)\big)}{(1+r_\Delta)^2} \,.
\end{align}
In the limit of small mass splitting between the HNL pair, $r_\Delta \ll 1$, the function simplifies to $\mathcal{B}_{PP'} \approx 2\mathcal{A}_{PP'}(m_N)$. We see that Eq.~\eqref{eq:Ndet_ISS} is effectively independent of $\phi_{e1}$.

The two plots in the centre of Fig.~\ref{fig:0vbb_DUNE_compare} show the value of $N_{\text{DUNE}}$ (using the approximate formula in Eq.~\eqref{eq:Ndet_HNLpair}) and $|m_{\beta\beta}^{\text{eff}}|^2$ as a function of $|\Theta_{e1}|^2$, for $m_N = 400$~MeV ($800$~MeV) and allowing the value of $\cos\phi_{e1}$ to vary. The two horizontal dashed lines indicate the measured experimental values $N_{\text{DUNE}}^{\text{exp}}$ and $|m_{\beta\beta}^{\text{exp}}|^2$, while the vertical dot-dashed line shows the value of $|\Theta_{e1}|^2$ compatible with the two measurements. 

Finally, the two plots to the right of Fig.~\ref{fig:0vbb_DUNE_compare} show the values of $1+\cos\phi_{e1}$ determined from the $0\nu\beta\beta$ decay and DUNE signals using Eqs.~\eqref{eq:cosphie1_0vbb} and \eqref{eq:Ndet_HNLpair}. It is evident that DUNE has little sensitivity to the value of $\cos\phi_{e1}$. The crossing point of these curves indicates the compatible values of $|\Theta_{e1}|^2$ and $\phi_{e1}$ (for given values of $m_\nu$, $m_N$ and $r_\Delta$). We also show the values of $1+\cos\Delta\phi_{e}$, in the lower plot being multiplied by a factor of $10^7$ to be visible. We see that $\cos\Delta\phi_{e} \approx -1$ is being probed in both scenarios.

In the scenario where the light neutrino mass does not saturate $0\nu\beta\beta$ decay and the HNL-like event rate in DUNE is large, we will now show that the value of $r_\Delta$ being probed by both experiments is approximately insensitive to the value of $\cos\phi_{e1}$. Using Eq.~\eqref{eq:Ndet_ISS}, the active-sterile mixing implied by an observation at DUNE is given approximately by
\begin{align}
    |\Theta_{e1}|^2\approx 
    2\times 10^{-7}
    &\bigg(\frac{N^{\text{exp}}_{\text{DUNE}}}{300}\bigg)^{1/2}\bigg(\frac{6.6\times 10^{21}}{N_{\text{POT}}}\bigg)^{1/2}\bigg(\frac{5~\text{m}}{\Delta \ell_{\text{det}}}\bigg)^{1/2} \nonumber\\
    \times&\bigg(\frac{7.3\times 10^3~\text{MeV}^2}{\mathcal{A}_{PP'}(m_N)}\bigg)^{1/2}\,,
\end{align}
where we have used $\mathcal{A}_{D_s\pi}(800~\text{MeV}) = 7.3\times 10^3~\text{MeV}^2$. With this active-sterile mixing, we can then solve Eq.~\eqref{eq:se1_0vbb} for $r_\Delta$. If the value of $N_{\text{DUNE}}^{\text{det}}$ is large, and therefore the implied value of $r_\Delta$ small, we obtain the approximate result:
\begin{align}
    r_{\Delta}\sim 1.5\times 10^{-3} \bigg(\frac{2\times 10^{-7}}{|\Theta_{e1}|^2}\bigg)\bigg(\frac{m_N}{800~\text{MeV}}\bigg)\bigg(\frac{10^{28}~\text{yr}}{T_{1/2}^{0\nu}}\bigg)^{1/2} \,.
\end{align}
This estimate is not possible if the light neutrino mass \textit{does} almost saturate the observed $0\nu\beta\beta$ decay half-life, because there is then much more freedom in the value of $\cos\phi_{e1}$. To better understand this region of the parameter space, we perform a statistical analysis of the experimental likelihoods in the next section.

\subsection{Statistical Analysis}

To incorporate statistical uncertainties, we perform a simple Bayesian analysis based on a combined likelihood for DUNE and LEGEND-1000. We treat both as simple counting experiments and to avoid running an ensemble of mock experiments, we take the continuous version of the Poisson distribution,
\begin{align}
    \label{eq:Poisson-lambda}
   \operatorname{Poisson}\left(n_{\mathrm{obs}} \mid \lambda_{\mathrm{sig}}(\boldsymbol{\theta})+\lambda_{\mathrm{bkg}}(\boldsymbol{\theta})\right) \propto \frac{\left(\lambda_{\mathrm{sig}}(\boldsymbol{\theta})+\lambda_{\mathrm{bkg}}(\boldsymbol{\theta})\right)^{n_{\mathrm{obs}}} e^{-\left(\lambda_{\mathrm{sig}}(\boldsymbol{\theta})+\lambda_{\mathrm{bkg}}(\boldsymbol{\theta})\right)}}{\Gamma\left(n_{\mathrm{obs}}+1\right)}\,.
\end{align}
Here, $n_\text{obs}$ is the `observed' number of events, expected in a hypothetical benchmark scenario (or $n_\text{obs} = 0$, if the experiment is assumed to see no signal). The expected number of signal and background events given a theoretically-predicted parameter choice $\boldsymbol{\theta}$ are $\lambda_\text{sig}(\boldsymbol{\theta})$ and $\lambda_\text{bkg}$, respectively.

For $0\nu\beta\beta$ decay, the number of signal events in the $1+2$ model, with the parameters $\boldsymbol{\theta} = \{m_\nu, m_N, r_\Delta, |\Theta_{e1}|^2, \phi_{e1}\}$, can be calculated as
\begin{align}
\label{eq:lambda_0vbb}
    \lambda^{0\nu}_\text{sig}(\boldsymbol{\theta}) = \frac{\ln 2 \cdot N_A \cdot \mathcal{E}}{m_A \cdot T_{1/2}^{0\nu}\left(\boldsymbol{\theta}\right)}\,,
\end{align}
where $N_{A}$ is Avogadro's number, $m_A$ is the average molar mass of the $0\nu\beta\beta$ decaying material ($m_A = 75.74$~g/mol for $^{76}$Ge), and $\mathcal{E}$ is the sensitive exposure of the detector, with $\mathcal{E} = 6632~\text{kg} \cdot$yr for LEGEND-1000~\cite{Agostini:2022zub, Agostini:2022bjh}. The half-life is calculated from Eq.~\eqref{eq:0vbb_rate} using the full interpolating formula discussed in Sec.~\ref{sec:0vbb}. The background event rate is $\lambda^{0\nu}_\text{bkg} = \mathcal{E}\cdot \mathcal{B}$, where $\mathcal{B}$ is the sensitive background, giving $\lambda^{0\nu}_\text{bkg} = 0.4$ events for LEGEND-1000. Although the $0\nu\beta\beta$ decay NMEs have considerable discrepancies between different nuclear structure calculations and it may be possible that the nuclear axial coupling is quenched~\cite{Deppisch:2016rox}, we assume that there are no theoretical uncertainties to explore the full potential of future $0\nu\beta\beta$ decay searches.

For DUNE, we use the sensitivity formula derived in Sec.~\ref{sec:DUNE}. For the expected number of signal events in the $1+2$ model, we incoherently add up the contribution Eq.~\eqref{eq:DUNE_sensitivity} from each HNL separately, i.e.,
\begin{align}
\label{eq:lambda_DUNE1}
    \lambda^\text{DUNE}_\text{sig}(\boldsymbol{\theta})
    = N_\text{sig}(m_N,|\Theta_{e1}|^2) 
    + N_\text{sig}\left(m_N(1+r_\Delta),|\Theta_{e2}|^2\right) \,,
\end{align}
where $|\Theta_{e2}|^2$ is given by Eq.~\eqref{eq:se2} as a function of $m_\nu$, $m_N$, $r_\Delta$ $|\Theta_{e2}|^2$ and $\cos\phi_{e1}$. We assume a negligible background for HNL-like events in the DUNE ND, $\lambda^\text{DUNE}_\text{bkg} = 0$.

\begin{table}
\centering
\renewcommand{\arraystretch}{1.25}
\setlength\tabcolsep{6pt}
\begin{tabular}{c|cccc|ccc} 
	\hline
	Scenario & $m_{N}$ [MeV] & $|\Theta_{e1}|^2$ & $r_\Delta$ & $m_\nu$ [eV]& 
	$\lambda_\text{DUNE}$ & $\lambda_{0\nu}$ & $T^{0\nu}_{1/2}$ [yr] \\
	\hline
	1 & 400 & $10^{-9.0}$ & $10^{-0.5}$ & $10^{-1.9}$ & 76.7 & 5.94 & $10^{27.8}$ \\ 
	2 & 400 & $10^{-9.0}$ & $10^{-0.5}$ & $10^{-2.5}$ & 76.7 & 2.73 & $10^{28.1}$ \\
	3 & 800 & $10^{-6.7}$ & $10^{-2.5}$ & $10^{-1.9}$ & 325  & 15.5 & $10^{27.4}$ \\
	4 & 800 & $10^{-6.7}$ & $10^{-2.5}$ & $10^{-2.5}$ & 325  & 12.3 & $10^{27.5}$ \\
	\hline
\end{tabular}
\caption{Benchmark scenarios for the light neutrino mass $m_\nu$, the HNL mass $m_N$, the active-sterile mixing strength $|\Theta_{e1}|^2$ and the HNL mass splitting $r_\Delta$ adopted in our statistical analysis of the $1+2$ model. In all cases, the HNL phase parameter is $\cos\phi_{e1} = 0$. Also given are the expected number of signal events at DUNE, $\lambda_\text{DUNE}$, and LEGEND-1000, $\lambda_{0\nu}$, as well as the corresponding half-life for the latter, $T_{1/2}^{0\nu}$.}
\label{tab:benchmarks}
\end{table}
In our analysis, we consider four benchmark scenarios, listed in Table~\ref{tab:benchmarks}. The first two benchmarks are chosen for an HNL mass $m_N = 400$~MeV with an active-sterile mixing strength just below the current bound from T2K, $|\Theta_{e1}|^2 < 1.5\times 10^{-9}$ at 90\% CL \cite{T2K:2019jwa}. The third and fourth benchmarks are at $m_N = 800$~MeV with an active-sterile mixing strength well below the current bound from CHARM, $|\Theta_{e1}|^2 < 4\times 10^{-7}$ at 90\% CL \cite{CHARMII:1994jjr}. The benchmark points are indicated in Fig.~\ref{fig:constraints}. The light neutrino mass is chosen to be either $m_\nu = 10^{-1.9}$~eV or $10^{-2.5}$~eV. These roughly correspond to the smallest and largest possible effective Majorana masses for the light neutrinos in the IO and NO, respectively. The benchmark scenarios are further distinguished by having a rather large HNL mass splitting $r_\Delta = \Delta m_N / m_N = 10^{-0.5}$ in scenarios 1 and 2 (being close to the seesaw line, the HNLs are not required to form a quasi-Dirac pair) and a small splitting $r_\Delta = 10^{-2.5}$ in scenarios 3 and 4. The benchmark scenarios are chosen to represent different parameter regions of interest that yield appreciable number of events, both at DUNE and LEGEND-1000, also listed in Table~\ref{tab:benchmarks}. As expected, the DUNE event rate is not affected by the light neutrino mass, being only sensitive to the incoherent sum of the HNL contributions. On the other hand, the $0\nu\beta\beta$ decay event rate is affected by $m_\nu$, as well as the HNL phase parameter $\cos\phi_{e1}$, which is chosen to be $\cos\phi_{e1} = 0$ in all scenarios, as the light neutrino and HNL contributions are coherently added.

In the following, we discuss four hypothetical experimental outcomes, namely, (A) both DUNE and LEGEND-1000 observe a signal, (B) DUNE observes signal events but not LEGEND-1000, (C) LEGEND-1000 observes signal events but not DUNE and (D) both DUNE and LEGEND-1000 see no signal events. We sample the parameter space using an MCMC to find the posterior parameter distribution for a given likelihood formed by the Poisson-Lambda distribution in Eq.~\eqref{eq:Poisson-lambda}, for DUNE and LEGEND-1000. We scan over the parameters $\log_{10}(|\Theta_{e1}|^2)$, $\log_{10}(r_\Delta)$ and $\cos\phi_{e1}$ in the range $[-12,-6]$, $[-4,0]$ and $[-1,1]$, respectively, using a flat prior, in analysis (a) and (b), i.e., $m_N$ is assumed to be known and fixed to the value in the benchmark scenario considered. In (c) and (d), $m_N$ is also scanned over the range $[0.1, 1.0]$~MeV.

As mentioned, in the analyses (a) and (b), DUNE is assumed to have made an observation of HNL events and we further assume that the HNL mass has been determined to a sufficient precision, via reconstruction using the kinematical properties of HNL decay final states. We have checked that a mass uncertainty of 10\% would not affect the conclusions of our analysis, but as an estimation of the mass precision at DUNE (and SHiP) is still outstanding, we have not included in our analysis. Given the fairly large number of expected events at DUNE in the benchmark scenarios, cf. Table~\ref{tab:benchmarks}, we expect that a kinematical reconstruction of the HNL mass should be feasible, conservatively at the 10\%-level \cite{Coloma:2023oxx}. This will be especially the case using two-body, fully visible decays such as $N\to \ell^\mp \pi^\pm$. As far as we know, an analysis of the mass precision at SHiP is still outstanding as well, but a reconstruction at a similar, if not better level, should be possible \cite{SHiP:2015vad}. Our analysis strongly motivates efforts to explore precision mass determination of HNLs on observation. 

For each benchmark point, we insert the expected event rate for LEGEND-1000 and DUNE, following the Asimov data set approach~\cite{Cowan:2010js}, setting the number of observed events $n_\text{obs}$ for each experiment equal to the number of expected events $\lambda_\text{sig}(\boldsymbol{\theta}) + \lambda_\text{bkg}$ in the given benchmark scenario. In case of no signal, we set $n_\text{obs} = \lambda_\text{bkg}$. By construction, assuming a signal will reconstruct the parameter values of the benchmark scenarios used, apart from statistical errors and parametric degeneracies. If no signal at an experiment is assumed, the recovered parameter space will reflect that, and it will not include the benchmark point itself.

\subsubsection{Signal at DUNE and Signal at LEGEND-1000 (A)}

\begin{figure}[t!]
	\centering
	\setkeys{Gin}{width=0.45\linewidth}
	\includegraphics[width=0.49\linewidth]{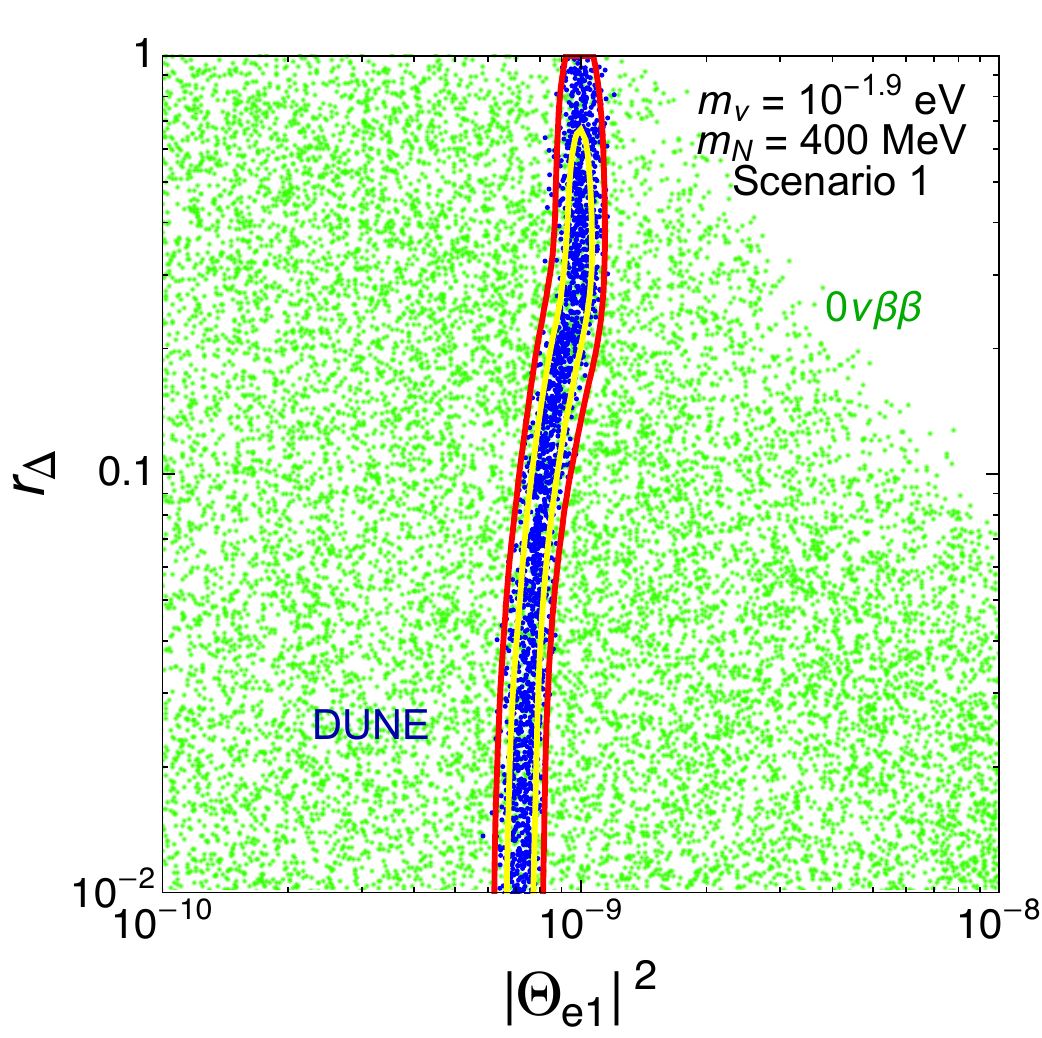} 
	\includegraphics[width=0.49\linewidth]{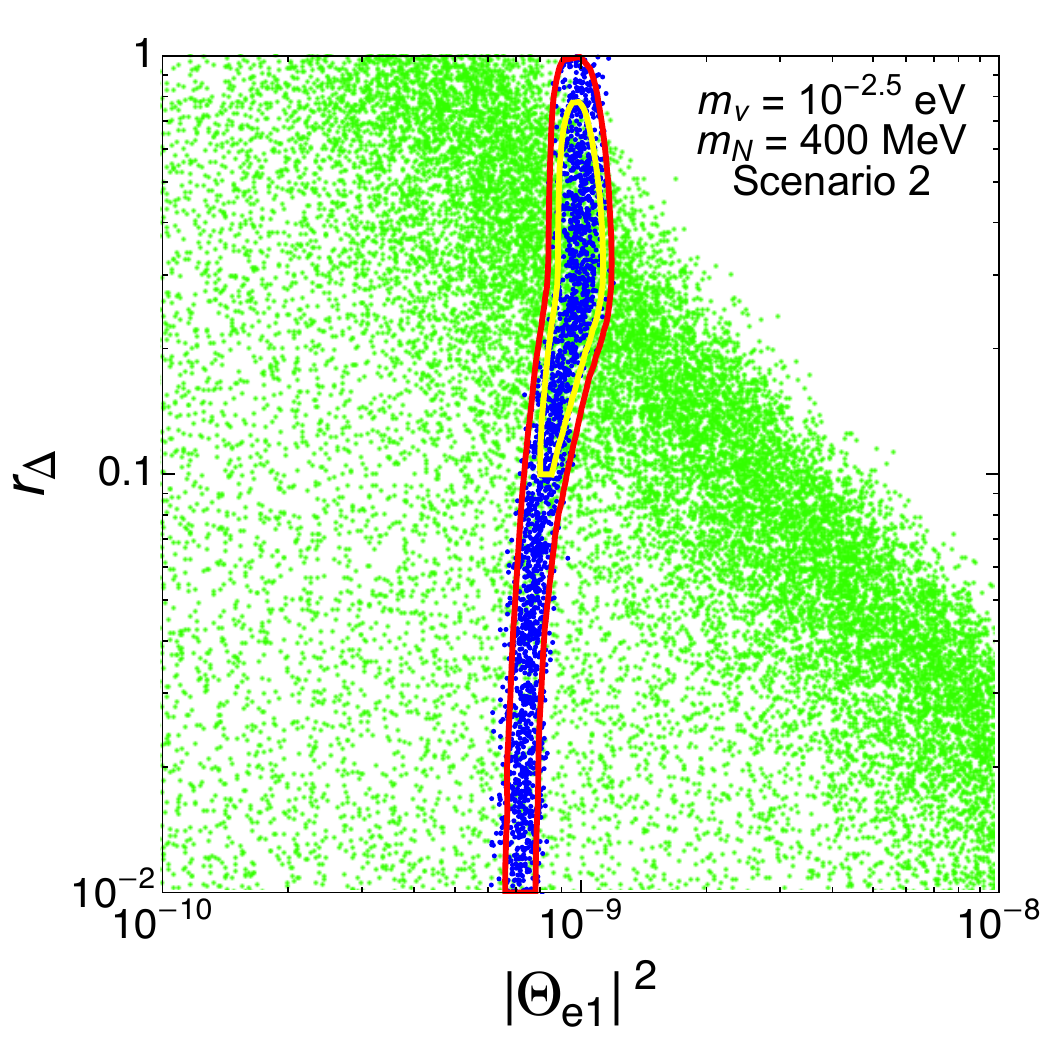}\\
	\includegraphics[width=0.49\linewidth]{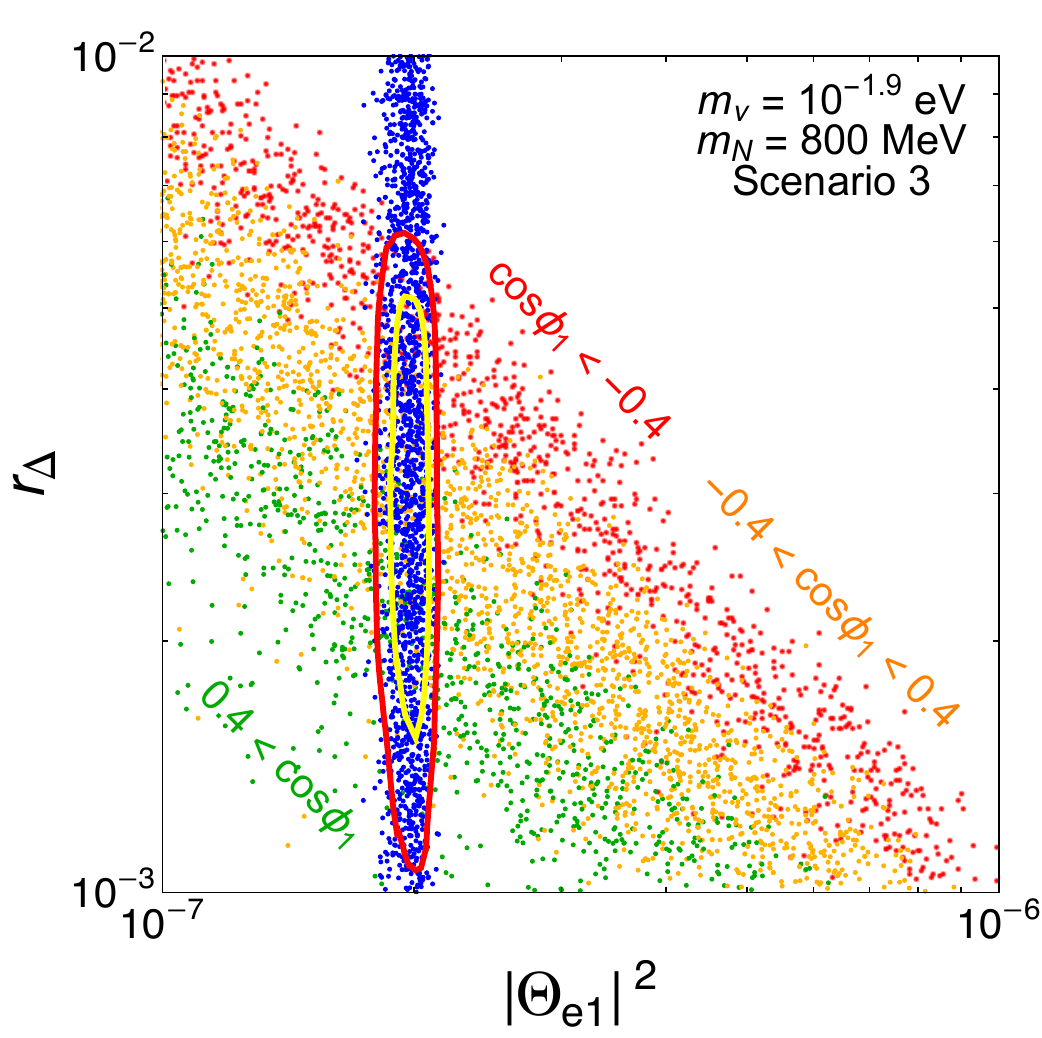} 
	\includegraphics[width=0.49\linewidth]{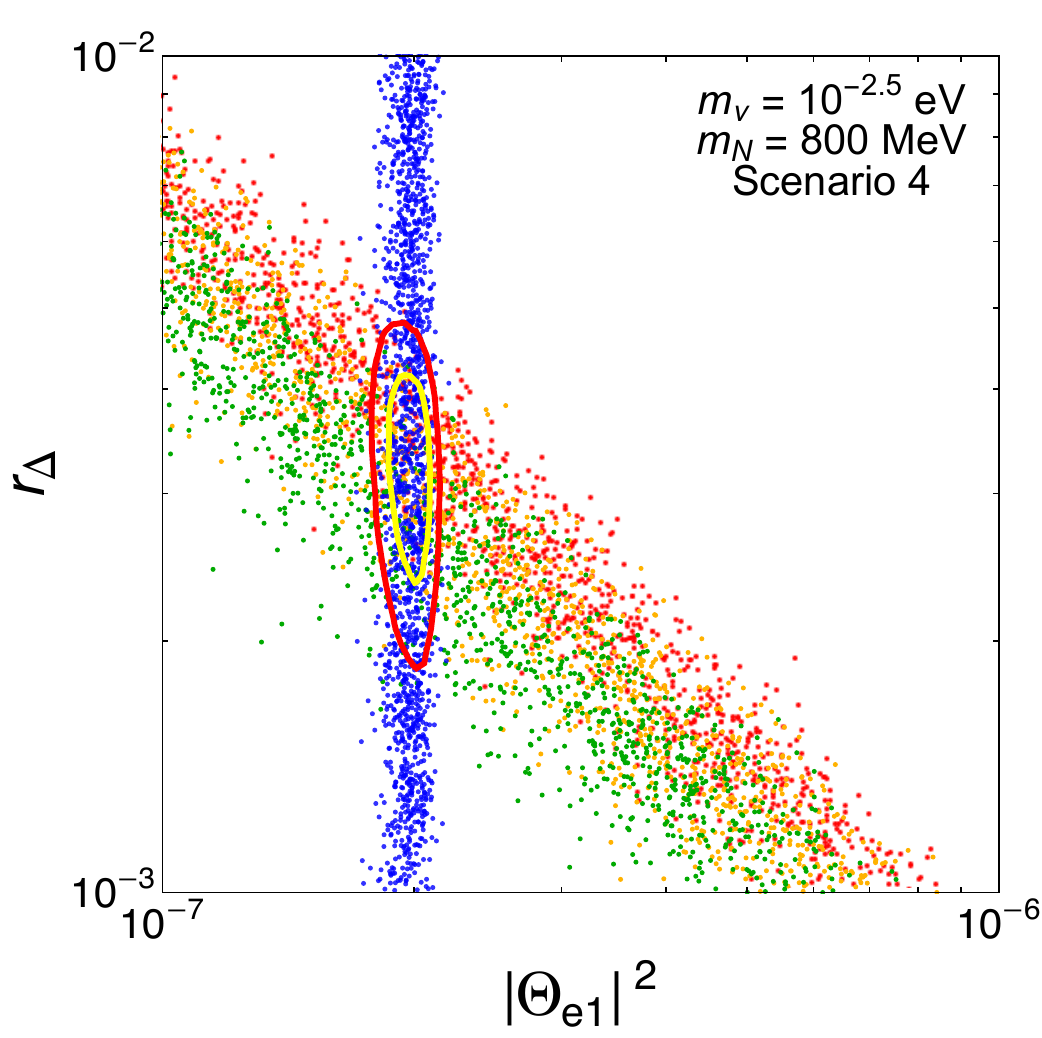}
	\caption{Posterior distribution marginalized to the $(|\Theta_{e1}|^2, r_\Delta)$ parameter plane assuming an observation at DUNE (blue points), LEGEND-1000 (diagonal green or coloured points) or both simultaneously (yellow and red contours representing the 68\% and 95\% credible regions) in the four benchmark scenarios. In the bottom plots, the LEGEND-1000 points are colour-coded according to the value of the HNL phase parameter $\cos\phi_{e1}$ as indicated.}
\label{fig:DUNEandDBD}
\end{figure}
In Fig.~\ref{fig:DUNEandDBD}, we illustrate the posterior distributions resulting from the MCMC, marginalized to the $(|\Theta_{e1}|^2, r_\Delta)$ parameter plane. In this scenario, we make the most optimistic assumption of an observation of the expected number of events at DUNE (near vertical blue points) and LEGEND-1000 (diagonal green or coloured points) as per Table~\ref{tab:benchmarks} for the four benchmark scenarios. In the top plots, the red, orange and green points for observed $0\nu\beta\beta$ decay are filtered according to the value of the HNL phase parameter $\cos\phi_{e1}$, namely $\cos\phi_{e1} < -0.4$, $-0.4 \leq \cos\phi_{e1} \leq 0.4$ and $\cos\phi_{e1}> 0.4$, respectively. The points thus indicate the posterior distributions assuming observation in either DUNE or LEGEND-1000 but not both. The combined posterior assuming observation in both is illustrated by the red and yellow contours indicating the 68\% and 95\% credible regions, respectively.

In all benchmark scenarios, the number of DUNE events is fairly large, resulting in a narrow, near vertical band, essentially fixing the value of $|\Theta_{e1}|^2$. The deviation from vertical in the top plots is due to the large mass splitting $r_\Delta \gtrsim 0.1$ and the resulting dependence on the mass of the second HNL $N_2$. In Scenarios 1 and 2 with $m_N = 400$~GeV (top plots), the number of events at LEGEND-1000 is, on the other hand, small, near the detection limit. This does not allow a definite determination of the model parameters, only setting a weak limit $r_\Delta \lesssim 1$ on the mass splitting. In Scenarios 3 and 4 (bottom plots), the number of expected LEGEND-1000 events is, instead, much larger, constraining the parameter space to a diagonal band. This reflects the fact that the contribution of the HNLs is constrained to be sufficiently small, either due to a small active-sterile mixing or a small mass splitting with a resulting quasi-Dirac nature of the HNLs. In combination with the DUNE observation this would allow a measurement of the mass splitting $r_\Delta \approx 10^{-2.5}$, with a higher precision in Scenario 4, where the light neutrino contribution is not saturating the $0\nu\beta\beta$ decay half life.

\begin{figure}[t!]
	\centering
	\setkeys{Gin}{width=0.45\linewidth}
	\includegraphics[width=0.49\linewidth]{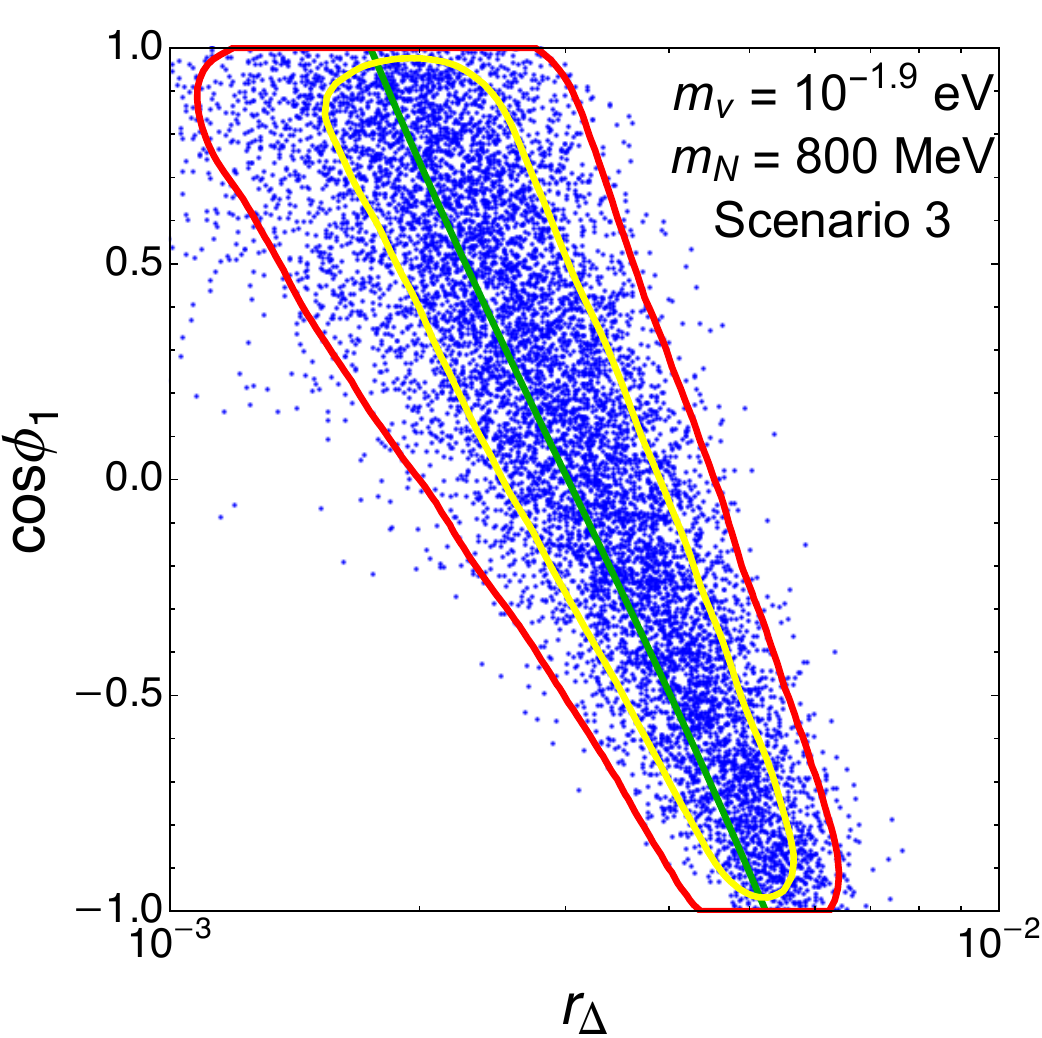}
	\caption{Posterior distribution (blue points and yellow/red contours) marginalized to the $(\cos\phi_{e1}, r_\Delta)$ plane assuming observations at both DUNE and LEGEND-1000 in benchmark Scenario 3. The green line shows the analytic relation in Eq.~\eqref{eq:cosphie1_0vbb}.}
\label{fig:cos-rD}
\end{figure}
\begin{figure}[t!]
	\centering
	\setkeys{Gin}{width=0.45\linewidth}
	\includegraphics[width=0.49\linewidth]{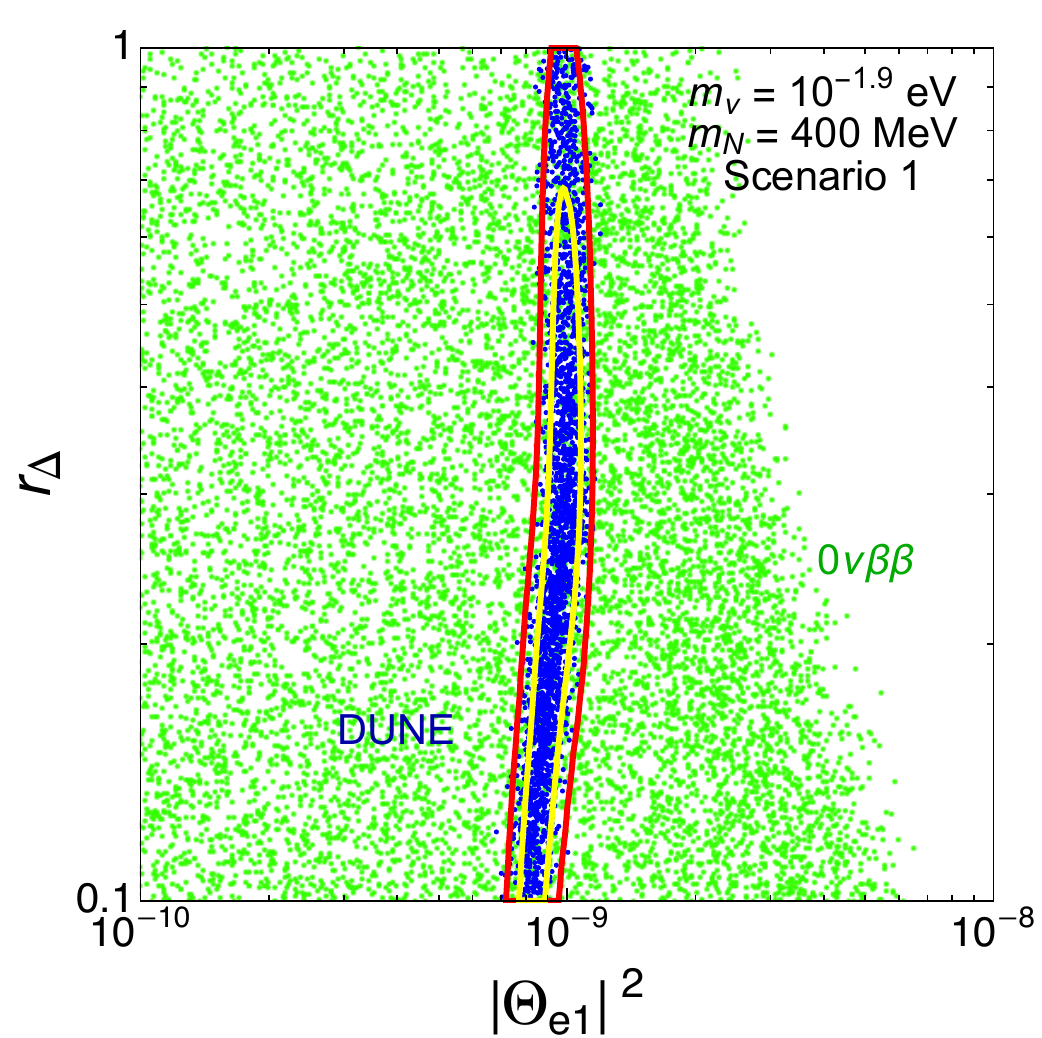} 
	\includegraphics[width=0.49\linewidth]{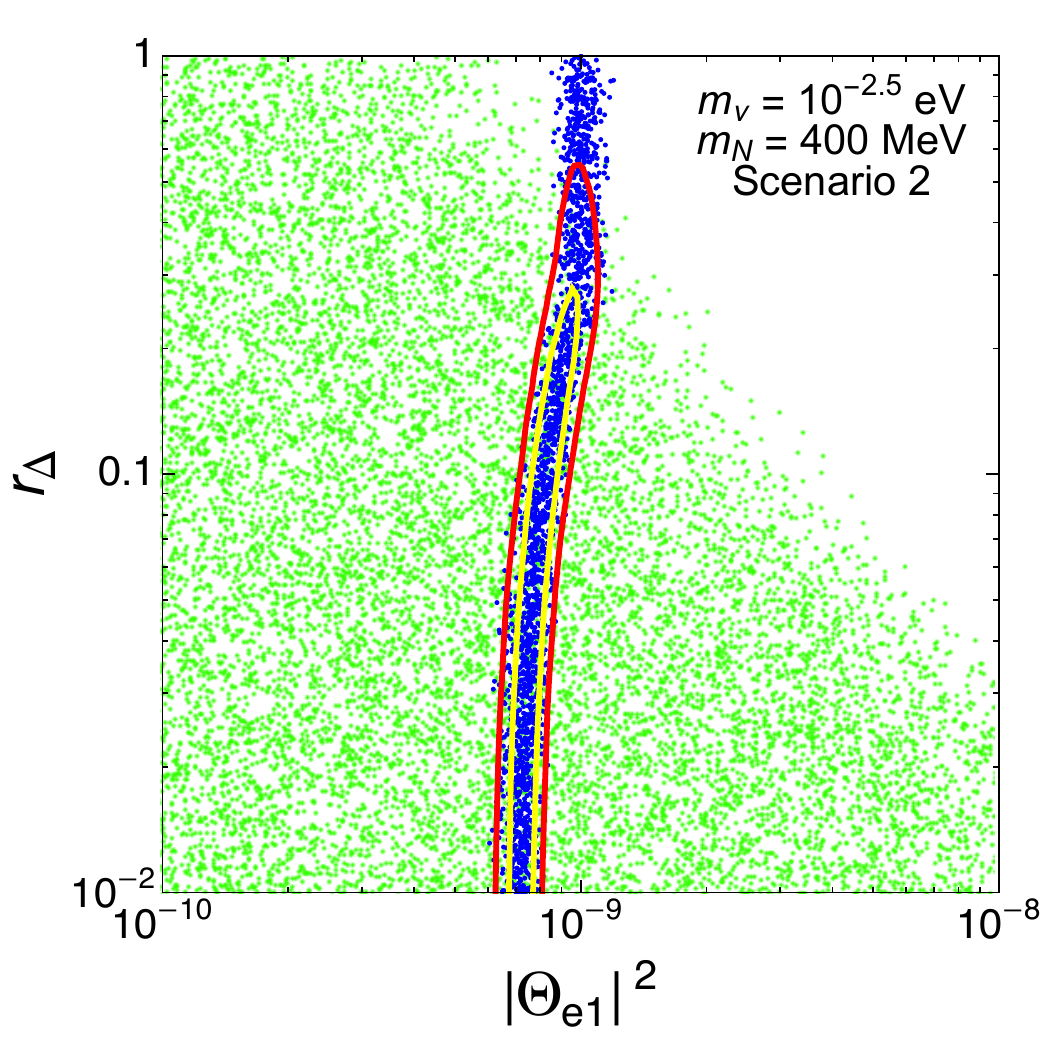}\\
	\includegraphics[width=0.49\linewidth]{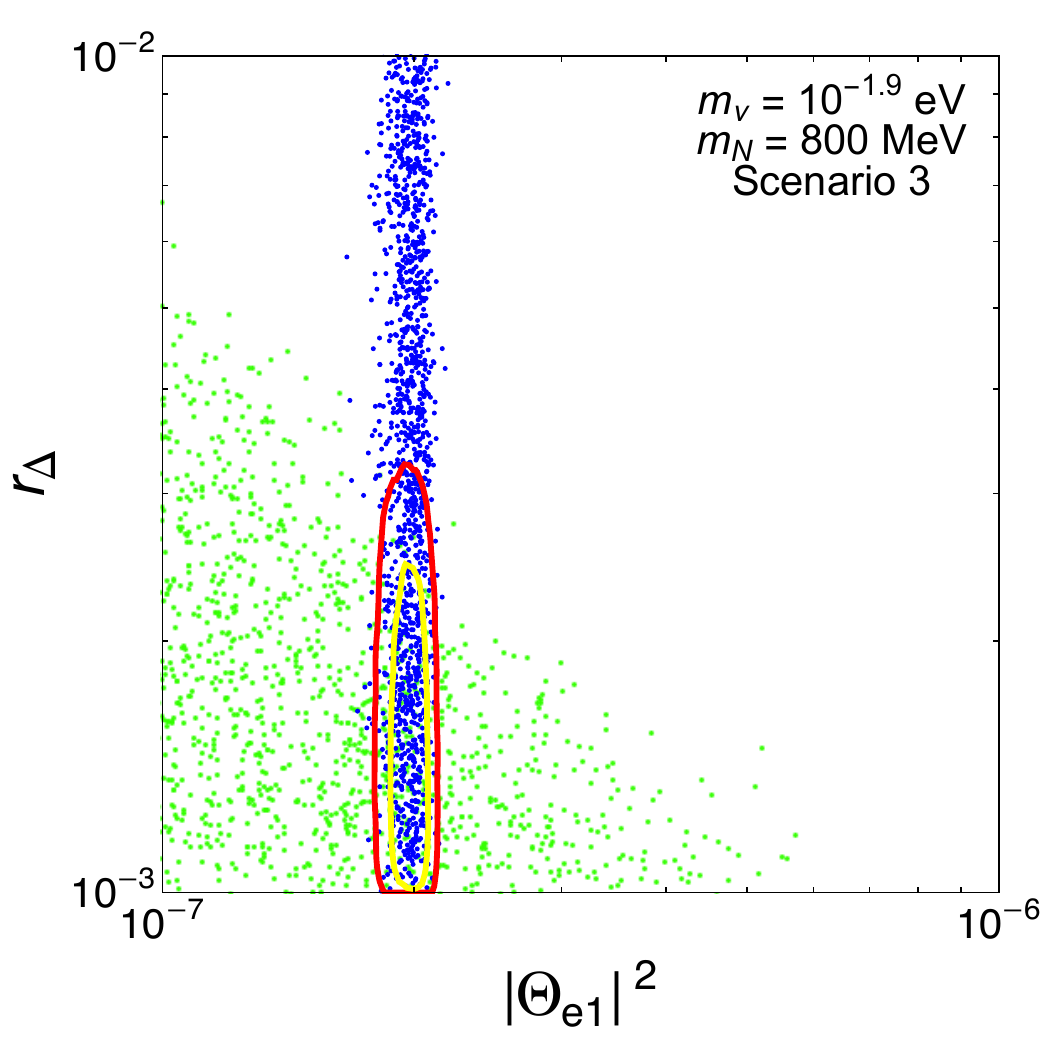} 
	\includegraphics[width=0.49\linewidth]{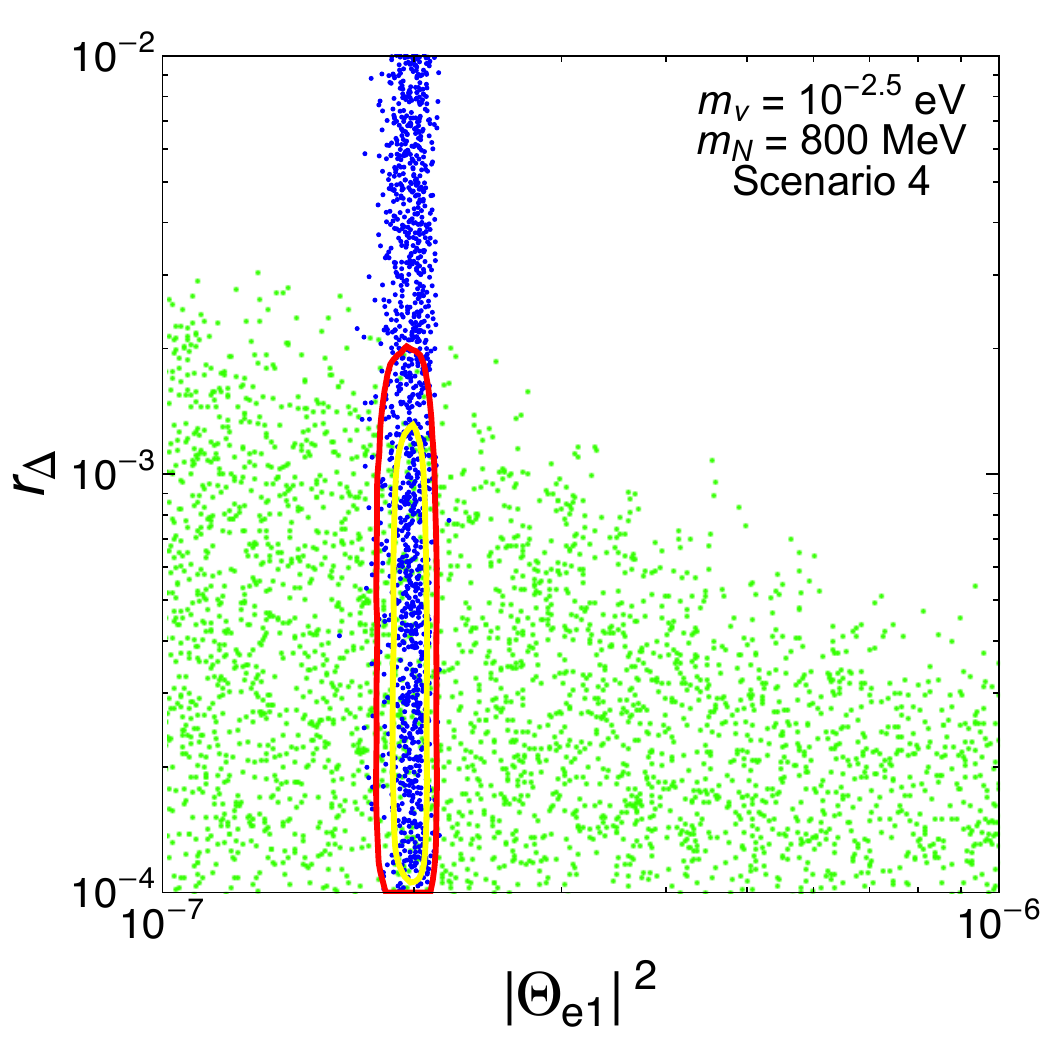}
	\caption{As Fig.~\ref{fig:DUNEandDBD}, but assuming a signal at DUNE with no events at LEGEND-1000 in the four benchmark scenarios.}
	\label{fig:NoDBD}
\end{figure}
As can be seen in Fig.~\ref{fig:DUNEandDBD}, DUNE is effectively insensitive to the light neutrino mass and HNLs with $r_\Delta < 0.1$, whereas $0\nu\beta\beta$ decay is sensitive to both the HNL mass splitting ratio and the active-sterile mixing. In addition, $0\nu\beta\beta$ decay is also sensitive to the HNL phase $\phi_{e1}$, namely, the interference term proportional to $\cos\phi_{e1}$ in Eq.~\eqref{eq:mbb_eff_simp} is important if the light and heavy neutrino contributions are of a similar size. In Fig.~\ref{fig:DUNEandDBD}~(bottom), the widths of the $0\nu\beta\beta$ bands depends on the light neutrino mass; when the light neutrino contribution saturates the $0\nu\beta\beta$ decay half-life, as in Scenario 3, the band of allowed points is much wider. This is in agreement with the analytical behaviour discussed in Sec.~\ref{sec:analytal_comp}. Increasing the HNL mass can be seen to shift the allowed region to smaller $|\Theta_{e1}|^2$ and $r_{\Delta}$ values. The resulting degeneracy between $\cos\phi_{e1}$ and $r_\Delta$ is displayed in Fig.~\ref{fig:cos-rD}, showing the posterior distribution in Scenario 3 marginalized to this parameter plane. The points and contours both illustrate the combined distribution assuming observation at both DUNE and LEGEND-1000. The central green line results from the analytical relation in Eq.~\eqref{eq:cosphie1_0vbb}.

\subsubsection{Signal at DUNE but No Signal at LEGEND-1000 (B)}

In Fig.~\ref{fig:NoDBD}, we analogously display the posterior distribution in $(|\Theta_{e1}|^2, r_\Delta)$, assuming an observation at DUNE (blue points) but no events at LEGEND-1000 (green points), for the four benchmark scenarios. The yellow and red contours show the 68\% and 95\% credible regions combining both observations. As mentioned earlier, in this case the number of $0\nu\beta\beta$ decay signal events is not set according to Table~\ref{tab:benchmarks} but $\lambda_{0\nu}^\text{sig} = 0$ is imposed. With regard to DUNE the assumptions are identical to (A) above and thus the blue point distributions are the same. 

With no $0\nu\beta\beta$ decay events, the corresponding regions extend to arbitrarily small values of $r_\Delta$ but they also partially overlap with those in Fig.~\ref{fig:DUNEandDBD} where $0\nu\beta\beta$ decay is assumed to be observed. This is due to the inherent statistical uncertainties but also because cancellations in the $0\nu\beta\beta$ decay rate can occur, especially for a large light neutrino mass $m_\nu$. With the signal at DUNE being agnostic whether LNV occurs or not, and no $0\nu\beta\beta$ decay observed, the nature of HNLs remains undetermined. They could be Majorana or Dirac, but with the given exclusion by LEGEND-1000, the corresponding mass splitting can be constrained, especially in Scenarios 3 and 4, limiting $r_\Delta \lesssim 10^{-3} - 10^{-2}$.

\subsubsection{No Signal at DUNE but Signal at LEGEND-1000 (C)}

\begin{figure}[t!]
	\centering
	\setkeys{Gin}{width=0.45\linewidth}
	\includegraphics[width=0.49\linewidth]{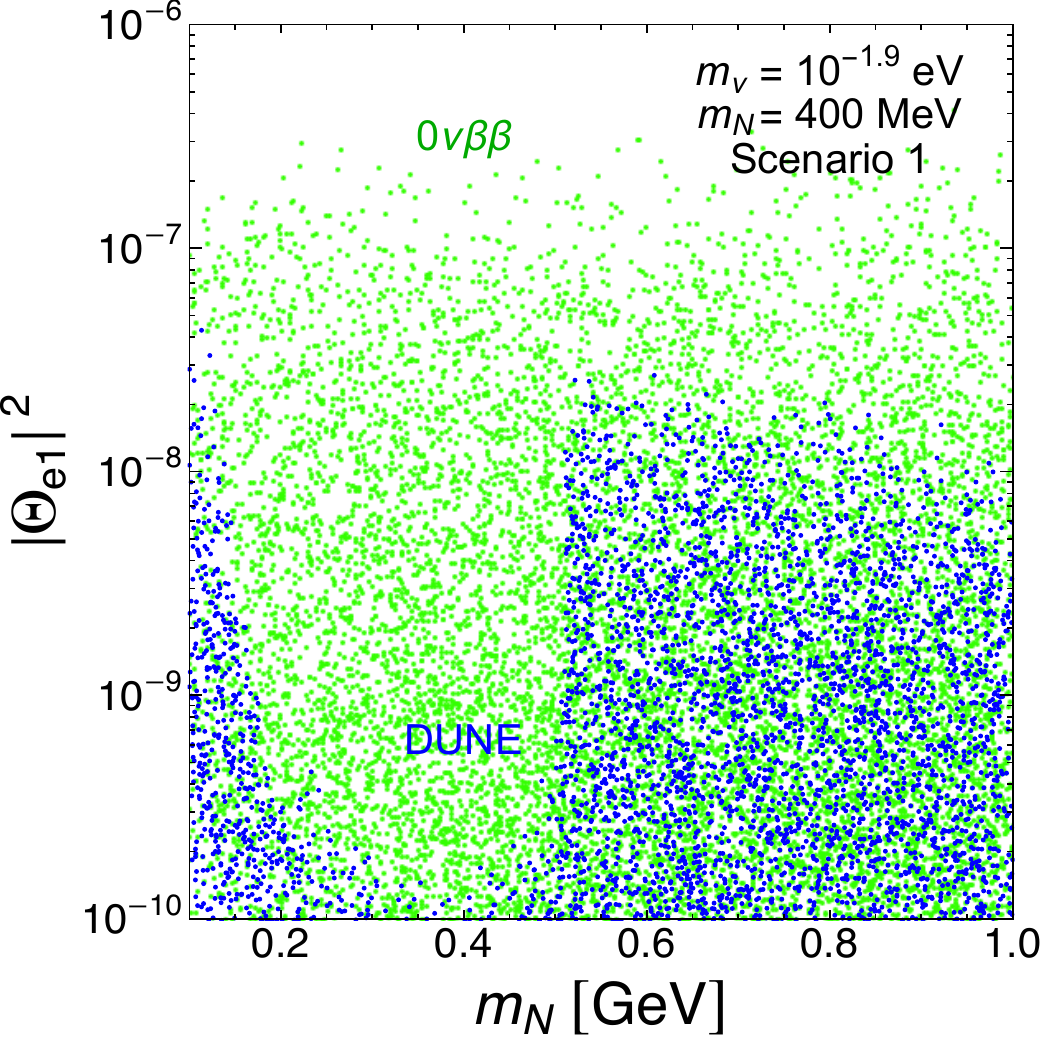} 
	\includegraphics[width=0.49\linewidth]{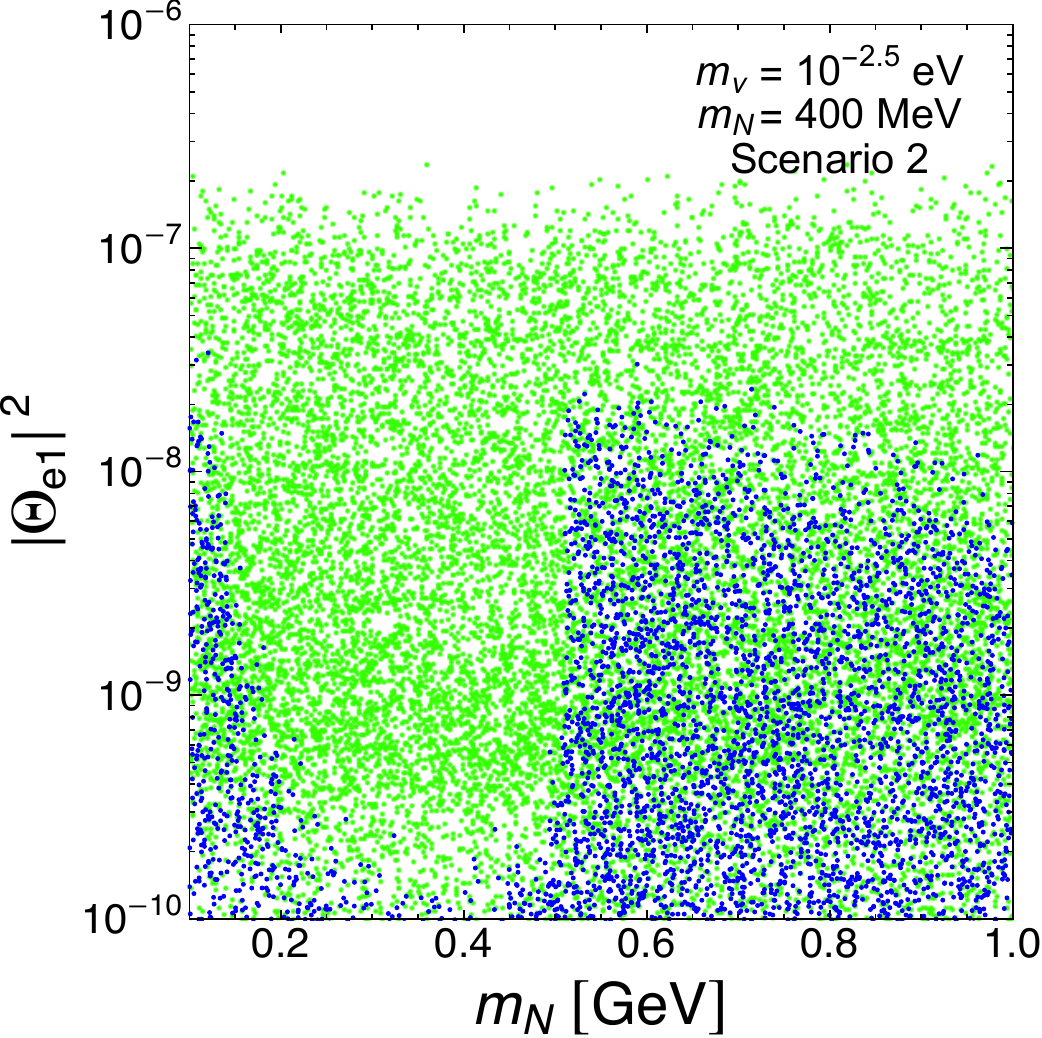}\\
	\includegraphics[width=0.49\linewidth]{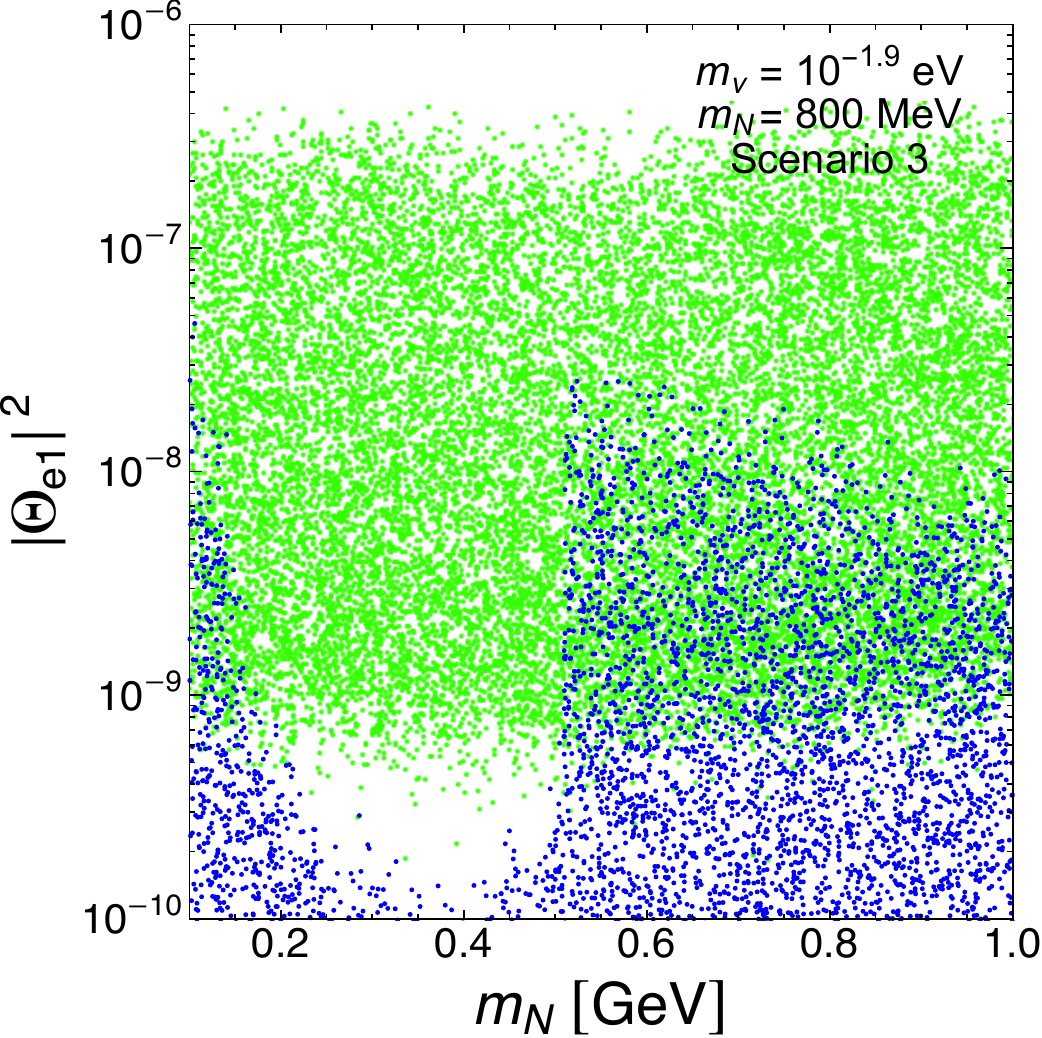} 
	\includegraphics[width=0.49\linewidth]{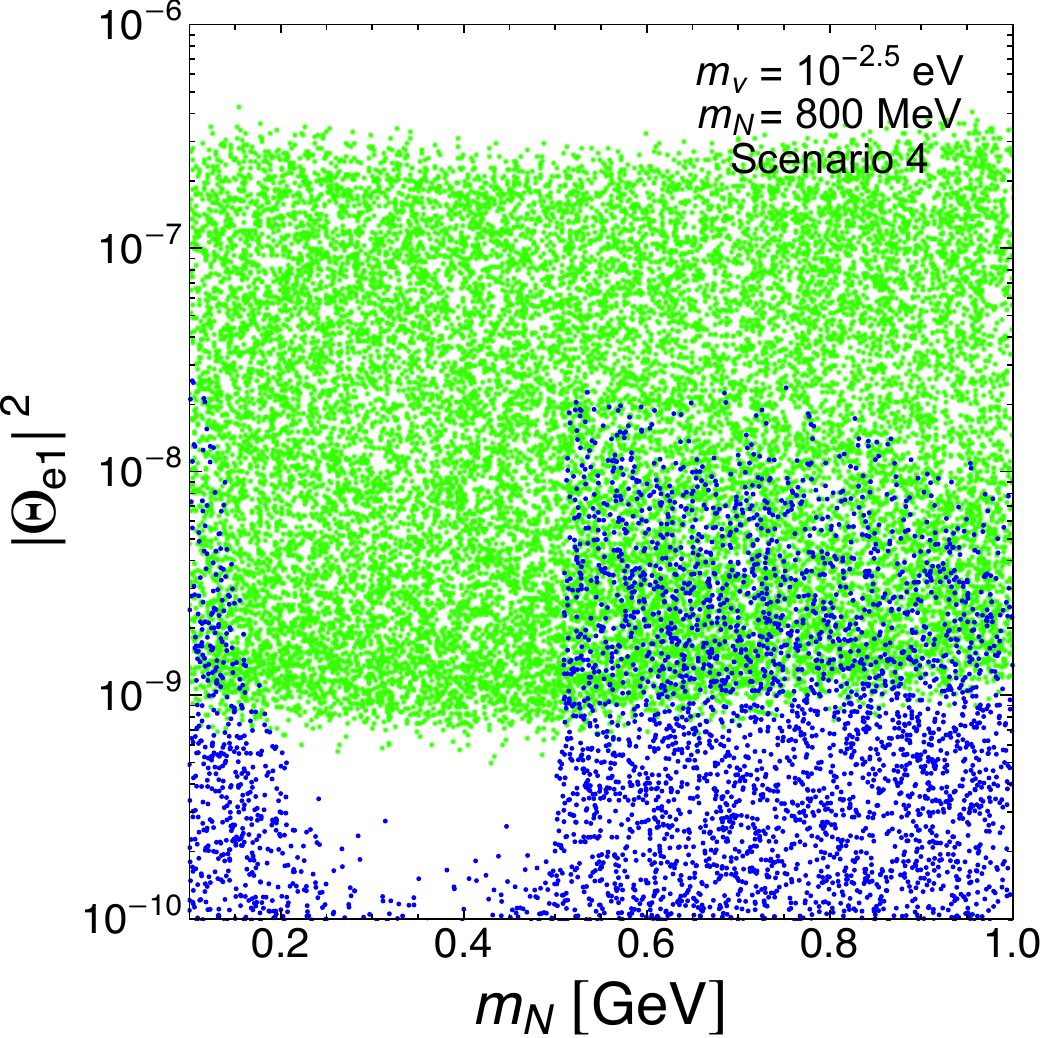}
	\caption{Posterior distribution marginalised in the $(m_{N}, |\Theta_{e1}|^2)$ plane assuming an observation at LEGEND-1000 but no signal for DUNE for the four benchmark scenarios. The green points represent the distribution for LEGEND-1000, the blue points for no signal at DUNE.}
	\label{fig:noDUNEmNss}
\end{figure}
\begin{figure}[t!]
	\centering
	\setkeys{Gin}{width=0.45\linewidth}
	\includegraphics[width=0.49\linewidth]{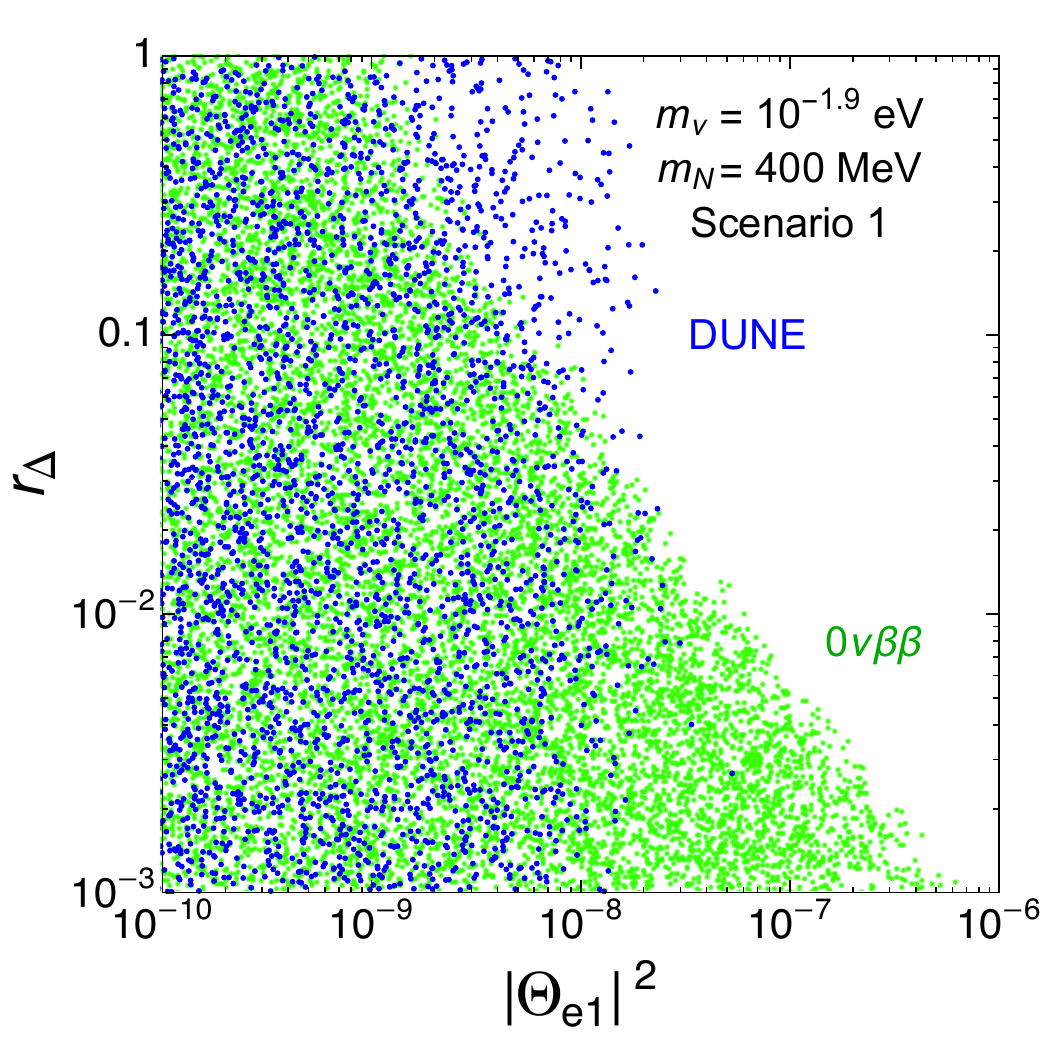} 
	\includegraphics[width=0.49\linewidth]{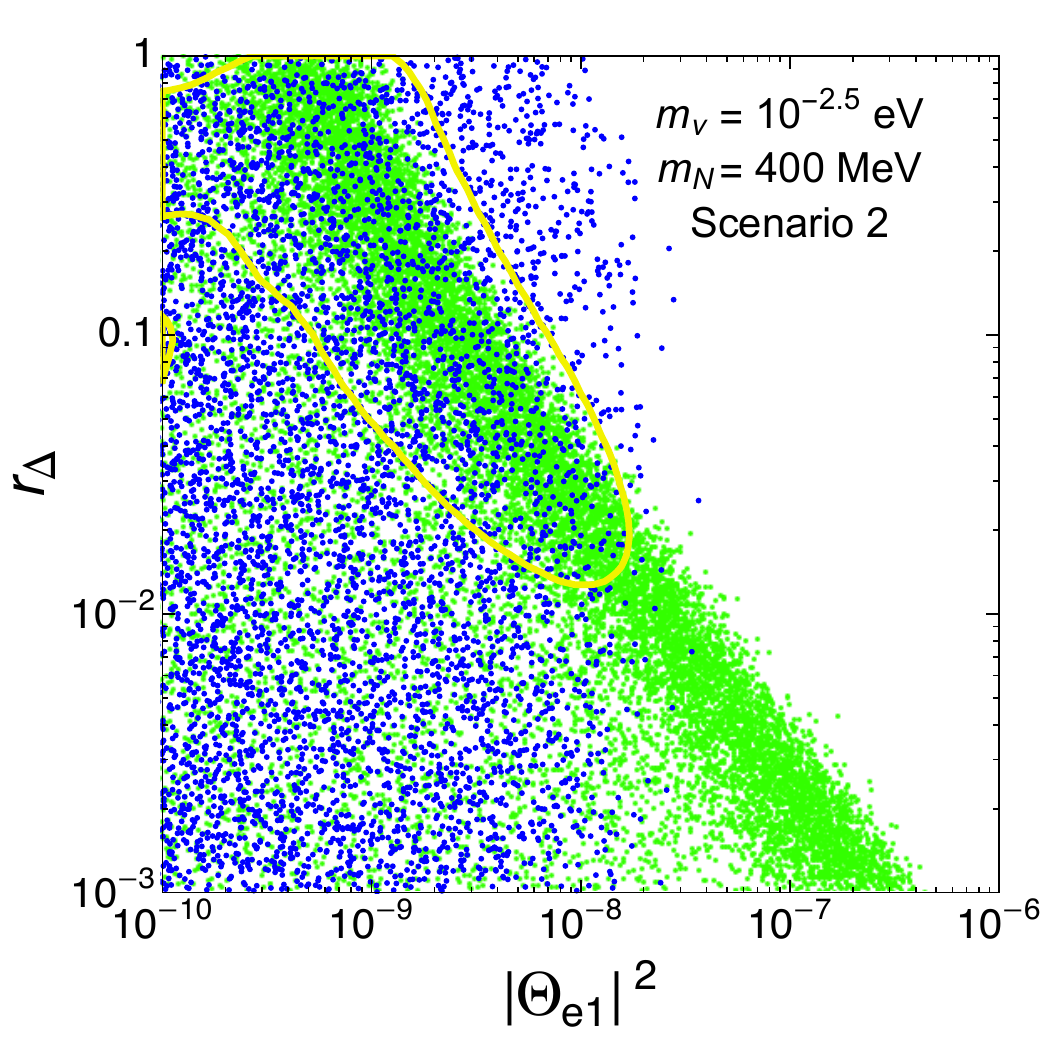}\\
	\includegraphics[width=0.49\linewidth]{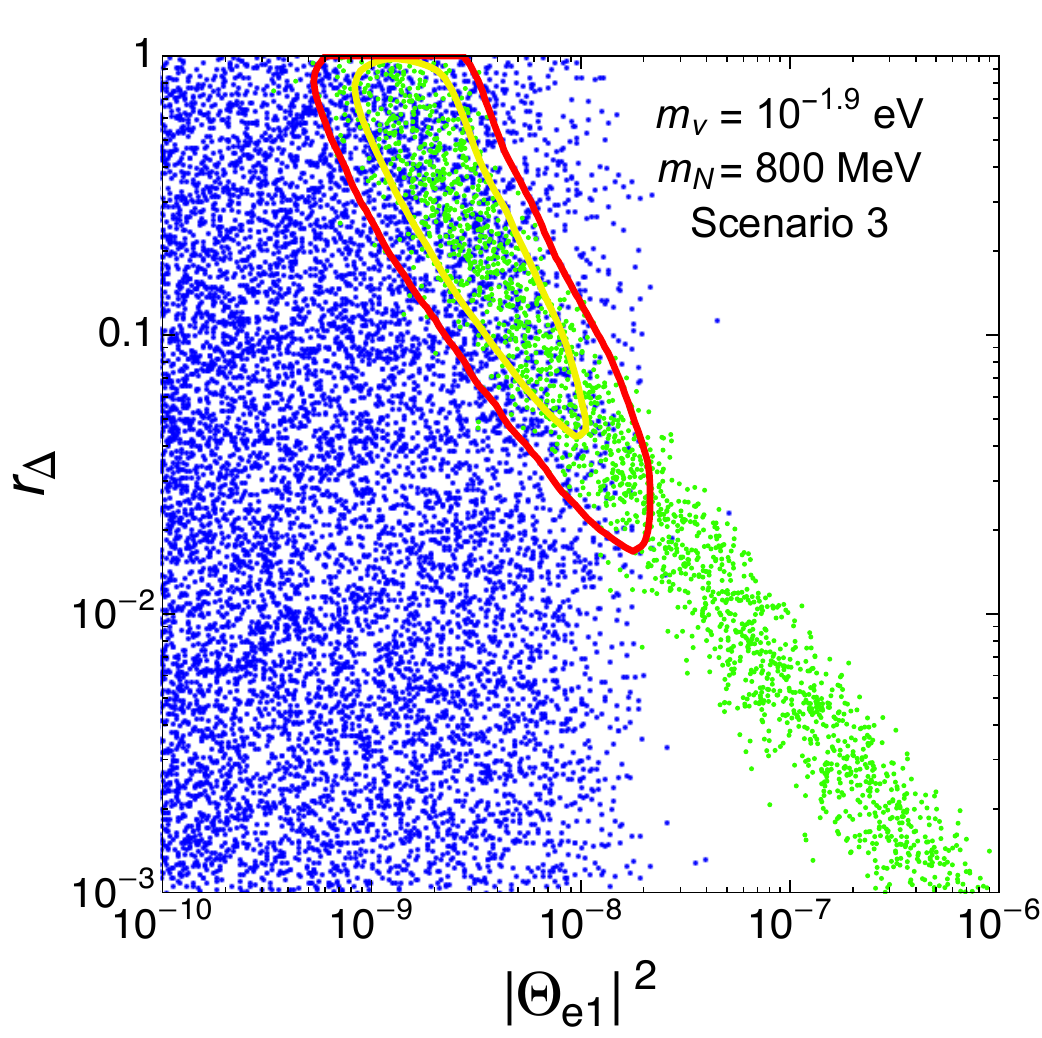} 
	\includegraphics[width=0.49\linewidth]{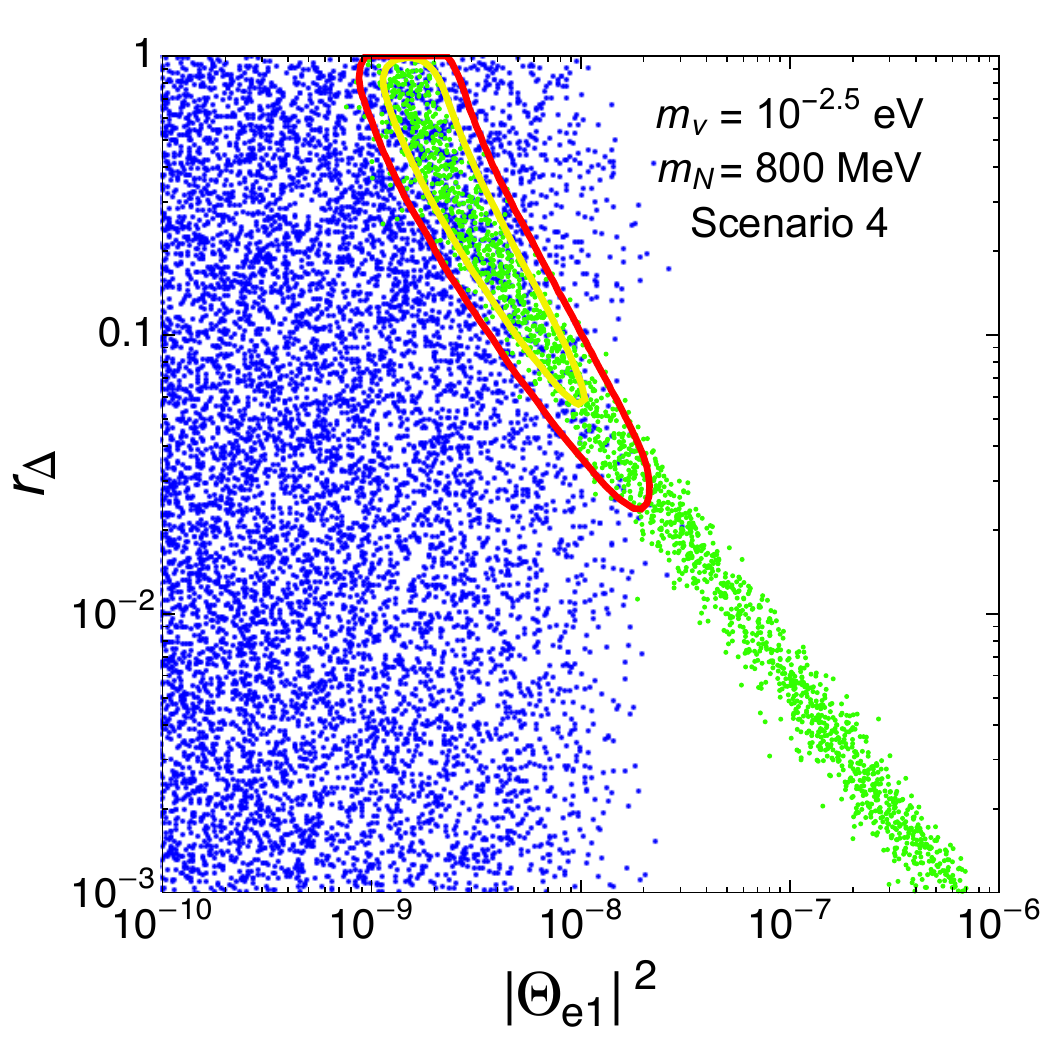}
	\caption{The same data set as Fig.~\ref{fig:noDUNEmNss} shows, but in the $(|\Theta_{e1}|^2, r_\Delta)$ plane for the four benchmark scenarios. The green points represent the distribution for LEGEND-1000, the blue points for no signal at DUNE. The yellow and red contours are the 68\% and 95\% credible regions combining both measurements respectively.}
\label{fig:noDUNEssr}
\end{figure}

In Figs.~\ref{fig:noDUNEmNss} and \ref{fig:noDUNEssr}, we show the points generated in the MCMC scans in the $(m_N, |\Theta_{e1}|^2)$ and $(|\Theta_{e1}|^2, r_\Delta)$ planes, respectively, assuming no HNL-like DUNE events (blue points) and a positive $0\nu\beta\beta$ decay signal (green points), for the four benchmark scenarios. In Fig.~\ref{fig:noDUNEmNss}, it can be seen that increasing the HNL mass shifts the $0\nu\beta\beta$ decay points to larger values of $|\Theta_{e1}|^2$. This is to be expected from the approximate formula of Eq.~\eqref{eq:se1_0vbb}, where the active-sterile mixing is linearly proportional to the HNL mass. We see that the light neutrino mass does not change the results of the MCMC scan significantly in the $(m_N, |\Theta_{e1}|^2)$ plane. The region compatible with an observation of $0\nu\beta\beta$ decay but not HNL-like events at DUNE is mostly determined by $0\nu\beta\beta$ decay for large HNL masses.

If there is no observation of HNL-like events at DUNE, there is still the chance to observe $0\nu\beta\beta$ decay for the benchmark points in the red 68\% credible contour regions in the $(|\Theta_{e1}|^2, r_\Delta)$ plane, shown in Fig.~\ref{fig:noDUNEssr}. The combined regions are now mostly determined by $0\nu\beta\beta$ decay, since a positive DUNE signal is no longer placing a stringent constraint on the parameter space. The combined regions are situated at large HNL mass splitting and relatively large active-sterile mixing, $|\Theta_{e1}|^2 \gtrsim 10^{-9}$.

\subsubsection{No signal at DUNE and No Signal at LEGEND-1000 (D)}

\begin{figure}[t!]
	\centering
	\setkeys{Gin}{width=0.45\linewidth}
	\includegraphics[width=0.49\linewidth]{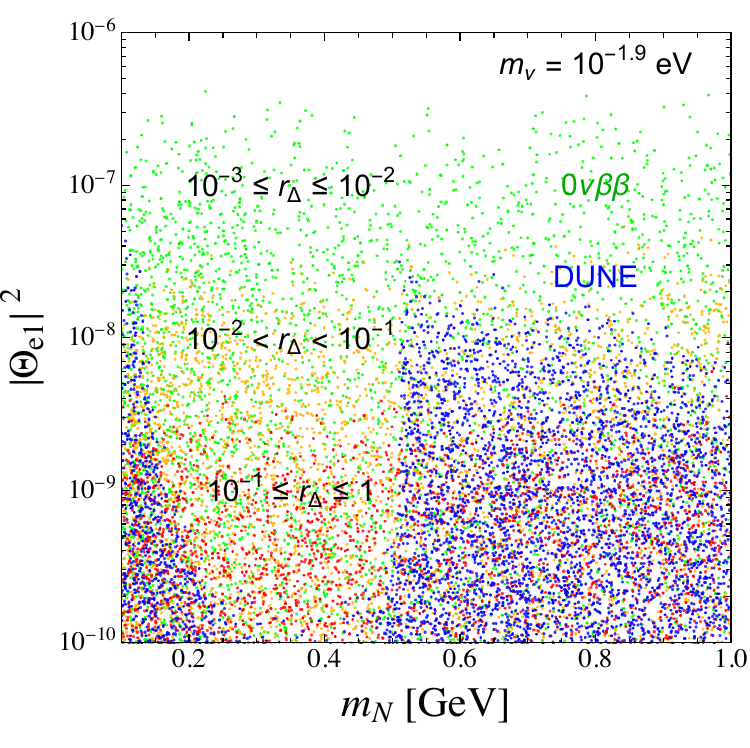}
	\includegraphics[width=0.49\linewidth]{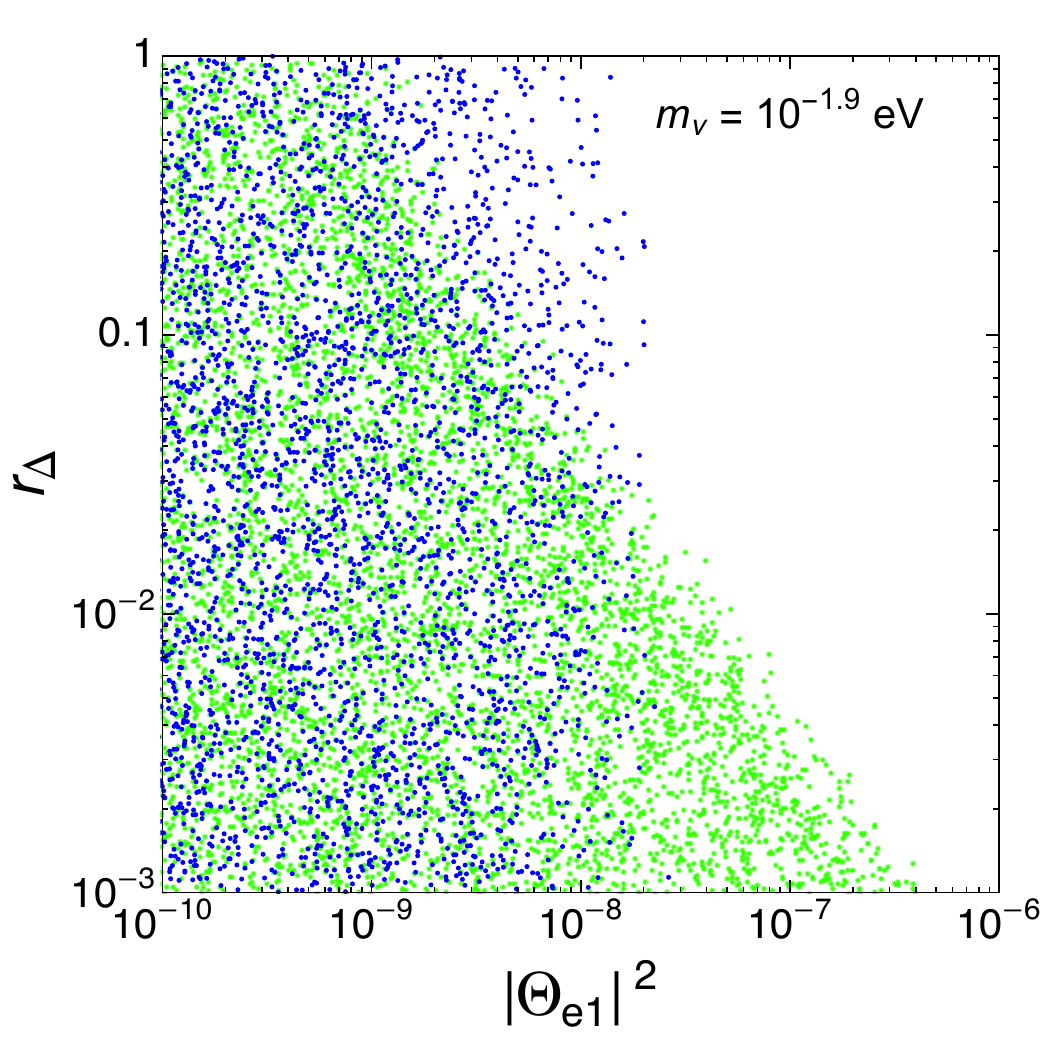}\\
	\includegraphics[width=0.49\linewidth]{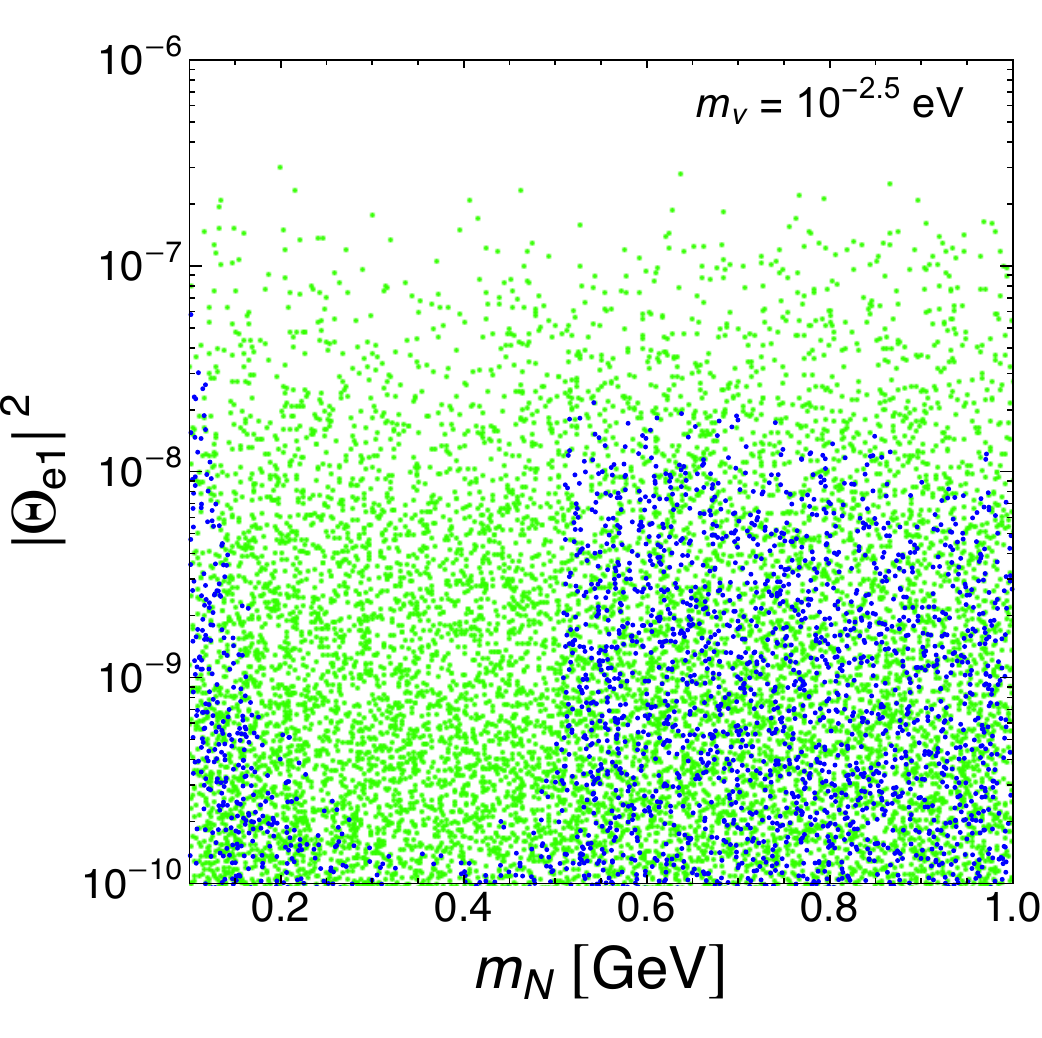}
	\includegraphics[width=0.49\linewidth]{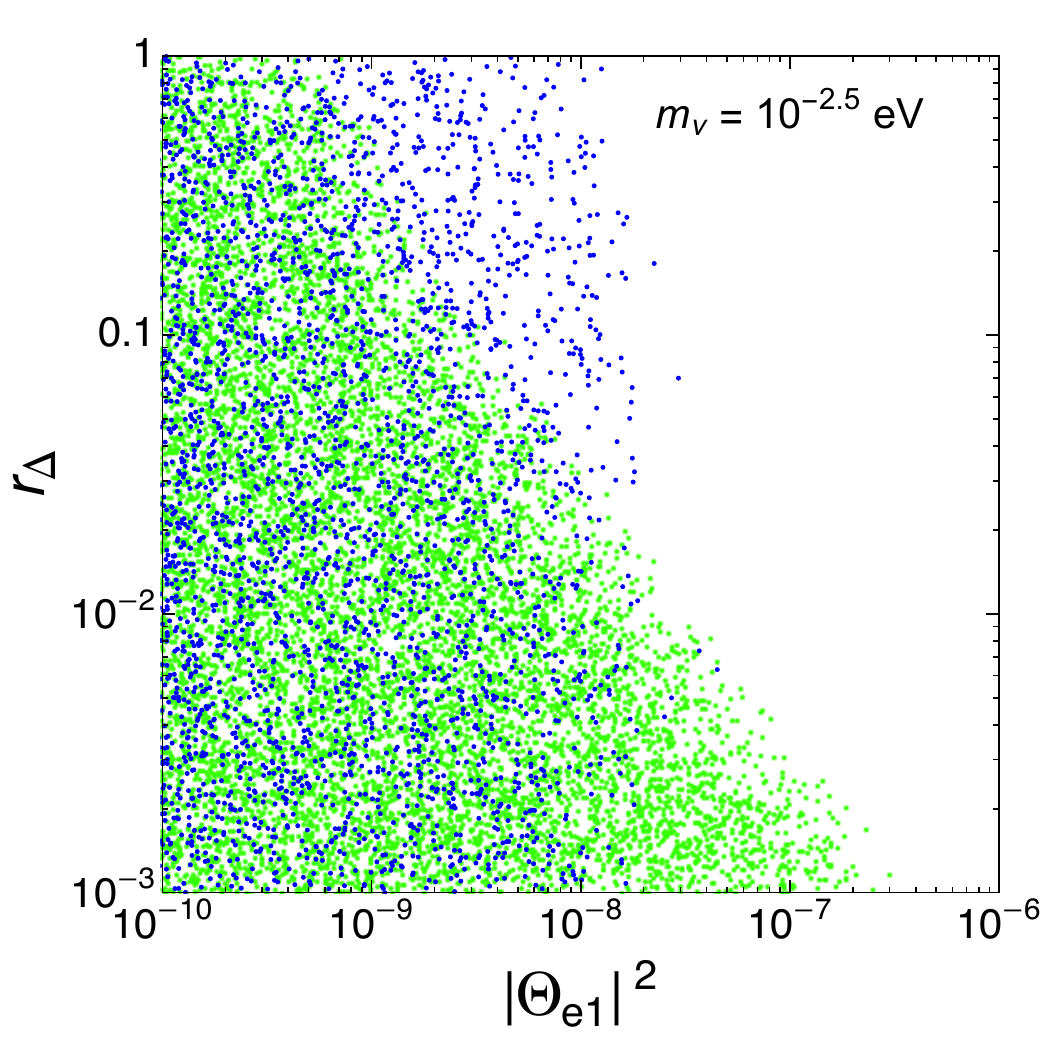}
	\caption{Posterior distribution marginalised in the $(m_{N}, |\Theta_{e1}|^2)$ plane (left) and $(|\Theta_{e1}|^2, r_\Delta)$ plane (right) assuming no observation at LEGEND-1000 nor DUNE for the four benchmark scenarios. The blue points represent the distribution for no signal at DUNE and the green points for no signal at LEGEND-1000. In the top left-hand plot, the points for no observation at LEGEND-1000 are colour-coded according to the value of the HNL mass splitting $r_\Delta$, as indicated.}
\label{fig:0both}
\end{figure}
Finally for the $1+2$ model, in Fig.~\ref{fig:0both}, we display the points produced in the MCMC scans in the $(m_N, |\Theta_{e1}|^2)$ plane (left) and $(|\Theta_{e1}|^2, r_\Delta)$ plane (right), assuming no DUNE (blue points) or $0\nu\beta\beta$ decay (green points) signals. In the top left-hand plot, where the light neutrino mass $m_\nu = 10^{-1.9}$~eV only just saturates a $0\nu\beta\beta$ decay half-life of $10^{28}$~yr, we indicate the value of the HNL mass splitting $r_\Delta$ for each point; green for $r_{\Delta} < 10^{-2}$, orange for $10^{-2} < r_{\Delta} < 10^{-1}$, and red for $10^{-1} < r_{\Delta} < 1$. It can be seen that the points with larger active-sterile mixing correspond to larger values of $r_\Delta$. The points with small $r_\Delta$ (red) cover the whole region in which no $0\nu\beta\beta$ decay signal is seen, while points with large $r_\Delta$ (red) only occupy the region where the active-sterile mixing is small.  In the plots to the right, it can be seen that the regions compatible with no DUNE or $0\nu\beta\beta$ decay signals extend from arbitrarily small $|\Theta_{e1}|^2$ values up to an upper limit determined by $r_\Delta$. For $0\nu\beta\beta$ decay, these upper limits are shifted slightly by the value of $m_\nu$.

Comparing the plots in the $(m_N, |\Theta_{e1}|^2)$ plane in Figs.~\ref{fig:DUNEandDBD} and \ref{fig:0both}, it can be seen that in some regions of the parameter space, $0\nu\beta\beta$ decay can either be observed or not observed. The is due to the fine-tuned nature of cancellations between the light neutrino and HNL pair needed to suppress the $0\nu\beta\beta$ decay rate; nevertheless, the specific arrangement of the parameters in the model needed for this cancellation are possible over much of the parameter space. 

\subsection{Signals at DUNE and LEGEND-1000 in the $3+2$ scenario}
\label{sec:DUNE_3+2}

\begin{figure}[t!]
	\centering
	\setkeys{Gin}{width=0.45\linewidth}
	\includegraphics[width=0.49\linewidth]{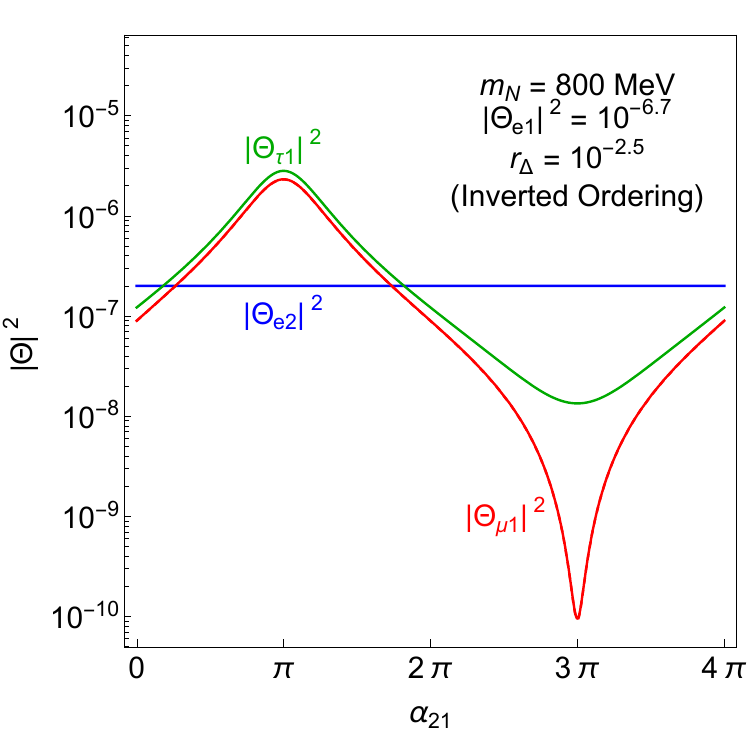}
	\includegraphics[width=0.49\linewidth]{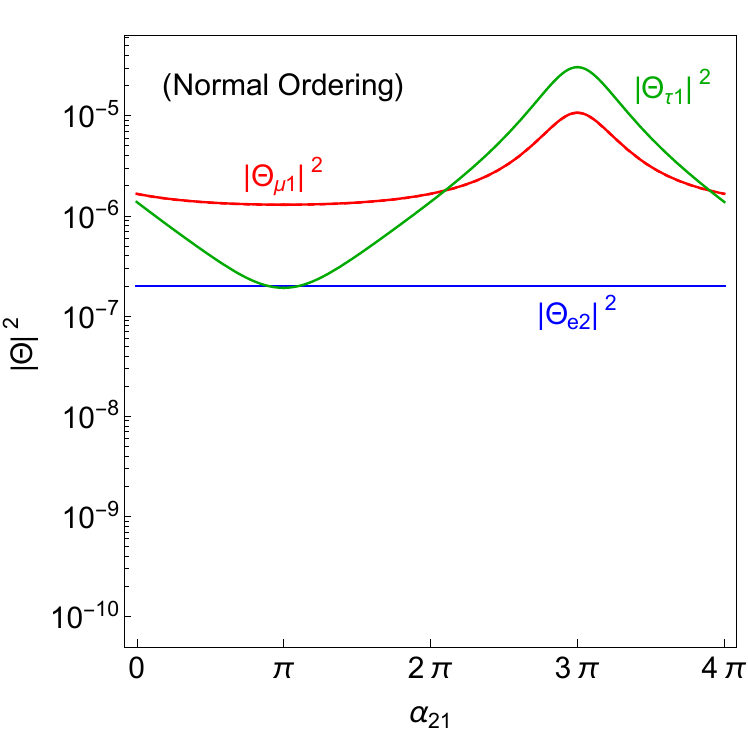} 
	\caption{Active-sterile mixing strengths $|\Theta_{e2}|^2$, $|\Theta_{\mu 1}|^2$ and $|\Theta_{\tau 1}|^2$ as a function of the light Majorana phase $\alpha_{21}$ in the $3+2$ model for IO (left) and NO (right). The other mixing strengths satisfy the relations that $|\Theta_{\ell 1}|^2 \approx |\Theta_{\ell 2}|^2$.}
\label{fig:mixing_alpha}
\end{figure}
In this final section, we further investigate the $m_N = 800$~MeV scenario, but now in the $3 + 2$ model with $m_\text{light} = 0$, i.e., the lightest active neutrino is massless. The benefit of the 3+2 scenario is that it provides a minimal and consistent framework to incorporate the correct light neutrino data and, accordingly, flavour-specific couplings of the HNL that can provide further information. The free parameter in the active neutrino sector is now the light Majorana phase $\alpha_{21}$ instead of the active neutrino mass $m_\nu$ in the simplified $1+2$ scenario. We include HNL production and decay in DUNE via mixing to all three neutrino flavours, as discussed at the end of Sec.~\ref{sec:DUNE}. Events involving HNL mixing with the tau neutrino are included though charged $\tau$ will not be produced for $m_N = 800$~MeV, as it is not kinematically possible to produce $\tau^\pm$ from the considered meson decay channels. The expected number of total DUNE signal events is given by
\begin{align}
\label{eq:lambda_DUNE}
    \lambda^\text{DUNE}_\text{sig}(\boldsymbol{\theta})
    &= N_\text{sig}\left(m_N, |\Theta_{e1}|^2, |\Theta_{\mu1}|^2, |\Theta_{\tau1}|^2\right) 
    \nonumber\\
    &+ N_\text{sig}\left(m_N(1+r_\Delta), |\Theta_{e2}|^2, |\Theta_{\mu2}|^2, |\Theta_{\tau2}|^2\right) \,,
\end{align}
where all the complex active-sterile mixing strengths $\Theta_{\alpha i}$ are written in terms of the single strength $\Theta_{e1}$, i.e., its squared magnitude $|\Theta_{e1}|^2$ and the phase $\phi_{e1}$, using Eq.~\eqref{eq:off-diag}. In Fig.~\ref{fig:mixing_alpha}, we plot $|\Theta_{e2}|^2$ and $|\Theta_{\mu 1}|^2$ as a function of $\alpha_{21}$ in the NO (left) and IO (right) cases for $|\Theta_{e1}|^2 = 10^{-6.7}$ and $r_\Delta = 10^{-2.5}$. This is in the regime $|\Theta_{e1}|^2 \gg |m_{ee}^\nu|/m_N$, where the ratio of mixing strengths is given by Eq.~\eqref{eq:sratio_ISS}; in the NO and IO scenarios, the only freedom is the Majorana phase $\alpha_{21}$, beyond the light oscillation data. For the latter we take the best fit values for the mixing angles and squared mass differences from \cite{deSalas:2020pgw} and the Dirac $CP$ phase is set at $\delta = 0$. For small mass splitting $r_\Delta$, $|\Theta_{e1}|^2 \approx |\Theta_{e2}|^2$ and $|\Theta_{\mu 1}|^2 \approx |\Theta_{\mu 2}|^2$ apply.

To perform the analysis, we consider the two benchmark scenarios shown in Table~\ref{tab:3+2benchmarks}; for the HNL mass $m_N = 800$~MeV, the active-sterile mixing strengths are chosen to lie just below the current experimental bounds. The light neutrino Majorana phase is $\alpha_{21} = 0$ throughout. This choice corresponds to the upper bounds on the effective Majorana masses for the light neutrinos for NO and IO, respectively. The value of $|\Theta_{e1}|^2$ is further constrained by the current muon- and tau-sterile mixing strength bounds and the mixing strengths can be one order of magnitude different between $|\Theta_{eN}|^2$ and $|\Theta_{\mu N}|^2$, $|\Theta_{\tau N}|^2$ as Fig.~\ref{fig:mixing_alpha} shows. With the choice of different values of $r_\Delta = 10^{-2.5}$ and $r_\Delta = 10^{-1.5}$ to keep within the LEGEND-1000 sensitivity, $|\Theta_{e1}|^2 = 10^{6.70}$ and $|\Theta_{e1}|^2 = 10^{7.64}$ keep the scenarios just below the active-sterile mixing constraints and give a similar number of total DUNE events for IO and NO, respectively. The electron (DUNE(e)) and muon (DUNE($\mu$)) only events incorporate channels where the electron and muon flavour can be tagged, i.e., with a single, charged lepton of the given flavour without any other leptons or missing energy (due to neutrinos of unknown flavour). They can be considered as separate observables to constrain the HNL phase parameter $\phi_{e1}$, which is chosen to be $\phi_{e1} = \pi/2 $ in both scenarios, to match the previous $\cos\phi_{e1} = 0 $ benchmark value.
\begin{table}[t!]
	\centering
	\renewcommand{\arraystretch}{1.25}
	\setlength\tabcolsep{6pt}
	\begin{tabular}{c|cc|ccccc} 
		\hline
		Ordering & $|\Theta_{e1}|^2$ & $r_\Delta$ & $T_{1/2}^{0\nu}$ [yr] & $\lambda_{0\nu}$ & $\lambda_\text{DUNE}$ & $\lambda_\text{DUNE}^{(e)}$ & $\lambda_\text{DUNE}^{(\mu)}$ \\ \hline
		IO & $10^{-6.7}$ & $10^{-2.5}$ & $10^{26.7}$ & 65.2 & 192 & 71.5 & 28.1 \\ 
		NO & $10^{-7.5}$ & $10^{-1.5}$ & $10^{27.1}$ & 28.2 & 193 & 12.3 & 95.4 \\
		\hline
	\end{tabular}
	\caption{Benchmark values in the $3+2$ NO and IO scenarios considered, for the HNL mass $m_N$, the active-sterile mixing strength $|\Theta_{e1}|^2$ and the HNL mass splitting $r_\Delta$. In addition, the HNL mass is $m_N = 800$~MeV, the light Majorana phase is $\alpha_{21} = 0$ and the HNL phase parameter is $\phi_{e1} = \pi/2$ throughout. Also given are the total expected number of events at DUNE, $\lambda_\text{DUNE}$, in addition to the DUNE electron and muon only events, $\lambda_\text{DUNE}^{(e)}$ and $\lambda_\text{DUNE}^{\mu)}$, and at LEGEND-1000, $\lambda_{0\nu}$, as well as the corresponding half-life $T_{1/2}^{0\nu}$.}
	\label{tab:3+2benchmarks}
\end{table}
\begin{figure}[t!]
	\centering
	\setkeys{Gin}{width=0.40\linewidth}
	\includegraphics[width=0.49\linewidth]{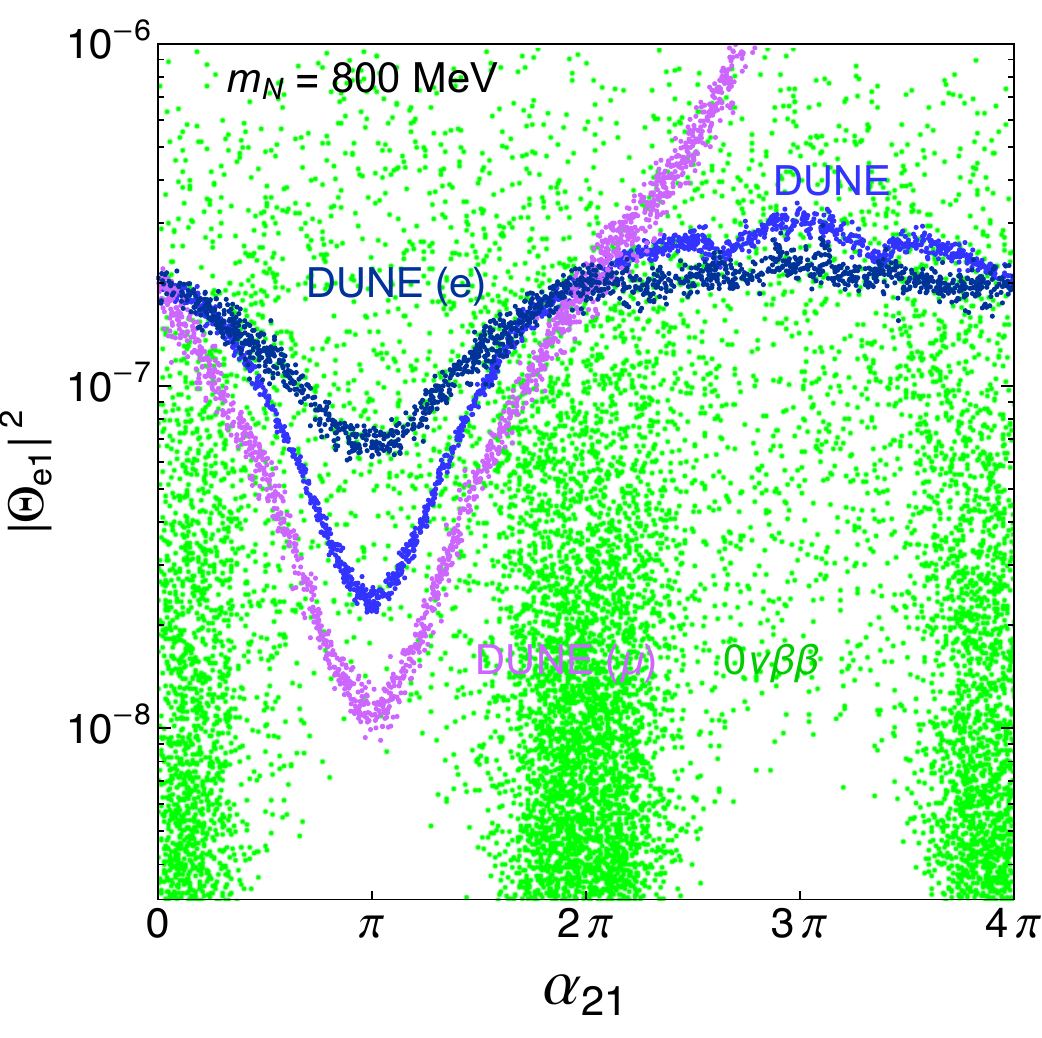}
	\includegraphics[width=0.49\linewidth]{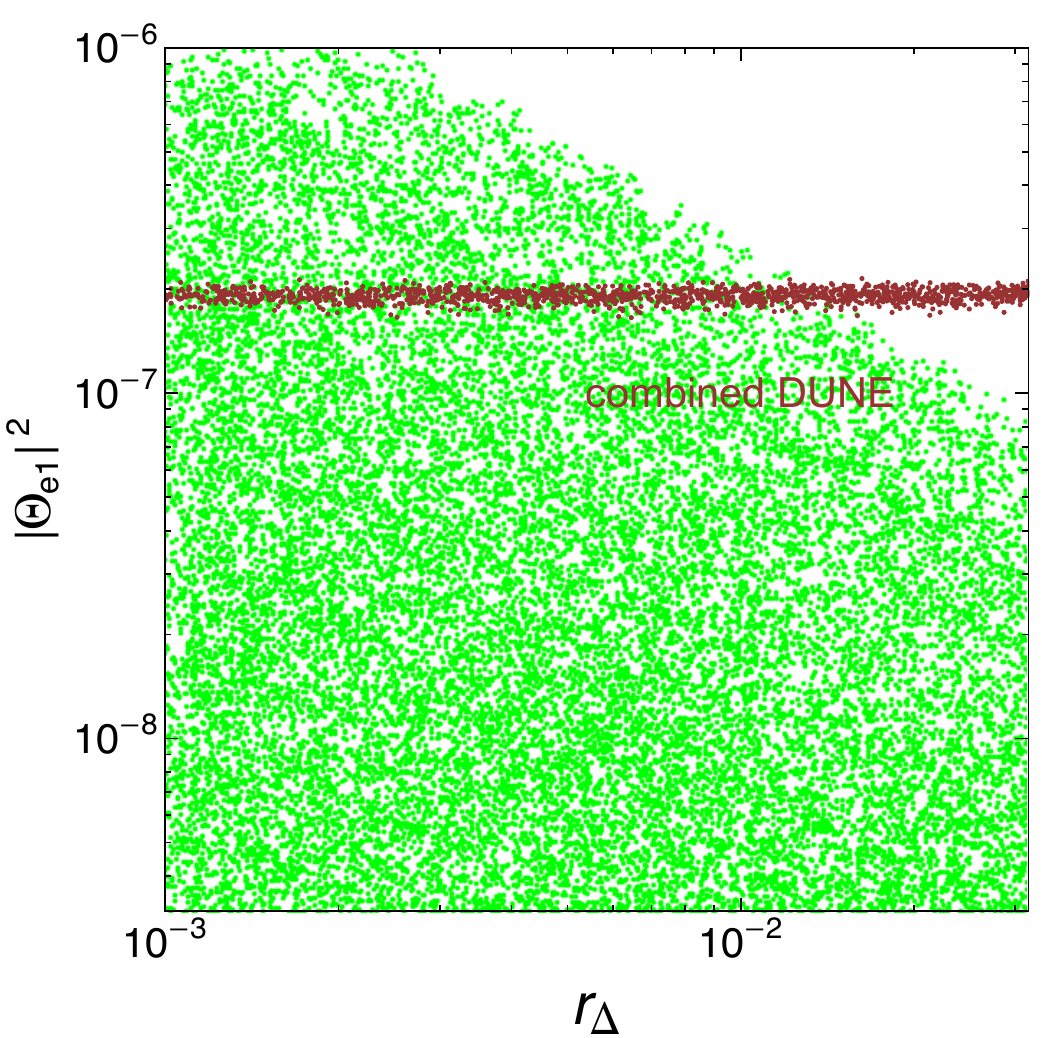} \\
	\includegraphics[width=0.49\linewidth]{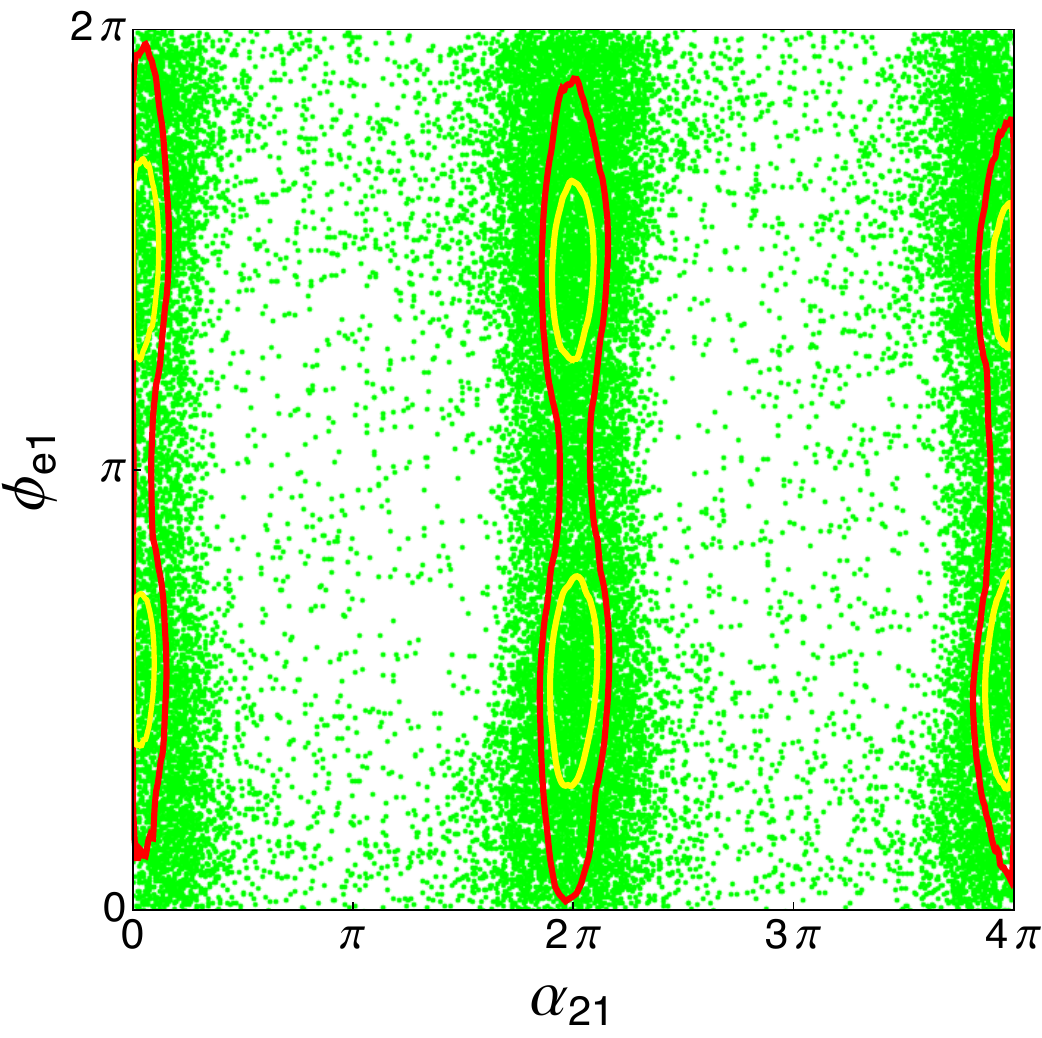}
	\includegraphics[width=0.49\linewidth]{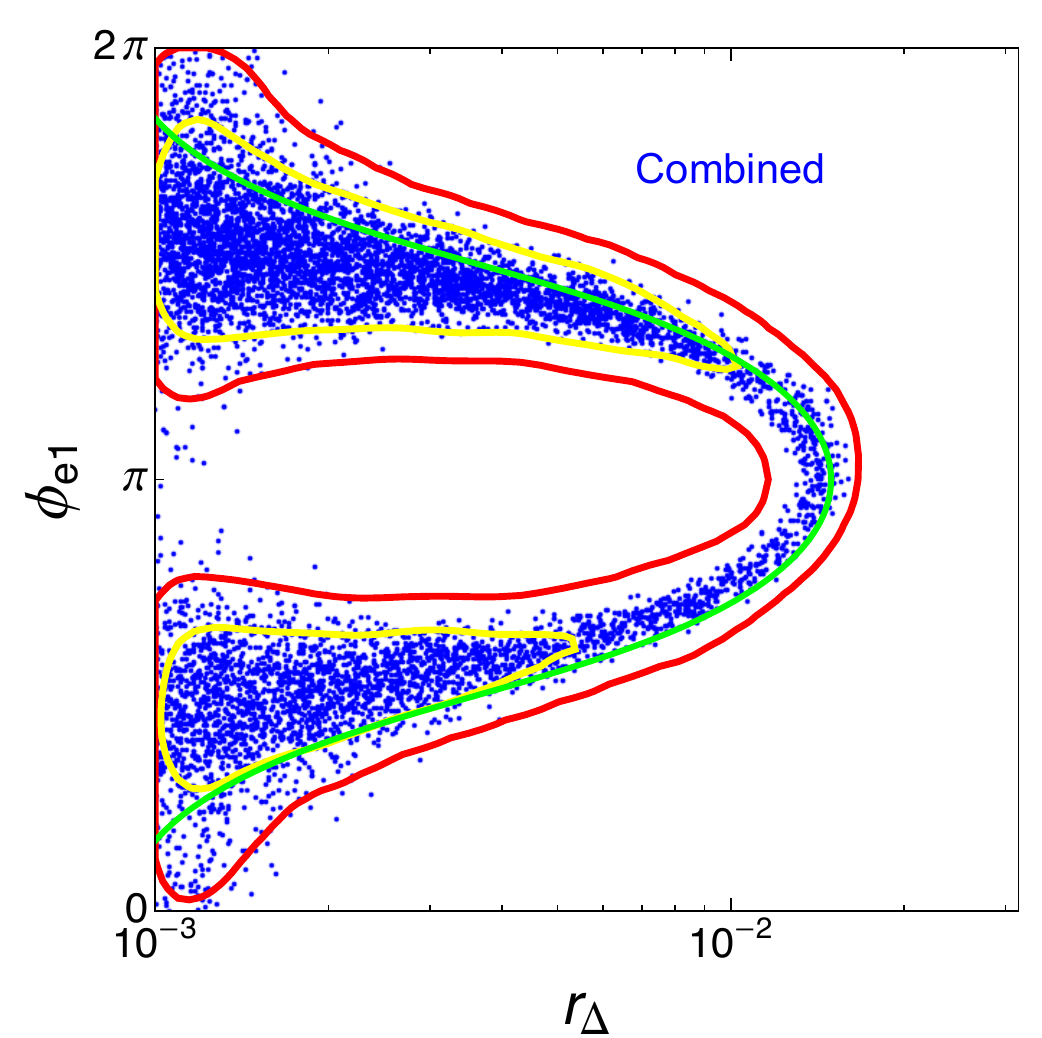}
	\caption{Posterior distribution marginalised in the $(\alpha_{21},{|\Theta_{e1}|}^2)$ plane (top left), $(r_\Delta, {|\Theta_{e1}|}^2)$ plane (top right), $(\alpha_{21}, \phi_{e1})$ plane (bottom left) and $(r_\Delta, \phi_{e1})$ plane (bottom right), assuming observations at both DUNE and LEGEND-1000 for the IO benchmark scenario. The blue points represent the total DUNE event, the pink points are for electron only events the and purple points for muon only events at DUNE. The green points represent LEGEND-1000 events. The green curve in the bottom right plot is the predicted relation between $r_\Delta$ and $\phi_{e1}$ by using Eq.~\eqref{eq:cosphie1_0vbb}. The yellow and red contours are the 68\% and 95\% credible regions combining both measurements respectively.}
\label{fig:muIO}
\end{figure}
Fig.~\ref{fig:muIO} shows the result of the MCMC scan for the IO scenario in the $(|\Theta_{e1}|^2, \alpha_{21})$, $(r_\Delta,|\Theta_{e1}|^2)$, $(\phi_{e1}, r_\Delta)$ and $(\alpha_{21}, \phi_{e1})$ planes with other parameters marginalized. We assume a combined observations of HNL-like events at both DUNE (dark blue points) and LEGEND-1000 (green points) with additional constraints from electron only (pale blue points) and muon only (purple points) events at DUNE, respectively. In the parameter space studied, DUNE is sensitive to the mixing strengths but insensitive to the light Majorana phase $\alpha_{21}$, whereas $0\nu\beta\beta$ decay is sensitive to $\alpha_{21}$ with a periodic feature but only weakly to the mixing strengths. In the top left plot, the combined analysis constrains the allowed space within two small circle which is the benchmark point and one of the periodic identical points of $\alpha_{21}$. The reason for no observation at $\alpha_{21} = 4\pi$ is clearly due to the constraint from muon only DUNE events. Probing individual flavours therefore has a strong impact on the information gained. While the HNL phase $\phi_{e1}$ cannot be probed individually, the bottom right plot demonstrates a degeneracy with the mass splitting $r_\Delta$ arising from the combination of all observables. This matches the predicted analytic relation between $r_\Delta$ and $\phi_{e1}$ in Eq.~\eqref{eq:cosphie1_0vbb} (green curve). 

\begin{figure}[t!]
	\centering
	\setkeys{Gin}{width=0.45\linewidth}
	\includegraphics[width=0.49\linewidth]{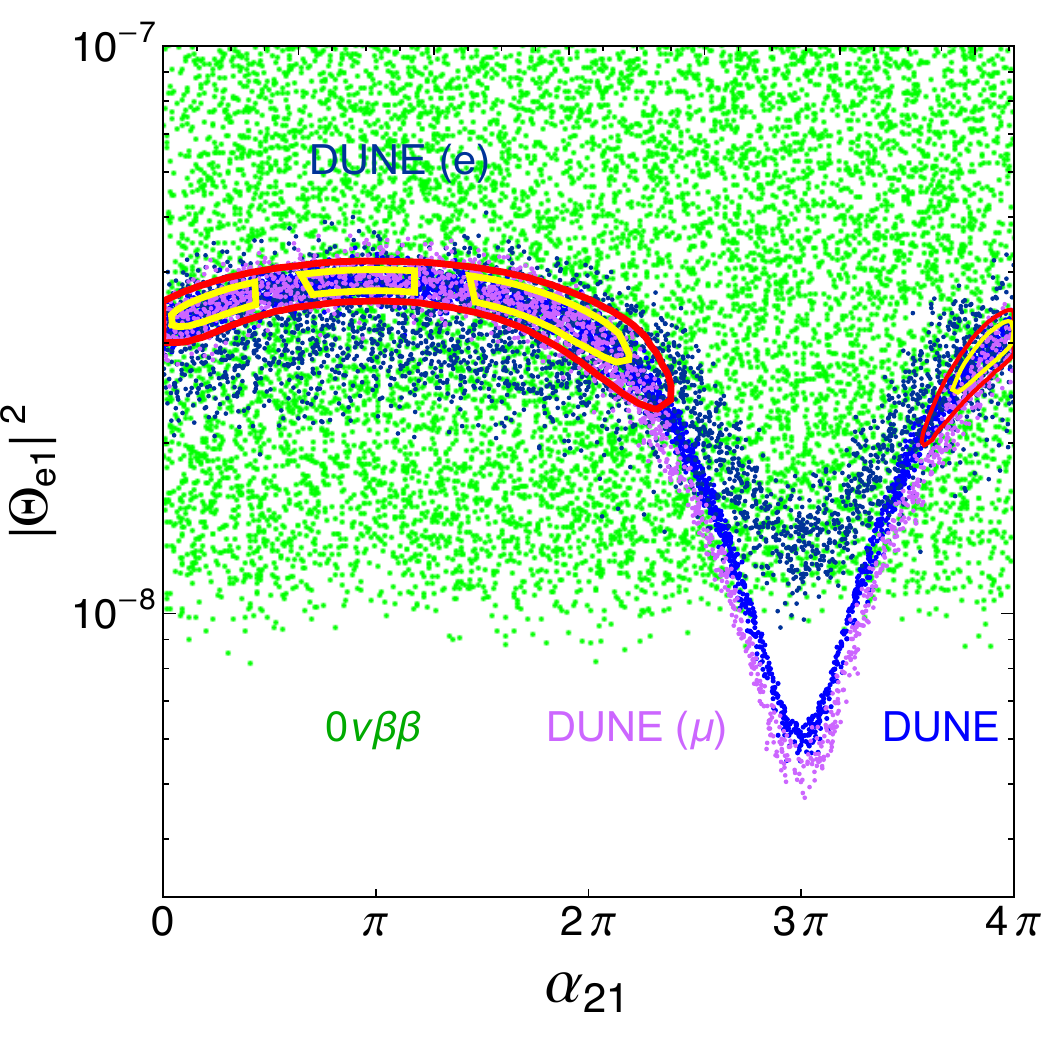}
	\includegraphics[width=0.49\linewidth]{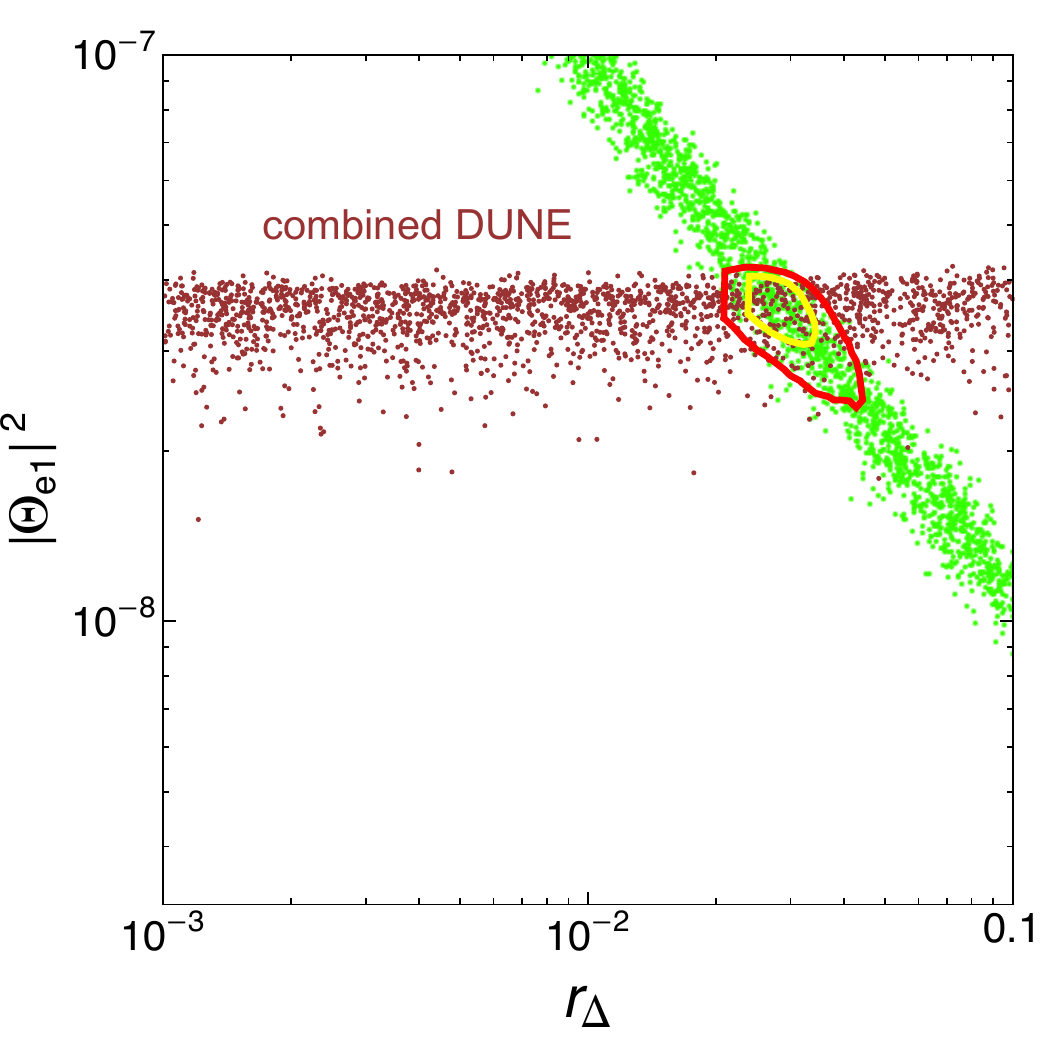}
	\caption{As Fig.~\ref{fig:muIO}, but showing the parameter planes $(\alpha_{21}, |\Theta_{e1}|^2)$ plane (left) and $(r_\Delta, |\Theta_{e1}|^2)$ plane (right) in the NO scenario.}
\label{fig:muNO}
\end{figure}
In Fig.~\ref{fig:muNO}, we likewise show the results in the NO scenario, highlighting the $(\alpha_{21}, |\Theta_{e1}|^2)$ and $(r_\Delta, |\Theta_{e1}|^2)$ parameter planes. The left plot demonstrates a similar feature at DUNE as in the IO scenario before, though the Majorana phase $\alpha_{21}$ cannot be pinpointed as precisely. Also, the observation of $0\nu\beta\beta$ decay results in a band giving a lower limit on the mixing strength, rather than the three vertical strips in Fig.~\ref{fig:muIO}~(top left). This is due to the saturation of the light effective mass in IO which enhances the impact of the Majorana phase in the light neutrino sector. The right plot agrees with the bottom right plot in Fig.~\ref{fig:DUNEandDBD}, allowing for a precise determination of both the mixing strength and the mass splitting. This is more favourable than in the IO scenario as there less chance of cancellation between the light and heavy neutrino contributions to occur in $0\nu\beta\beta$ decay for the given LEGEND-1000 sensitivity. As a result of the lack of interference, neither $\alpha_{21}$ nor $\phi_{e1}$ can be probed, though. 

\section{Conclusions}
\label{sec:conclusions}

In this paper, we have investigated the phenomenology of a pair of heavy neutral leptons (HNLs) in the MeV to GeV mass range, which can generate the observed masses and mixing of the light active neutrinos but also impact neutrinoless double beta ($0\nu\beta\beta$) decay and direct searches for HNLs at fixed target experiments. In particular, we have shown how the combination of a positive $0\nu\beta\beta$ decay signal at LEGEND-1000 and HNL-like events in the DUNE near detector (ND) can constrain the active-sterile mixing and mass splitting of the HNL pair. Additionally, we have considered the model implications of only one or neither of the signals being observed.

In Sec.~\ref{sec:model}, we introduced a phenomenological parametrisation of the $3+2$ model, which adds a gauge-singlet Weyl fermion pair to the SM and is sufficient to generate masses for two light neutrinos at tree level. Using the tree level relations between the light neutrino and HNL parameters in the limit of small active-sterile mixing, we derived useful ratio formulae making it possible to express all active-sterile mixing parameters in terms of a single active-sterile mixing and CP phase (in addition to the light neutrino data and HNL masses). This is in contrast to the Casas-Ibarra parametrisation, which uses a complex angle as the free parameter. This parametrisation is generalised for an arbitrary number $\mathcal{N}_S$ of additional HNL states in \ref{app:phenom_param}, with the number of free parameters increasing accordingly. We also illustrated the behaviours of the active-sterile mixing parameters in the simple $1+2$ model, which covers the relevant limits of the $3+2$ model. We saw how either of the HNL pair can effectively decouple, with one HNL giving a mass to the light neutrino via the standard seesaw, and how both active-sterile mixing strengths to the HNLs can become large in the inverse seesaw limit.

Using the phenomenological model, in Sec.~\ref{sec:0vbb} we examined the $0\nu\beta\beta$ decay process, carefully taking into account the exchange of HNLs in the 100 MeV to 1 GeV mass range. In this mass region, around the Fermi momentum $k_F$, the exchange of HNLs is difficult to treat using chiral effective field theory methods. However, the low and high-mass regions are well understood and so a carefully chosen interpolation formula can be used to good approximation in the intermediate region. In the simple $1+2$ model, we gave a schematic illustration of how the HNL pair can either interfere constructively or destructively with the light neutrino contribution. Thus, given a precise knowledge of the neutrino masses and whether they have a NO or IO, the observation of a $0\nu\beta\beta$ decay half-life $T_{1/2}^{0\nu}$ that is incompatible with the light neutrino contribution could be used to put stringent constraints on the HNL pair. Conversely, if the light neutrino contribution \textit{just} saturates the observed half-life, the $1+2$ model parameter space is less constrained.

In Sec.~\ref{sec:DUNE}, we covered the production and decay of long-lived HNLs in the DUNE experiment in the phenomenological model. Using \textsc{Pythia}, we performed a simulation of mesons produced from $120$~GeV protons on target, generating their production fractions and momentum profiles. All possible decays of these mesons to HNLs via the active-sterile mixing $|U_{eN}|^2$ are used to simulate the production of HNLs in the rest frame of the mesons. The HNLs are then boosted to the lab frame and required to decay to charged tracks inside the ND, which we take to have a conical cross section. This analysis allows us to estimate the sensitivity of DUNE to Majorana HNLs, which can either be detected via LNC or LNV decay modes. This is also relevant for quasi-Dirac neutrinos; for the mass splitting of interest, the oscillations between the HNL pair are averaged out and the quasi-Dirac pair appears to be a pair of Majorana states. 

In Sec.~\ref{sec:compare}, we examined the complementary of $0\nu\beta\beta$ decay and DUNE in constraining the HNL parameter space. We first compared a measurement of $0\nu\beta\beta$ decay and HNL-like events at DUNE analytically, finding the implied mass splitting $r_\Delta$ in the limit where the light neutrino mass does not saturate the observed $0\nu\beta\beta$ decay half-life and the HNL event rate is large, so the HNL contribution is tightly constrained. We then explored in more detail the parameter space of the simple $1+2$ model by performing a Markov chain Monte Carlo scan over the statistical likelihoods of signals at LEGEND-1000 and DUNE, given the HNL pair hypothesis. We found that if both signals are seen, the mass splitting between the HNL pair is well constrained, with values $r_\Delta \sim 0.1$ implied for HNL masses around $m_N = 400$~MeV and $r_\Delta \sim 3\times10^{-3}$ for $m_N = 800$~MeV, shown in Fig.~\ref{fig:DUNEandDBD} as contours at the 68\% and 95\% confidence levels. If one signal is observed but not the other, we showed how the mass splitting is generally less constrained in Figs.~\ref{fig:NoDBD} and \ref{fig:noDUNEssr}. Finally, we found the regions of the parameter space in Fig.~\ref{fig:0both} where neither $0\nu\beta\beta$ decay nor HNL-like events are observed. Considering a more general $3+2$ scenario, including all three active neutrinos, we determined the prospect of probing individual flavours at DUNE, providing additional constraints on the parameter space. 

In this work, we have demonstrated how two very different probes of HNLs can constrain an intriguing region of the HNL parameter space where the active-sterile mixing is just above or touching the seesaw floor, $|U_{\alpha N}|^2 = m_\nu/m_N$. We therefore have the exciting prospect for $0\nu\beta\beta$ decay and DUNE to not only confirm (as their principal experimental aims) the mass ordering of the light neutrinos and their Majorana nature, but also that their masses are generated by the presence of an HNL pair with a small mass splitting. 

\section*{Acknowledgments}

P.~D.~B. has received support from the European Union's Horizon 2020 research and innovation programme under the Marie Sk\l{}odowska-Curie grant agreement No 860881-HIDDeN. F.~F.~D. acknowledges support from the Science and Technology Facilities Council, part of U.K. Research and Innovation, Grant No. ST/T000880/1. M.~R. thanks ICTP, Trieste for hospitality while this work was being completed. The work of F.~F.~D. was performed in part at the Aspen Center for Physics, which is supported by National Science Foundation grant PHY-1607611. This work was partially supported by a grant from the Simons Foundation.

\appendix

\section{Extended Phenomenological Parametrisation}
\label{app:phenom_param}

In this appendix, we generalise the phenomenological parametrisation in Sec.~\ref{sec:model} to the $\mathcal{N}_A +\mathcal{N}_S$ scenario, i.e., the SM with an arbitrary number $\mathcal{N}_A$ of active fields extended to include $\mathcal{N}_S$ gauge-singlet Weyl fermion fields. Now, the relation in Eq.~\eqref{eq:Mnu=0} can be written as (at first-order in the active-sterile mixing strengths $\Theta_{\alpha \kappa}$),
\begin{align}
\label{eq:Mnu=0_general}
	m_{\alpha\beta}^{\nu} 
	+ m_N \Theta_{\alpha 1}\Theta_{\beta 1} 
	+ m_N (1+r^{21}_\Delta) \Theta_{\alpha 2}\Theta_{\beta 2} 
	+ \sum_{\kappa=3}^{n} m_N (1+r^{\kappa 1}_\Delta) \Theta_{\alpha \kappa}\Theta_{\beta \kappa} 
	= 0\,,
\end{align}
where $\Theta_{\alpha \kappa} \equiv |\Theta_{\alpha \kappa}|e^{i\phi_{\alpha \kappa}/2}$ and $r^{\kappa1}_{\Delta} \equiv (m_{N_\kappa}-m_{N})/m_N$. It is now possible to take the diagonal elements $(\alpha = \beta)$ and rearrange the expression above to find
\begin{align}
\label{eq:app_diag}
	\frac{\Theta_{\alpha 2}}{\Theta_{\alpha 1}} = \pm i\sqrt{\frac{1+x_{\alpha\alpha}^{\alpha}}{1+r_\Delta}}\,,
\end{align}
where the parameter $x_{\alpha\beta}^\rho$ is given by
\begin{align}
	x_{\alpha\beta}^\rho 
	\equiv \frac{m_{\alpha\beta}^{\nu}}{m_N \Theta_{\rho 1}^2} 
	+ \sum_{\kappa=3}^{n} (1+r^{\kappa 1}_\Delta) 
	  \frac{\Theta_{\alpha \kappa}\Theta_{\beta \kappa}}{\Theta_{\rho 1}^2}\,.
\end{align}
Additionally, it is possible to take the off-diagonal elements $(\alpha \neq \beta)$ to find
\begin{align}
\label{eq:app_off-diag}
	\frac{\Theta_{\beta 1}}{\Theta_{\alpha 1}} 
	= \frac{x_{\alpha\beta}^{\alpha}\pm \sqrt{(x_{\alpha\beta}^{\alpha})^2 - x_{\alpha\alpha}^{\alpha}x_{\beta\beta}^\alpha}\sqrt{1+x_{\alpha\alpha}^{\alpha\phantom{!}}}}{x_{\alpha\alpha}^{\alpha}}
	\equiv y_{\alpha\beta}^{\alpha}\,.
\end{align}
Combining these two results, we can write
\begin{align}
\label{eq:app_combined}
	\frac{\Theta_{\beta 2}}{\Theta_{\alpha 1}} 
	= \pm i\sqrt{\frac{(y_{\alpha\beta}^{\alpha})^2+x_{\alpha\alpha}^{\alpha}}{1+r_\Delta}}\,,
\end{align}
which reduces to Eq.~\eqref{eq:app_diag} for $\alpha = \beta$ because $y_{\alpha\alpha}^{\alpha} = 1$. Thus, for a given flavour $\alpha$, it is possible to completely determine the mixing to the first and second HNL, $\Theta_{\beta 1}$ ($\alpha\neq\beta$) and $\Theta_{\beta 2}$ ($\beta = e,\mu,\tau$), in terms of $\Theta_{\alpha 1}$ (the convenient choice for $0\nu\beta\beta$ decay is $\alpha=e$). 

Above, the constraint from the upper-left $\mathcal{N}_A\times \mathcal{N}_A$ sub-block of $\mathcal{M}_\nu$ has made it possible to eliminate $\mathcal{N}_A(\mathcal{N}_A-1)$ parameters among the $(\mathcal{N}_A+\mathcal{N}_S+1)(\mathcal{N}_A+\mathcal{N}_S)$ describing the complex symmetric mass matrix $\mathcal{M}_\nu$. Additionally, $\mathcal{N}_A$ phases can be eliminated via a redefinition of the charged lepton fields and $\mathcal{N}_S^2$ mixing angles and phases eliminated via an unphysical rotation among the sterile states. Consequently, the number of active-sterile mixing strengths and CP phases that can be eliminated is given by $\mathcal{N}_A(\mathcal{N}_A - 1) + \text{min}(\mathcal{N}_A,\mathcal{N}_S)$, i.e., two parameters in the $1+2$ scenario ($|\Theta_{e2}|$ and $\phi_{e2}$) and 10 parameters in the $3+2$ scenario ($|\Theta_{\beta 1}|$ and $\phi_{\beta 1}$ for $\beta = \mu,\tau$ and $|\Theta_{\beta 2}|$ and $\phi_{\beta 2}$ for $\beta = e,\mu,\tau$). In the $3+3$ scenario, it becomes possible to eliminate 12 parameters; i.e., an additional active-sterile mixing and CP phase such as $|\Theta_{\beta 3}|$ and $\phi_{\beta 3}$ for a single flavour.

In the $3+2$ model, the ratio in Eq.~\eqref{eq:app_diag} has branch points at
\begin{align}
\label{eq:branch_points}
	\Theta_{\alpha 1} 
	= |\Theta_{\alpha 1}| e^{i\phi_{\alpha 1}/2}
	= \pm i\sqrt{\frac{m_{\alpha\alpha}^\nu}{m_N}}\,.
\end{align}
This corresponds to the seesaw limit in which the heavier HNL decouples and the lighter HNL has a mixing $|\Theta_{\alpha 1}| = \sqrt{|m_{\alpha\alpha}^{\nu}|/m_N}$ and CP phase $\phi_{\alpha 1} = \phi_{\alpha\alpha}^{\nu}\pm \pi$, where we have defined $m_{\alpha\alpha}^{
\nu} \equiv |m_{\alpha\alpha}^{
\nu}|e^{i\phi_{\alpha\alpha}^\nu}$. In the $1+2$ model, with $m_{\alpha\alpha}^{\nu} \to m_\nu$, we instead have the requirement that $\phi_{\alpha 1} = \pm \pi$. As a matter of convention, we take the positive square root in Eqs.~\eqref{eq:app_combined} and \eqref{eq:branch_points} and negative square root Eq.~\eqref{eq:app_off-diag}. Having defined the Dirac phases in Eq.~\eqref{eq:Wij} and the Majorana phases in Eq.~\eqref{eq:maj_phases} to lie in the ranges $[0,2\pi]$ and $[0,4\pi]$, respectively, the CP phases $\phi_{\alpha \kappa} = \phi_{\kappa} - 2\eta_{\alpha i}$ are also taken to be in the range $[0,4\pi]$. This extended range takes into account both values of the square root sign; for example, in the inverse seesaw limit of the $3+2$ model, Eq.~\eqref{eq:app_off-diag} tends to
\begin{align}
	\frac{\Theta_{\beta 1}}{\Theta_{\alpha 1}} 
	= \frac{m_{\alpha\beta}^{\nu}\pm \sqrt{(m_{\alpha\beta}^{\nu})^2 - m_{\alpha\alpha}^{\nu}m_{\beta\beta}^\nu}}{m_{\alpha\alpha}^{\nu}} 
	= \frac{\sqrt{m_2}\,U_{\beta 2}\mp\sqrt{m_3}\,U_{\beta 3}}
	  {\sqrt{m_2}\,U_{\alpha 2}\mp\sqrt{m_3}\,U_{\alpha 3}}\,,
\end{align}
where the second equality is valid for the NO case. Both signs of the square root are taken into account by allowing the physically-relevant Majorana phase $\alpha_{21}$ to lie in the range $[0,4\pi]$.

\section{Phenomenological vs. Minimal Casas-Ibarra Pa\-ra\-metrisation}
\label{app:pheno_vs_CI_param}

In this appendix we compare the phenomenological parametrisation in Sec.~\ref{sec:model} with the minimal Casas-Ibarra parametrisation used to study the $3+2$ model in~\cite{Donini:2012tt}.

Firstly, we note that in the phenomenological approach of this work, the six active-sterile mixing strengths in the $3+2$ model can be written in terms of the elements of the light neutrino mass matrix $m_{\alpha\beta}^{\nu} \equiv \sum_{i} m_i U_{\alpha i}U_{\beta i}$ (depending on two light neutrino masses, three mixing angles, a Dirac CP phase and single Majorana phase for $m_{\text{light}} = 0$), the masses of the two HNLs $m_{N_1} = m_N$ and $m_{N_2} = m_N(1+r_\Delta)$ with $r_{\Delta} = \Delta m_N /m_N$, a single active-sterile mixing $|\Theta_{\alpha 1}|$ and CP phase $\phi_{\alpha 1}$, i.e.,
\begin{align}
	U_{\beta N_1} \approx |\Theta_{\alpha 1}|e^{i\phi_{\alpha 1}/2} y_{\alpha \beta}^{\alpha}\,,\quad 
	U_{\beta N_2} \approx 
	i|\Theta_{\alpha 1}|e^{i\phi_{\alpha 1}/2}
	\sqrt{\dfrac{(y_{\alpha\beta}^{\alpha})^2+x_{\beta\beta}^{\alpha}}{1+r_\Delta}}\,,
\end{align}
for $\beta = e,\mu,\tau$, where the factors $x_{\alpha\beta}^{\rho}$ and $y_{\alpha\beta}^{\rho}$ are functions of $|\Theta_{\alpha 1}|$ and $\phi_{\alpha 1}$ and are given in Eqs.~\eqref{eq:diag} and \eqref{eq:off-diag}, respectively. A useful choice for the comparison of $0\nu\beta\beta$ decay and direct searches at DUNE is $\alpha = e$.

In the scheme of~\cite{Donini:2012tt}, the neutrino mass matrix is diagonalised similarly (in the NO case with $m_1 = 0$) as
\begin{align}
   U^\dagger \mathcal{M}_{\nu} U^{*} 
   = \text{diag}(0,m_2,m_3,m_{N_1},m_{N_2})\,,\quad 
   U = \begin{pmatrix}
		U_{\nu\nu} & U_{\nu N} \\
		U_{N\nu}   & U_{NN}
   \end{pmatrix}\,,
\end{align}
where now the sub-blocks of the $5\times 5$ mixing matrix $U$ are parametrised as
\begin{align}
	U_{\nu\nu} &= 
	U_{\mathrm{PMNS}}
	\begin{pmatrix}
		1 & 0 \\
		0 & H
	\end{pmatrix}\,,\quad 
	U_{\nu N} = -i U_{\text{PMNS}}
	\begin{pmatrix}
		0 \\
		H m_l^{1/2} R^{T} m_h^{-1/2}
	\end{pmatrix}\,, \\
	U_{N\nu} &= 
	-i
	\begin{pmatrix}
		0 & \bar{H} m_h^{-1/2} R^* m_l^{1/2}
	\end{pmatrix}\,, \quad 
	U_{NN} = \bar{H} \,.
\end{align}
Here, $m_l$ and $m_h$ are the $2\times 2$ matrices,
\begin{align}
	m_l = 
	\begin{pmatrix}
		m_2 & 0   \\
		0   & m_3
	\end{pmatrix}, \quad 
	m_h = 
	\begin{pmatrix}
		m_{N_1} & 0 \\
		0       & m_{N_2}
\end{pmatrix}\,,
\end{align}
containing the masses of the light and heavy states, respectively. The $2\times 2$ matrix $R$ is described by a single complex angle,
\begin{align}
	R = 
	\begin{pmatrix}
		 \cos(\theta_{45}+i\gamma_{45}) & \sin(\theta_{45}+i\gamma_{45}) \\
		-\sin(\theta_{45}+i\gamma_{45}) & \cos(\theta_{45}+i\gamma_{45})
	\end{pmatrix}\,.
\end{align}
Finally, the $2\times 2$ matrices $H$ and $\bar{H}$ can be written in terms of the matrices above as
\begin{align}
	H       = \left[I_2 + m_l^{1/2} R^{T} m_h^{-1} R^* m_l^{1/2}\right]^{-1/2}, \quad 
	\bar{H} = \left[I_2 + m_h^{-1/2} R^* m_l R^{T} m_h^{-1/2}\right]^{-1/2}\,,
\end{align}
where $I_2$ is the $2\times 2$ identity matrix. The active-sterile mixing contained in $U_{\nu N}$ is therefore determined by the light neutrino masses, mixing angles and phases (a Dirac phase and single Majorana phase for $m_1 = 0$) contained in $m_l$ and $U_{\text{PMNS}}$, the HNL masses $m_{N_1} = m_N$ and $m_{N_2} = m_N(1+r_\Delta)$ and the two parameters $\theta_{45}$ and $\gamma_{45}$. 

The difference between the two parametrisations is the choice of $(|\Theta_{e1}|,\phi_{e1})$ or $(\theta_{45},\gamma_{45})$ as free parameters. The former, which has a direct physical interpretation, is useful for considering the experimental observation of an HNL that couples to a flavour $\alpha$. In that scenario, the active-sterile mixing $|\Theta_{\alpha 1}|^2$ and HNL mass $m_N$ are measured quantities, and all other active-sterile mixing strengths are determined for particular values of $\phi_{e1}$ and $r_{\Delta}$.


\bibliographystyle{JHEP}
\bibliography{references}
\end{document}